\documentclass[twocolumn]{aastex62}

\usepackage{graphicx}
\usepackage{amssymb}
\usepackage{aas_macros}
\usepackage{amsmath}

\newcommand{\teff}{\mbox{$T_{\rm eff}$}}
\newcommand{\logg}{\mbox{$\log g_\star$}}
\newcommand{\vsini}{\mbox{$v \sin I_\star$}}
\newcommand{\mictrb}{\mbox{$\xi_{\rm t}$}}
\newcommand{\mactrb}{\mbox{$v_{\rm mac}$}}

\newcommand{\kms}{\mbox{km\,s$^{-1}$}}
\newcommand{\ms}{\mbox{m\,s$^{-1}$}}

\graphicspath{{./}{figures/}}

\shorttitle{WASP-85\,A\,\lowercase{b}, WASP-116\,\lowercase{b}, and WASP-149\,\lowercase{b}}
\shortauthors{D. J. A. Brown et al.}

\begin{document}

\title{Three transiting planet discoveries from the Wide Angle Search for Planets: WASP-85\,A\,\lowercase{b}; WASP-116\,\lowercase{b}, and WASP-149\,\lowercase{b}\footnote{based on observations (under proposal 089.C-0151(A)) made using the HARPS high resolution {\'e}chelle spectrograph mounted on the ESO 3.6-m  at the ESO La Silla observatory, and the IO:O camera on the 2.0-m Liverpool Telescope under program PL12B13.}}

\correspondingauthor{D. J. A. Brown}
\email{d.j.a.brown@warwick.ac.uk}

\author[0000-0003-1098-2442]{D. J. A. Brown}
\affiliation{Department of Physics, University of Warwick, Coventry CV4 7AL, UK}
\affiliation{Centre for Exoplanets and Habitability, University of Warwick, Coventry CV4 7AL, UK}

\author{D. R. Anderson}
\affiliation{Astrophysics Group, School of Physical \& Geographical Sciences, Lennard-Jones Building, Keele University, Staffordshire, ST5 5BG, UK}

\author{A. P. Doyle}
\affiliation{Department of Physics, University of Warwick, Coventry CV4 7AL, UK}

\author{E. Gillen}
\affiliation{Astrophysics Group, Cavendish Laboratory, J.J. Thomson Avenue, Cambridge CB3 0HE, UK}

\author{P. F. L. Maxted}
\affiliation{Astrophysics Group, School of Physical \& Geographical Sciences, Lennard-Jones Building, Keele University, Staffordshire, ST5 5BG, UK}

\author{B. Smalley}
\affiliation{Astrophysics Group, School of Physical \& Geographical Sciences, Lennard-Jones Building, Keele University, Staffordshire, ST5 5BG, UK}

\author{J. McCormac}
\affiliation{Department of Physics, University of Warwick, Coventry CV4 7AL, UK}
\affiliation{Centre for Exoplanets and Habitability, University of Warwick, Coventry CV4 7AL, UK}

\author{J. M. Almenera}
\affiliation{Observatoire Astronomique de l'Universit{\'e} de Gen{\`e}ve, Chemin des Maillettes 51, 1290 Sauverny, Switzerland}

\author{J. Prieto-Arranz}
\affiliation{Instituto de Astrof\'isica de Canarias (IAC), V{\'i}a L{\'a}ctea s/n, 38205, La Laguna, Spain}
\affiliation{Departamento de Astrofísica, Universidad de La Laguna, Spain}

\author{M. Deleuil}
\affiliation{Aix Marseille Universit{\'e}, CNRS, LAM (Laboratoire d'Astrophysique de Marseille) UMR 7326, 13388, Marseille, France}

\author{R. F. D\'iaz}
\affiliation{Instituto de Astronom\'ia y F\'isica del Espacio (IAFE-CONICET), Buenos Aires, Argentina}

\author{E. Foxell}
\affiliation{Department of Physics, University of Warwick, Coventry CV4 7AL, UK}
\affiliation{Centre for Exoplanets and Habitability, University of Warwick, Coventry CV4 7AL, UK}

\author{G. H{\'e}brard}
\affiliation{Institut d'Astrophysique de Paris, UMR7095 CNRS, Universit{\'e} Pierre \& Marie Curie, 98bis boulevard Arago, 75014 Paris, France}
\affiliation{Observatoire de Haute Provence, CNRS/OAMP, 04870 St Michel l'Observatoire, France}

\author{M. Lendl}
\affiliation{Space Research Institute, Austrian Academy of Sciences, Schmiedlstr. 6, 8042, Graz, Austria}
\affiliation{Observatoire Astronomique de l'Universit{\'e} de Gen{\`e}ve, Chemin des Maillettes 51, 1290 Sauverny, Switzerland}

\author{L. Delrez}
\affiliation{Astrophysics Group, Cavendish Laboratory, J.J. Thomson Avenue, Cambridge CB3 0HE, UK}

\author{M. Gillon}
\affiliation{Astrobiology Research Unit, Universit{\'e} de Li{\`e}ge, All{\'e}e du 6 Ao{\^u}t, 4000 Li{\'e}ge (Sart-Tilman), Belgium}

\author{E. Jehin}
\affiliation{Space Sciences, Technologies and Astrophysics Research (STAR) Institute, Universit{\'e} de Li{\`e}ge, 19C All{\'e}e du 6 Ao{\^u}t, B-4000 Li{\'e}ge, Belgium}

\author{K. W. F. Lam}
\affiliation{Department of Physics, University of Warwick, Coventry CV4 7AL, UK}
\affiliation{Centre for Exoplanets and Habitability, University of Warwick, Coventry CV4 7AL, UK}
\affiliation{Zentrium f\"ur Astronomie und Astrophysik, Technische Universit\"at Berlin, Hardenbergstr. 36, 10623 Berlin, Germany}

\author{A. H. M. J. Triaud}
\affiliation{Institute of Astronomy, Madingley Road, Cambridge, CB3 0HA, UK}
\affiliation{School of Physics \& Astronomy, University of Birmingham, Edgbaston, Birmingham B15 2TT, UK}

\author{O. D. Turner}
\affiliation{Astrophysics Group, School of Physical \& Geographical Sciences, Lennard-Jones Building, Keele University, Staffordshire, ST5 5BG, UK}
\affiliation{Observatoire Astronomique de l'Universit{\'e} de Gen{\`e}ve, Chemin des Maillettes 51, 1290 Sauverny, Switzerland}

\author{D. J. Armstrong}
\affiliation{Department of Physics, University of Warwick, Coventry CV4 7AL, UK}
\affiliation{Centre for Exoplanets and Habitability, University of Warwick, Coventry CV4 7AL, UK}

\author{F. Bouchy}
\affiliation{Observatoire Astronomique de l'Universit{\'e} de Gen{\`e}ve, Chemin des Maillettes 51, 1290 Sauverny, Switzerland}

\author{A. Collier Cameron}
\affiliation{SUPA, School of Physics and Astronomy, University of St Andrews, North Haugh, St Andrews, Fife KY16 9SS, UK}

\author{D. Pollacco}
\affiliation{Department of Physics, University of Warwick, Coventry CV4 7AL, UK}
\affiliation{Centre for Exoplanets and Habitability, University of Warwick, Coventry CV4 7AL, UK}

\author{F. Faedi}
\affiliation{Department of Physics, University of Warwick, Coventry CV4 7AL, UK}

\author{Y. G{\'o}mez Maqueo Chew}
\affiliation{Instituto de Astronom\'ia, Universidad Nacional Aut\'onoma de M\'exico, Circuito Exterior S/N, Ciudad Universitaria, Coyoac\'an, 04510, Ciudad de M\'exico, M\'exico}

\author{L. Hebb}
\affiliation{Hobart and William Smith Colleges, Department of Physics, Geneva, NY 14456, USA}

\author{C. Hellier}
\affiliation{Astrophysics Group, School of Physical \& Geographical Sciences, Lennard-Jones Building, Keele University, Staffordshire, ST5 5BG, UK}

\author{M. Neveu-VanMalle}
\affiliation{Astrophysics Group, Cavendish Laboratory, J.J. Thomson Avenue, Cambridge CB3 0HE, UK}
\affiliation{Observatoire Astronomique de l'Universit{\'e} de Gen{\`e}ve, Chemin des Maillettes 51, 1290 Sauverny, Switzerland}

\author{E. Pall{\'e}}
\affiliation{Instituto de Astrof\'isica de Canarias (IAC), V{\'i}a L{\'a}ctea s/n, 38205, La Laguna, Spain}
\affiliation{Departamento de Astrofísica, Universidad de La Laguna, Spain}

\author{D. Queloz}
\affiliation{Astrophysics Group, Cavendish Laboratory, J.J. Thomson Avenue, Cambridge CB3 0HE, UK}
\affiliation{Observatoire Astronomique de l'Universit{\'e} de Gen{\`e}ve, Chemin des Maillettes 51, 1290 Sauverny, Switzerland}

\author{D. Segransan}
\affiliation{Observatoire Astronomique de l'Universit{\'e} de Gen{\`e}ve, Chemin des Maillettes 51, 1290 Sauverny, Switzerland}

\author{S. Udry}
\affiliation{Observatoire Astronomique de l'Universit{\'e} de Gen{\`e}ve, Chemin des Maillettes 51, 1290 Sauverny, Switzerland}

\author{R. G. West}
\affiliation{Department of Physics, University of Warwick, Coventry CV4 7AL, UK}
\affiliation{Centre for Exoplanets and Habitability, University of Warwick, Coventry CV4 7AL, UK}



\begin{abstract}

We report the discovery of three new transiting planets by the Wide Angle Search for Planets: WASP-85\,A\,b, WASP-116\,b, and WASP-149\,b. Through combined analysis of photometric lightcurves and radial velocity observations, we determine key orbital and physical parameters for these planetary systems. WASP-85\,b orbits its host star every $2.66$\,days, and has a mass of $1.25$\,$M_{\rm Jup}$ and a radius of $1.25$\,$R_{\rm Jup}$. The host star is of G5 spectral type, with magnitude $V=11.2$, and lies $141$\,pc distant. The system has a K-dwarf binary companion, WASP-85\,B, at a separation of $\approx1.5$\,\arcsec. The close proximity of this companion leads to contamination of our photometry, decreasing the apparent transit depth that we account for during our analysis. We find a stellar effective temperature of \teff$=5685$\,K, and super-solar metallicity (${\rm [Fe/H]}=0.08$\,dex) from analysis of spectroscopic observations of the host star, but our MCMC fit to the dilution-corrected photometry suggests a significantly hotter star of $6150$\,K. We find a long-term trend in the binary position angle, indicating a misalignment between the binary and planetary orbital planes. Analysis of the Ca\,{\sc ii}\,H+K lines shows strong emission that implies that both binary components are strongly active. WASP-116\,b is a warm, mildly inflated super-Saturn, with a mass of $0.59$\,$M_{\rm Jup}$ and a radius of $1.43$\,$R_{\rm Jup}$. It was discovered orbiting a metal-poor (${\rm [Fe/H]}=-0.28$\,dex), cool (\teff$=5950$\,K) G0 dwarf every $6.61$\,days. WASP-149\,b is a typical hot Jupiter, orbiting a G6 dwarf with a period of $1.33$\,days. The planet has a mass and radius of $1.05$\,$M_{\rm Jup}$ and $1.29$\,$R_{\rm Jup}$, respectively. The stellar host has an effective temperature of \teff$=5750$\,K and has a 
metallicity of ${\rm [Fe/H]}=0.16$\,dex. WASP photometry of the system is contaminated by a nearby star, but our follow-up photometry are unaffected; we therefore corrected the depth of the WASP transits using the measured dilution. WASP-149 lies inside the `Neptune desert' identified in the planetary mass-period plane by \citet{2016AA...589A..75M}.

WASP and \textit{K2} observations of the WASP-85 system show clear variability, indicative of rotational modulation caused by stellar activity. We model the modulation visible in the \textit{K2} lightcurve of WASP-85 using a simple three-spot model consisting of two large spots on WASP-85\,A, and one large spot on WASP-85\,B, finding rotation periods of $13.1\pm0.1$\,days for WASP-85\,A and $7.5\pm0.03$\,days for WASP-85\,B. We estimate stellar inclinations of $I_{\rm A}=66.8^o\pm0.7$ and $I_{\rm B}=39.7^o\pm0.2$, and constrain the obliquity of WASP-85\,A\,b to be $\psi<27^o$. We therefore conclude that WASP-85\,A\,b is very likely to be aligned.

\end{abstract}



\section{Introduction}
\label{sec:intro}
Though the science of transiting exoplanets has been pushed forward by space-based missions such as CoRoT \citep{2006cosp...36.3749B}, \textit{Kepler} \citep{2010Sci...327..977B}, \textit{K2} \citep{2014PASP..126..398H}, \textit{TESS} \citep{2015JATIS...1a4003R}, and the upcoming PLATO \citep{2014ExA....38..249R,2016AN....337..961R}, the contribution of ground-based surveys cannot be understated. Projects such as HATnet \citep{2002PASP..114..974B}, TrES \citep{2004ApJ...613L.153A}, XO Project \citep{2005PASP..117..783M}, WASP \citep{2006PASP..118.1407P}), KELT \citep{2007PASP..119..923P}, and QES \citep{2013AcA....63..465A} have collectively discovered large numbers of exoplanets orbiting bright stars ($8.5 \lesssim V \lesssim 12.5$); these are particularly useful, as their brightness opens up the possibility of detailed follow-up studies to determine planetary masses, spin-orbit alignment angles, and atmospheric compositions. WASP (the Wide Angle Search for Planets) is by far the most successful of these surveys, with more than 150 published planet discoveries to date, but has recently concluded science operations in both the Northern and Southern hemispheres.

Owing to the limitations imposed by observing through the Earth's atmosphere, the focus for the majority of ground-based surveys has been deep transit events rather than the more shallow transits that tend to be the objective of satellite missions. Deep transits implies either small planets around small stars (as targeted by the MEarth \citep{2012AJ....144..145B}, TRAPPIST \citep{2013EPJWC..4703001G} and SPECULOOS \citep{2017haex.bookE.130B} surveys), or giant planets around solar-type stars. WASP focused on the latter, which continue to challenge formation and evolution theories. A significant diversity of physical properties has been observed, particular for planetary radius where it seems that a broad range of values is possible for the same planetary mass. In part this is due to different environmental conditions; bloated planets (e.g. WASP-12, \citealt{2009ApJ...693.1920H}; WASP-21, \citealt{2010AA...519A..98B}; WASP-54, \citealt{2013AA...551A..73F}; WASP-102, \citealt{2016arXiv160804225F}; WASP-127, \citealt{2017AA...599A...3L}), for example, tend to be preferentially found on short-period orbits or around particularly active stars, both of which produce a strong irradiation environment that can lead to an inflated planetary radius \citep{1996ApJ...459L..35G, 2011ApJS..197...12D}. Internal mechanisms, for example atmospheric circulation \citep[e.g.][]{2002AA...385..166S}, enhanced atmospheric opacities \citep{2007ApJ...668L.171B}, Ohmic heating \citep{2010ApJ...714L.238B, 2012ApJ...757...47H, 2013ApJ...763...13W}, or tidal energy dissipation \citep[e.g][]{2003ApJ...592..555B, 2009ApJ...702.1413M, 2010ApJ...713..751I} can also play a role in inflating a planet's radius. However, there is no single mechanism that can explain the full diversity of giant planet radii. Moreover, planets with different masses respond differently to many of the influencing factors \citep{2012AA...540A..99E}, such that exploring a variety of planetary mass regimes is important to a full understanding of planet formation and migration. If we are to continue exploring the conditions under which planets form and evolve, then it is vital that we continue to expand the sample of well-characterized, transiting planets around bright stars, and that we continue to explore underpopulated regions of planetary parameter space. Ground-based transiting surveys are invaluable in this endeavour.

A study of F, G, and K stars in the Sloan Digital Sky Survey (SDSS) showed that approximately $43\pm2$\,percent of solar-type stars have binary companions with periods of $<1000$\,days \citep{2014ApJ...788L..37G}. This supports earlier results by \citet{2010ApJS..190....1R} ($46\pm2$\,percent solar type stars in binaries), \citet{1991AA...248..485D}, and \citet{1976ApJS...30..273A}, among others. However, the binary fraction seems to be lower for exoplanet systems \citep{2012AA...542A..92R} , indicating a suppression of planet formation \citep{2014ApJ...791..111W}. There is a strong selection effect acting against binary systems in planet search programs, that \citeauthor{2012AA...542A..92R} acknowledge and \citeauthor{2014ApJ...791..111W} correct for. The presence of a companion star to the planet host introduces the problem of light from said star contaminating either the photometric observations (diluting the transit depth), spectroscopic observations (introducing a second set of spectral lines, and thus a second cross-correlation function peak), or both. The level of contamination depends on several factors - the resolution of the instrument, aperture size, fibre diameter, seeing, the separation of the stars, and the magnitude difference between the stellar components. These factors vary considerably from system to system, and the presence of a binary companion to an exoplanet candidate star tends to reduce the likelihood of that candidate becoming the target of follow-up observations. Nevertheless, there are several examples of hot Jupiter exoplanets in S-type orbits around binary stars, including WASP-70\,A\,b \citep{2014MNRAS.445.1114A}; WASP-77\,A\,b \citep{2013PASP..125...48M}; WASP-94\,A\,b and B\,b \citep{2014AA...572A..49N}; Kelt-2\,A\,b \citep{2012ApJ...756L..39B}, and Kepler-14\,A\,b \citep{2011ApJS..197....3B}.

In this paper we present the discovery of three new transiting exoplanets by WASP: WASP-85\,A\,b, WASP-116\,b, and WASP-149\,b. WASP-85\,A\,b is a hot Jupiter orbiting the brighter, solar-type component of a close visual binary, BD+07$^{\circ}$2474, that has an orbital period of $\sim3000$\,years. The binary companion is cooler than the host star but has a similar magnitude in the V-band. This companion contaminates both the photometric and spectroscopic data for the system, which thus require additional analysis compared to the standard WASP procedure. WASP-116\,b is a warm, mildly inflated super-Saturn orbiting an extremely metal-poor early G-dwarf, while WASP-149\,b is a hot Jupiter orbiting a metal-rich late G-dwarf.  

In Section\,\ref{sec:obs} we describe the photometric and spectroscopic observations of these systems. We then examine the characteristics of the host stars in Section\,\ref{sec:star}, and inSection\,\ref{sec:orbit} we discuss the methods by which we derive system parameters. Results for the three systems are presented in Sections\,\ref{sec:85model}, \ref{sec:116model}, and \ref{sec:149model}. We discuss various interesting aspects of our systems in Section\,\ref{sec:discuss}, and conclude by summarising our results in Section\,\ref{sec:summary}.

\section{Observations}
\label{sec:obs}
For a detailed account of the WASP telescopes, observing strategy, data reduction, and candidate identification and selection procedures, see \citet{2006PASP..118.1407P, 2008MNRAS.385.1576P} and \citet{cameron2007}.

\subsection{WASP-85}
\label{sec:85wasp}
BD+07$^{\circ}$2474 lies near the celestial equator, and was thus observed by both WASP (located at the Observatorio del Roque de los Muchachos on La Palma, Spain) and WASP-South (located at the South African Astronomical Observatory near Sutherland, South Africa). These observations resulted in $20936$ data, spanning the period 2008-02-05 to 2011-03-29.

The system was first identified as a planet candidate from WASP data in 2008. Initial analysis revealed transit like features with an apparent period of $\sim1.59$\,days. This signal was subsequently identified in both the SuperWASP and WASP-South photometry independently, and the star was selected for photometric and spectroscopic follow-up observations. Subsequent to these observations, the secondary peak in the periodogram, at $\sim2.66$\,days, was found to be the correct period.

The radius of the synthetic aperture ($48$\,\arcsec) used to extract the flux of BD+07$^{\circ}$2474 is much greater than the maximum binary separation ($1.8$\,\arcsec; see Section\,\ref{sec:binary}), such that the WASP lightcurve includes flux contributions from both stellar components of the binary. The additional light from the companion (sometimes referred to as 'third light') dilutes the transits in the WASP lightcurve, making them appear more shallow.

\subsubsection{Spectroscopic follow-up}
\label{sec:85spec}
Initial spectroscopic reconnaissance was carried out between 2008 and 2010 using the high efficiency mode (HE mode; $R=40,000$) of the SOPHIE spectrograph \citep{2008SPIE.7014E..0JP, 2009AA...505..853B} mounted on the $1.93$-m telescope of the Observatoire de Haute-Provence (OHP), resulting in the acquisition of $11$ spectra. Radial velocities (RVs) were derived through cross-correlation with a spectral mask suitable for a star of G2 spectral type. The separation of the binary components is smaller than the fibre diameter of SOPHIE ($3$\,\arcsec), and thus the RVs obtained from this instrument are contaminated by the contribution of the companion. These data show a sinusoidal variation in radial velocity (RV) with a period of $\sim2.66$\,days, in disagreement with the period initially identified in the WASP lightcurves.

Further spectroscopic observations were made using the fibre-fed CORALIE spectrograph ($R=50,000$) at the Euler-Swiss telescope at ESO's La Silla observatory, and using HARPS ($R=100,000$) at the ESO $3.6$\,m telescope, also at La Silla. A total of $31$ observations were made using CORALIE between 2009 January 03 and 2014 June 24. For details of the instrument and data reduction procedure, see \citet{queloz2000proc} and \citet{2008ApJ...675L.113W}. RVs were derived using cross-correlation with a spectral mask suitable for a G2-type star, and confirmed the $\sim2.66$\,day period as the correct one. Unfortunately, while the CORALIE guiding camera is able to identify the presence of both stellar components, the aperture of the spectrograph ($2$\,\arcsec) is insufficiently small to access a single stellar component at a time. The CORALIE spectra are therefore also contaminated by light from the companion star. The level of contamination is seeing dependent owing to the close similarity of the aperture size and binary separation, rendering a correction to the RVs both uncertain and difficult to make.

Eight observations of the brighter binary component (hereafter WASP-85\,A), and five observations of the fainter companion star (hereafter WASP-85\,B) were made using HARPS. The small spectrograph aperture ($1$\arcsec\ diameter) and good seeing allowed the components to be observed separately, with little contamination from the other star. RVs were extracted in the same manner as for the CORALIE observations. 

\begin{figure}
\gridline{\fig{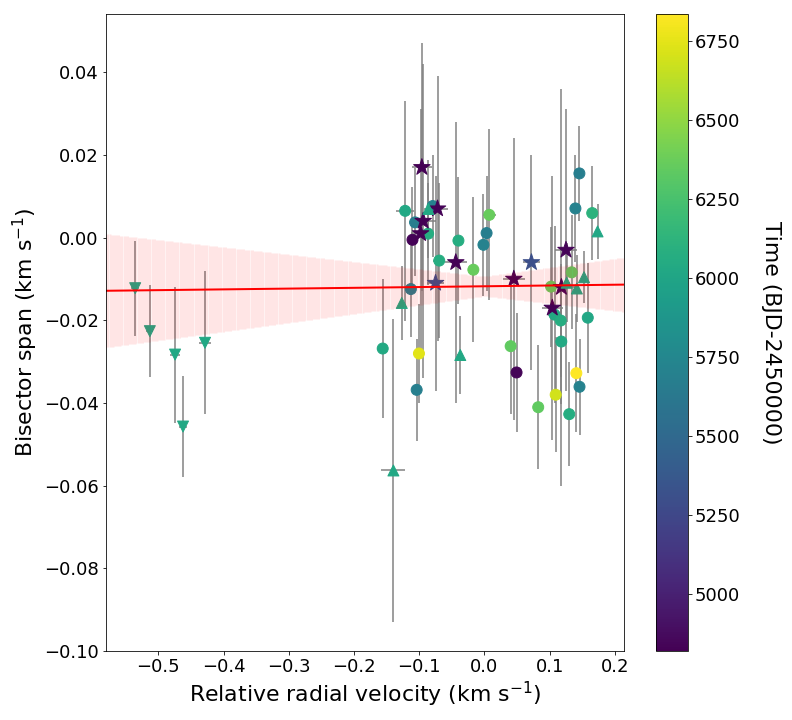}{0.48\textwidth}{}}
\gridline{\fig{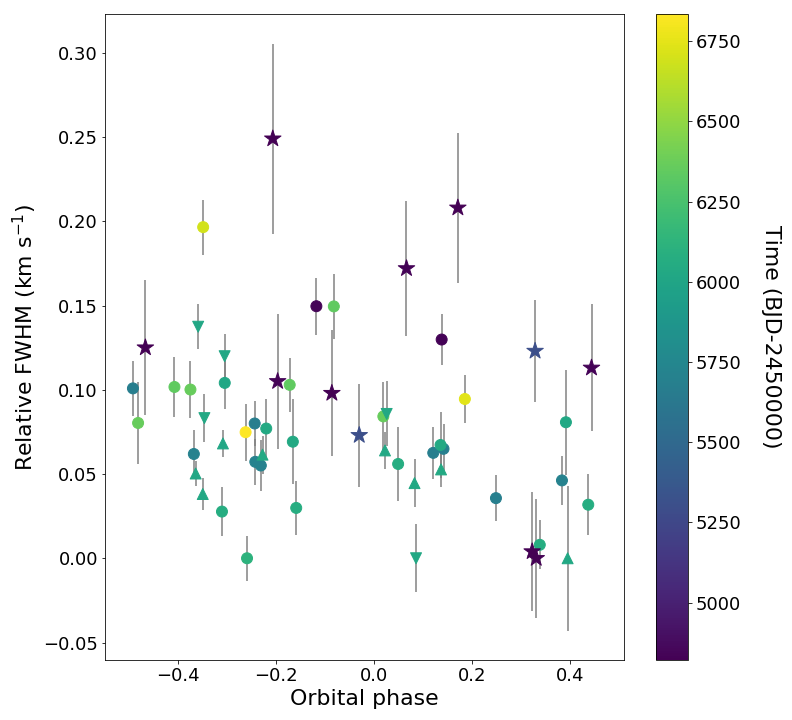}{0.48\textwidth}{}}
	\caption{\textit{Upper panel: }Radial velocity bisector span plotted as a function of relative radial velocity for the WASP-85 system. The uncertainties in the bisector measurements are taken to be $2.0\times\sigma_{\rm RV}$. CORALIE data are denoted by circles, SOPHIE data by stars. HARPS data are denoted by upwards triangles for WASP-85\,A, and by downwards triangles for WASP-85\,B. The shading indicates the observation date. The solid line represents an odr linear fit to the data for WASP-85\,A, with the shaded region indicating the $1\sigma$ limits to the fit. No statistically significant correlation is found. \textit{Lower panel: } Radial velocity full width at half maximum (FWHM) as a function of orbital phase. The uncertainties in the FWHMs are taken to be $2.35\times\sigma_{\rm RV}$. The FWHM values have been offset from the minimum value for each data set to allow comparison. Legend as for the upper panel. There is no clear variation with orbital phase.}
	\label{fig:bisector85}
\end{figure}

We ruled out possible false positive sources of the RV variation (e.g. background eclipsing binaries or stellar activity) by examining both the bisector spans and full widths at half maximum (FWHM) of our cross-correlation functions (CCFs; Figure\,\ref{fig:bisector85}). Typically, an anti-correlation between the bisector and RV data indicates that stellar activity is producing a false positive detection through distortions of the stellar absorption lines \citep{2001AA...379..279Q,2013AA...557A..93F}; we tested for this by conducting both a Spearman Rank Correlation test and a Pearson Rank Correlation test on the relative RV and bisector data for WASP-85\,A. We obtain a Spearman Rank Correlation coefficient of $-0.11$ with a p-value of $0.47$, and a Pearson Rank Correlation coefficient of $-0.13$ with a p-value of $0.35$, insufficient to reject the null hypothesis of no correlation. As a further test, we carried out a linear fit to the data using orthogonal distance regression (odr). This returned a gradient of $0.002$ with a significance of less than $0.5\sigma$. We conclude that there no statistically significant correlation between RV and bisector for WASP-85\,A.

We also plot the FWHM of our radial velocity observations as a function of orbital phase, checking for variations in phase that might indicate a false positive \citep{2011IAUS..276..549S}. None are seen (see Figure\,\ref{fig:bisector85}). We confirm which binary component is the planet host by checking for phasing of the HARPS radial velocity observations. The observations of WASP-85\,A show variations in phase with both the photometry, and with the CORALIE and SOPHIE observations, while the observations of WASP-85\,B show variations that are not correlated with the phase determined from the lightcurve, and show no other significant periodicity.

We list our RV data in Table\,\ref{tab:85RVs}, along with the bisector spans, which measure the asymmetry of the cross-correlation functions (CCFs), and the FWHMs of the CCFs. Conservatively, the uncertainties on the bisectors were taken to be twice the uncertainty on the RV, while those on the FWHMs were taken to be $2.35$ times the RV uncertainty. There is no indication of any time-dependent variation in our RV data.

\begin{deluxetable*}{lllllll}
\tabletypesize{\scriptsize}
\caption{Radial velocity data of WASP-85\,A and WASP-85\,B, obtained using HARPS, and of the two components combined, obtained using SOPHIE and CORALIE. \label{tab:85RVs}}
\centering
\tablehead{
		\colhead{${\rm BJD}_{\rm TDB}$} & \colhead{RV} & \colhead{$\sigma_{\rm RV}$} & \colhead{Bisector} & \colhead{$\sigma_{\rm bis}$} & \colhead{FWHM} & \colhead{$\sigma_{\rm FWHM}$} \\
		\colhead{$-2450000$} & \colhead{} & \colhead{} & \colhead{} & \colhead{} & \colhead{} & \colhead{} \\
		\colhead{(days)} & \colhead{(\kms)} & \colhead{(\kms)} & \colhead{(\kms)} & \colhead{(\kms)} & \colhead{(\kms)} & \colhead{(\kms)}
}
\startdata
		\multicolumn{7}{l}{\textit{SOPHIE}} \\
		4820.64118 & 13.579 & 0.024 & -0.012 & 0.048 & 10.100 & 0.056 \\
		4821.64438 & 13.367 & 0.019 & 0.004 & 0.038 & 10.059 & 0.045 \\
		4822.60638 & 13.506 & 0.017 & -0.010 & 0.034 & 9.976 & 0.040 \\
		4823.61708 & 13.565 & 0.016 & -0.017 & 0.032 & 9.949 & 0.038 \\
		4824.70288 & 13.365 & 0.015 & 0.017 & 0.030 & 9.855 & 0.035 \\
		4824.72498 & 13.363 & 0.015 & 0.001 & 0.030 & 9.851 & 0.035 \\
		4834.64438 & 13.417 & 0.017 & -0.006 & 0.034 & 10.023 & 0.040 \\
		4835.64718 & 13.389 & 0.016 & 0.007 & 0.032 & 9.964 & 0.038 \\
		4836.60168 & 13.586 & 0.017 & -0.003 & 0.034 & 9.956 & 0.040 \\
		5304.44298 & 13.533 & 0.013 & -0.006 & 0.026 & 9.924 & 0.031 \\
		5305.39608 & 13.386 & 0.013 & -0.011 & 0.026 & 9.974 & 0.031 \\ \\
		\multicolumn{7}{l}{\textit{CORALIE}} \\
		4834.834967 & 13.42322 & 0.00643 & -0.00053 & 0.01286 & 8.91982 & 0.01511 \\
		4836.811531 & 13.58281 & 0.00719 & -0.03262 & 0.01438 & 8.93960 & 0.01690 \\
		5675.670857 & 13.67921 & 0.00577 & 0.01552 & 0.01154 & 8.87000 & 0.01356 \\
		5676.638783 & 13.42716 & 0.00652 & 0.00368 & 0.01304 & 8.85258 & 0.01532 \\
		5677.666565 & 13.53687 & 0.00696 & 0.00109 & 0.01382 & 8.89084 & 0.01636 \\
		5679.634306 & 13.42072 & 0.00577 & -0.01245 & 0.01154 & 8.82570 & 0.01356 \\
		5684.661489 & 13.42964 & 0.00617 & -0.03682 & 0.01234 & 8.85497 & 0.01450 \\
		5706.548896 & 13.45416 & 0.00613 & 0.00763 & 0.01226 & 8.83622 & 0.01441 \\
		5707.572297 & 13.67309 & 0.00644 & 0.00705 & 0.01288 & 8.84507 & 0.01513 \\
		5712.520574 & 13.53203 & 0.00614 & -0.00174 & 0.01228 & 8.85190 & 0.01443 \\
		5715.509353 & 13.67937 & 0.00582 & -0.03608 & 0.01164 & 8.84725 & 0.01368 \\
		6020.746523 & 13.69205 & 0.00663 & -0.01937 & 0.01326 & 8.89409 & 0.01558 \\
		6030.562877 & 13.41197 & 0.01331 & 0.00647 & 0.02662 & 8.87074 & 0.03128 \\
		6031.594135 & 13.65143 & 0.00757 & -0.02514 & 0.01514 & 8.86701 & 0.01779 \\
		6031.738294 & 13.64104 & 0.01067 & -0.01863 & 0.02134 & 8.85919 & 0.02507 \\
		6032.538879 & 13.37756 & 0.00838 & -0.02683 & 0.01676 & 8.85724 & 0.01969 \\
		6067.600608 & 13.44722 & 0.00618 & 0.00095 & 0.01236 & 8.79800 & 0.01452 \\
		6068.533616 & 13.66361 & 0.00625 & -0.04270 & 0.01250 & 8.81772 & 0.01469 \\
		6069.488518 & 13.46423 & 0.00933 & -0.00556 & 0.01866 & 8.84598 & 0.02193 \\
		6070.519073 & 13.49368 & 0.00767 & -0.00073 & 0.01534 & 8.82180 & 0.01802 \\
		6071.591742 & 13.65052 & 0.00686 & -0.02006 & 0.01372 & 8.81989 & 0.01612 \\
		6116.471681 & 13.69886 & 0.00569 & 0.00593 & 0.01138 & 8.79002 & 0.01337 \\
		6339.779756 & 13.66741 & 0.00689 & -0.00840 & 0.01378 & 8.89293 & 0.01619 \\
		6341.810736 & 13.61604 & 0.00753 & -0.04102 & 0.01506 & 8.89167 & 0.01770 \\
		6342.674625 & 13.57377 & 0.00815 & -0.02626 & 0.01630 & 8.93949 & 0.01915 \\
		6362.859062 & 13.54098 & 0.01032 & 0.00550 & 0.02064 & 8.87031 & 0.02425 \\
		6365.798530 & 13.63578 & 0.00726 & -0.01186 & 0.01452 & 8.89014 & 0.01706 \\
		6366.844086 & 13.51656 & 0.00874 & -0.00777 & 0.01748 & 8.87421 & 0.02054 \\
		6697.826971 & 13.64283 & 0.00695 & -0.03800 & 0.01390 & 8.98653 & 0.01633 \\
		6741.736317 & 13.43313 & 0.00603 & -0.02803 & 0.01206 & 8.88454 & 0.01417 \\
		6833.496459 & 13.67453 & 0.00717 & -0.03279 & 0.01434 & 8.86483 & 0.01685 \\ \\
		\multicolumn{7}{l}{\textit{HARPS star A}} \\
		6020.588912 & 13.74735 & 0.00316 & -0.00958 & 0.00632 & 7.59818 & 0.00743 \\
		6020.736685 & 13.76810 & 0.00337 & 0.00149 & 0.00674 & 7.61603 & 0.00792 \\
		6028.594310 & 13.73631 & 0.00405 & -0.01231 & 0.00810 & 7.58601 & 0.00952 \\
		6029.582170 & 13.55755 & 0.00467 & -0.02840 & 0.00934 & 7.61188 & 0.01097 \\
		6029.742186 & 13.50857 & 0.00598 & 0.00692 & 0.01196 & 7.59248 & 0.01405 \\
		6030.572124 & 13.45497 & 0.01833 & -0.05632 & 0.03666 & 7.54788 & 0.04308 \\
		6031.573927 & 13.72069 & 0.00478 & -0.01092 & 0.00956 & 7.60938 & 0.01123 \\
		6032.541680 & 13.46784 & 0.00449 & -0.01581 & 0.00898 & 7.60061 & 0.01055 \\ \\
		\multicolumn{7}{l}{\textit{HARPS star B}} \\
		6020.602442 & 13.07700 & 0.00573 & -0.01224 & 0.01146 & 7.90081 & 0.01347 \\
		6020.745898 & 13.09952 & 0.00558 & -0.02263 & 0.01106 & 7.88332 & 0.01311 \\
		6028.603233 & 13.15039 & 0.00614 & -0.04569 & 0.01228 & 7.84657 & 0.01443 \\
		6029.591232 & 13.13818 & 0.00822 & -0.02829 & 0.01644 & 7.84906 & 0.01932 \\
		6029.750461 & 13.18411 & 0.00863 & -0.02544 & 0.01726 & 7.76341 & 0.02028 \\
\enddata
\end{deluxetable*}

\subsubsection{Photometric follow-up}
\label{sec:85phot}
Follow-up photometric observations to solidify the planet's ephemeris, and check for transit timing shifts, were carried out using the James Gregory Telescope (JGT) at the University of St Andrews Observatory, the robotic TRAnsiting Planets and PlanetesImals Small Telescope (TRAPPIST; \citealt{2011Msngr.145....2J}) at La Silla, the Euler-Swiss telescope using EulerCam \citep{2012AA...544A..72L}, also at La Silla, and the Liverpool Telescope (LT; \citealt{2004SPIE.5489..679S}) on La Palma using the IO:O instrument. These observations are summarised in Table\,\ref{tab:phot}. The resolutions and pixel scales of the telescopes and instruments used for photometric follow-up were insufficient to distinguish the companion star. All lightcurves are therefore a blended combination of light from the two stellar components, and exhibit diluted transit depths.

\subsubsection{\textit{K2} observations}
\label{sec:85K2}
In the course of our follow-up campaign for the WASP-85 system, the first set of proposed \textit{K2} field coordinates were announced. Cross-checking the coordinates of WASP-85 with these fields showed that the system would be visible to \textit{K2} during Campaign\,1, and would in fact fall on silicon. We submitted a proposal\footnote{\textit{K2} proposal GO1041} to acquire observations of WASP-85; the system was subsequently selected for observation and assigned the EPIC identification number $201862715$. Four other \textit{K2} Campaign\,1 proposals that included WASP-85 were also selected.

\begin{figure}
	\includegraphics[width=0.48\textwidth]{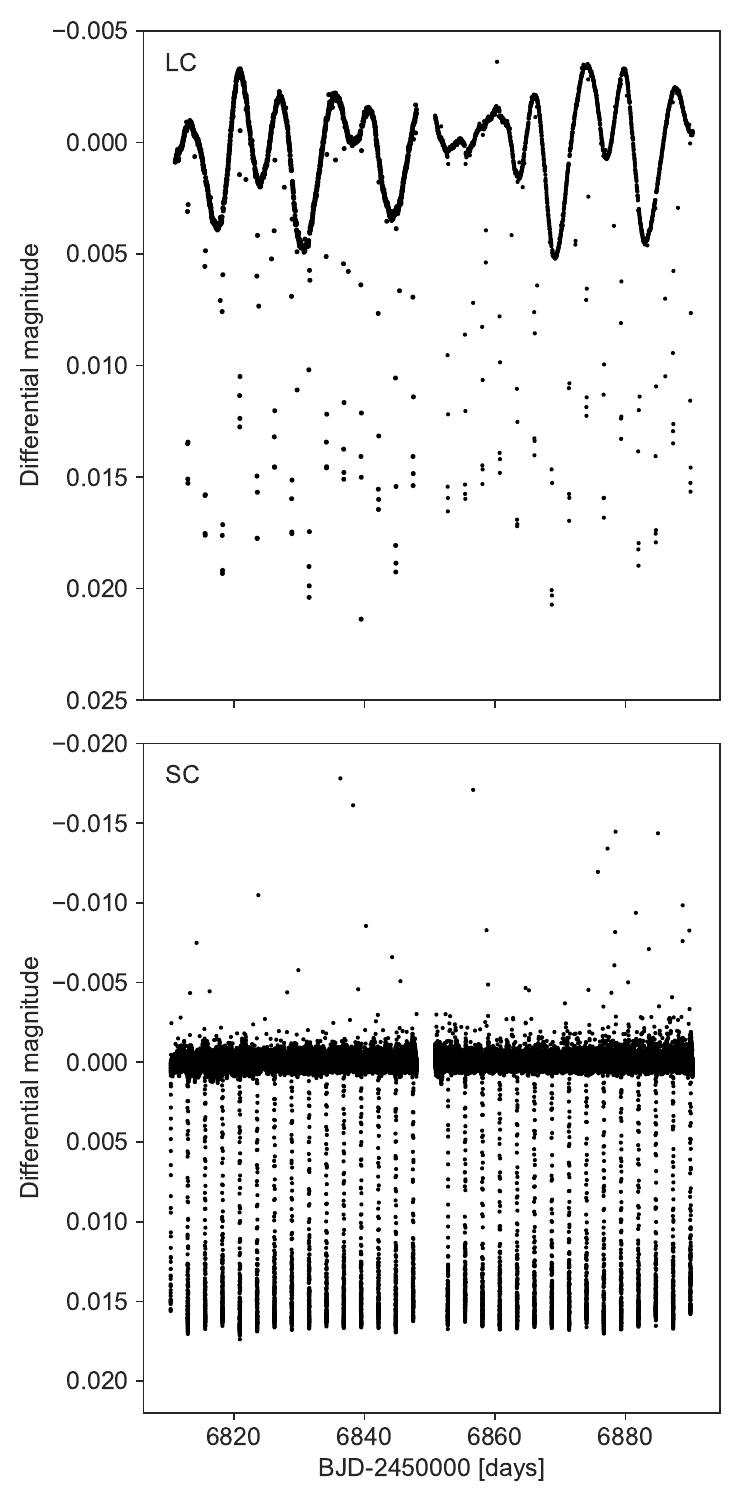}
	\caption{\textit{Upper panel: }\textit{K2} long cadence photometry of EPIC201862715 / BD+07$^{\circ}$2474 / WASP-85, as produced by the \textsc{everest} pipeline. The planetary transit signatures are clearly visible, as is variation arising from stellar activity. This stellar activity signal is of lower amplitude than the planetary transits.
	\textit{Lower panel: }The \textit{K2} short cadence lightcurve, as produced using the SFF procedure detailed in \citet{2016AJ....151..150M}.}
	\label{fig:K2}
\end{figure}

Pointing drift is the strongest error source in \textit{K2} data \citep{2014PASP..126..948V}, and several pipelines exist to produce corrected lightcurves \citep[e.g][]{2014PASP..126..948V, 2015AA...579A..19A,2016MNRAS.459.2408A, 2016AJ....152..100L}. We use the short cadence lightcurve produced by \citet{2016AJ....151..150M}\footnote{Note that these authors reference an earlier version of this work in their analysis} for our analysis, as shown in Figure\,\ref{fig:K2}. The transit depth is approximately $0.016$\,magnitudes, less than the transit depth observed in the WASP photometry. Table\,\ref{tab:results}. 
This is to be expected given the large pixel scale of the Kepler spacecraft's CCD ($3.98\arcsec$ per pixel) and its point-spread function (FWHM of $6\arcsec$), which mean that the stellar components of the system will be blended together (as is the case for our other photometry). The depth discrepancy is as expected given the Kepler passband, the peak transmission of which roughly coincides with the peak transmission of the Johnson\,R filter \citep{2009IAUS..253..121R}.

\subsection{WASP-116}
\label{sec:116obs}
WASP-116 also lies close to the celestial equator, and like WASP-85\,A\,B was observed by both SuperWASP and WASP-South. $22051$ photometric data were obtained between 2008-07-30 and 2010-12-27. A candidate planetary transit signal was first identified in a combined analysis of both sets of photometry, and the system was selected for photometric and spectroscopic follow-up to confirm the presence of an orbiting body, and characterise the signal.

\subsubsection{Spectroscopic follow-up}
\label{sec:116spec}
Spectroscopic follow-up of the candidate planet was initiated in 2013 as part of the long-term WASP follow-up program using SOPHIE. The first set of data confirmed the $6.6$\,day period seen in the WASP photometry, showing clear sinusoidal variation on that period, and with the predicted phase. Further observations were made using SOPHIE to constrain both the RV semi-amplitude, thereby allowing for rejection of false positive scenarios, and the orbital eccentricity. All SOPHIE data were obtained in HE mode. Several observations were also made with CORALIE. The optical fibre of CORALIE was replaced in 2014 November, leading to an offset in the instrument's zero point. One of our RV data for WASP-116 was obtained after the fibre replacement; this is clearly indicated in Table\,\ref{tab:116RVs}, and was excluded from our analysis.

All SOPHIE and CORALIE data were reduced using their respective standard data reduction pipelines. Our spectroscopic observations show that the candidate planet hosting star is a single point source (within the resolution of the instruments). We list our RV data, bisector spans, and CCF FWHMs for WASP-116 in Table\,\ref{tab:116RVs}. We note that one of the SOPHIE data, marked with $^\ast$ in Table\,\ref{tab:116RVs}, has an uncertainty $1.4$ times the standard deviation of the SOPHIE dataset. We therefore exclude this datum from our analysis.

Both the Spearman Rank Correlation and Pearson Rank Correlation tests return a correlation coefficient of $-0.14$ with a p-value of $0.39$. A linear, odr fit to the data for WASP-116 returned a gradient of $-0.15$ that is significant at the $1.5\sigma$ level. These tests suggest a possible correlation between the bisector and RVs, but the evidence is marginal and insufficient to reject our null hypothesis that the two parameters are uncorrelated. We therefore proceed under the assumption that the signal we have detected is due to a planetary transit.

\begin{figure}
\gridline{\fig{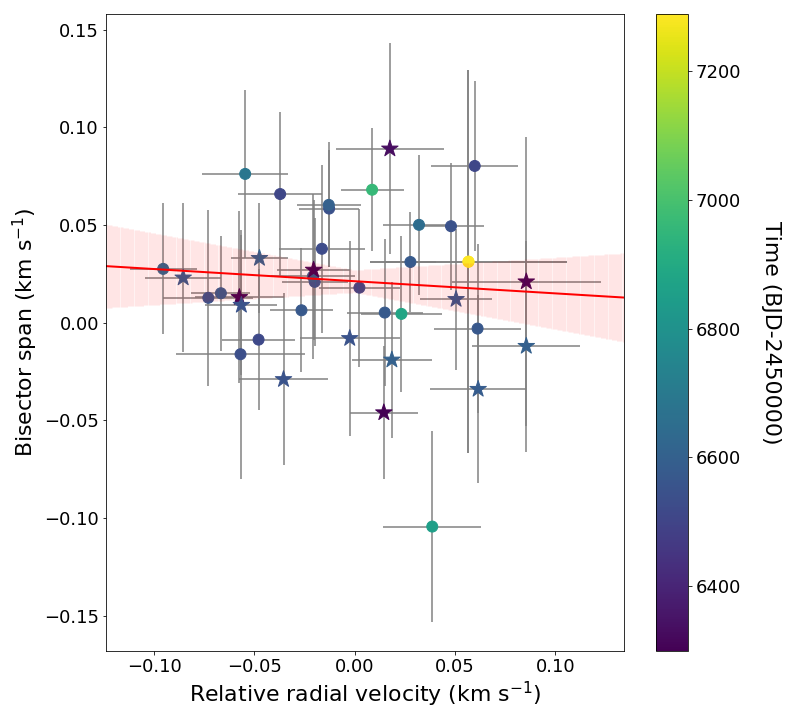}{0.48\textwidth}{}}
\gridline{\fig{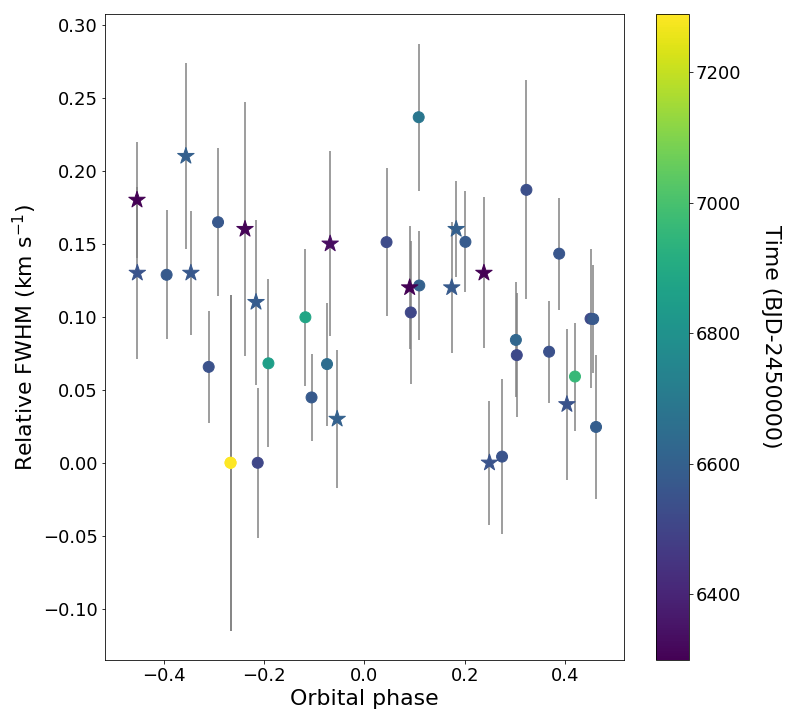}{0.48\textwidth}{}}
	\caption{\textit{Upper panel: }Radial velocity bisector span plotted as a function of relative radial velocity for the WASP-116 system. The uncertainties in the bisector measurements are taken to be $2.0\times\sigma_{\rm RV}$. CORALIE data are denoted by circles, SOPHIE data by stars. The shading indicates the observation date. The solid line represents an odr linear fit to the data, with the shaded region indicating the $1\sigma$ limits to the fit. No statistically significant correlation is found. \textit{Lower panel: } Radial velocity full width at half maximum (FWHM) as a function of orbital phase. The uncertainties in the FWHMs are taken to be $2.35\times\sigma_{\rm RV}$. The FWHM values have been offset from the minimum value for each data set to allow comparison. Legend as for the upper panel. There is no clear variation with orbital phase.}
	\label{fig:bisector116}
\end{figure}

We also find no significant correlation between FWHM and the orbital phase, and the colour-mapping of our data with time (see Figure\,\ref{fig:bisector116}) shows no indication of a time-dependent signal.

\begin{deluxetable*}{lllllll}
\tabletypesize{\scriptsize}
\caption{Radial velocity data of WASP-116, obtained using SOPHIE and CORALIE. The uncertainty on the bisectors is taken to be twice the uncertainty on the RVs. The uncertainty on the FWHMs is taken to be $2.35$ times the RV uncertainty.\label{tab:116RVs}}
\centering
\tablehead{
		\colhead{${\rm BJD}_{\rm TDB}$} & \colhead{RV} & \colhead{$\sigma_{\rm RV}$} & \colhead{Bisector} & \colhead{$\sigma_{\rm bis}$} & \colhead{FWHM} & \colhead{$\sigma_{\rm FWHM}$} \\
		\colhead{$-2450000$} & \colhead{} & \colhead{} & \colhead{} & \colhead{} & \colhead{} & \colhead{} \\
		\colhead{(days)} & \colhead{(\kms)} & \colhead{(\kms)} & \colhead{(\kms)} & \colhead{(\kms)} & \colhead{(\kms)} & \colhead{(\kms)}
		}
\startdata
		\multicolumn{7}{l}{\textit{SOPHIE}} \\
		$6299.24328$	& $-11.046$	& $0.018$	& $0.027$	& $0.036$	& $10.05$	& $0.042$ \\
		$6300.22228$	& $-11.083$	& $0.022$	& $0.013$	& $0.044$	& $10.06$	& $0.052$ \\
		$6302.26078$	& $-11.011$	& $0.017$	& $-0.046$	& $0.034$	& $10.11$	& $0.040$ \\
		$6310.29808$	& $-10.940$	& $0.037$	& $0.021$	& $0.074$	& $10.09$	& $0.087$ \\
		$6330.27918$\tablenotemark{$\ast$}	& $-10.990$	& $0.072$	& $0.270$	& $0.144$	& $10.02$	& $0.169$ \\
		$6331.26068$	& $-11.008$	& $0.027$	& $0.089$	& $0.054$	& $10.08$	& $0.063$ \\
		$6551.59778$	& $-11.082$	& $0.018$	& $0.009$	& $0.036$	& $9.93$	& $0.042$ \\
		$6552.61848$	& $-11.061$	& $0.022$	& $-0.029$	& $0.044$	& $9.97$	& $0.052$ \\
		$6553.56688$	& $-11.028$	& $0.025$	& $-0.008$	& $0.050$	& $10.06$	& $0.059$ \\
		$6567.49948$	& $-10.975$	& $0.018$	& $0.012$	& $0.036$	& $10.06$	& $0.042$ \\
		$6577.55268$	& $-11.111$	& $0.019$	& $0.023$	& $0.038$	& $10.05$	& $0.045$ \\
		$6581.58488$	& $-10.964$	& $0.024$	& $-0.034$	& $0.048$	& $10.04$	& $0.056$ \\
		$6597.45098$	& $-11.073$	& $0.014$	& $0.033$	& $0.028$	& $10.09$	& $0.033$ \\
		$6600.49948$	& $-10.940$	& $0.027$	& $-0.012$	& $0.054$	& $10.14$	& $0.063$ \\
		$6602.49448$	& $-11.007$	& $0.020$	& $-0.019$	& $0.040$	& $9.96$	& $0.047$ \\ \\
		\multicolumn{7}{l}{\textit{CORALIE}} \\
		$6508.8620326$	& $-10.92283$	& $0.02181$	& $0.08021$	& $0.04362$	& $8.70999$	& $0.05125$ \\
		$6510.8804376$	& $-11.01980$	& $0.02096$	& $0.06583$	& $0.04192$	& $8.81294$	& $0.04926$ \\
		$6518.8915136$	& $-11.03056$	& $0.01808$	& $-0.00868$	& $0.03616$	& $8.78371$	& $0.04249$ \\
		$6519.8641576$	& $-10.98031$	& $0.02019$	& $0.01792$	& $0.04038$	& $8.80865$	& $0.04745$ \\
		$6523.7873716$	& $-10.99901$	& $0.02150$	& $0.03782$	& $0.04300$	& $8.86109$	& $0.05053$ \\
		$6538.8584086$	& $-11.03957$	& $0.03191$	& $-0.01614$	& $0.06382$	& $8.89689$	& $0.07499$ \\
		$6545.7705586$	& $-10.99540$	& $0.01488$	& $0.05839$	& $0.02976$	& $8.78606$	& $0.03497$ \\
		$6547.8947706$	& $-10.93466$	& $0.01628$	& $0.04937$	& $0.03256$	& $8.77566$	& $0.03826$ \\
		$6551.7635496$	& $-11.05561$	& $0.02249$	& $0.01273$	& $0.04498$	& $8.71422$	& $0.05285$ \\
		$6565.7431286$	& $-11.00259$	& $0.01635$	& $0.02089$	& $0.03270$	& $8.85319$	& $0.03842$ \\
		$6567.8578506$	& $-10.92141$	& $0.02155$	& $-0.00302$	& $0.04310$	& $8.87476$	& $0.05064$ \\
		$6572.8046436$	& $-11.00918$	& $0.01584$	& $0.00635$	& $0.03168$	& $8.80846$	& $0.03722$ \\
		$6573.7927496$	& $-10.96768$	& $0.01883$	& $0.00517$	& $0.03766$	& $8.83873$	& $0.04425$ \\
		$6575.7040016$	& $-10.95489$	& $0.01275$	& $0.03106$	& $0.02550$	& $8.75474$	& $0.02996$ \\
		$6577.7332716$	& $-11.04935$	& $0.01469$	& $0.01507$	& $0.02938$	& $8.86127$	& $0.03452$ \\
		$6592.6820456$	& $-11.00353$	& $0.02103$	& $0.02368$	& $0.04206$	& $8.73453$	& $0.04942$ \\
		$6603.5777036$	& $-10.99557$	& $0.01597$	& $0.06029$	& $0.03194$	& $8.83141$	& $0.03753$ \\
		$6624.6910646$	& $-11.07812$	& $0.01675$	& $0.02754$	& $0.03350$	& $8.79412$	& $0.03936$ \\
		$6648.6532706$	& $-10.95060$	& $0.01791$	& $0.05007$	& $0.03582$	& $8.77758$	& $0.04209$ \\
		$6689.5415156$	& $-11.03724$	& $0.02143$	& $0.07613$	& $0.04286$	& $8.94664$	& $0.05036$ \\
		$6852.8902586$	& $-10.94400$	& $0.02443$	& $-0.10434$	& $0.04886$	& $8.77811$	& $0.05741$ \\
		$6879.8304196$	& $-10.95937$	& $0.01998$	& $0.00450$	& $0.03996$	& $8.80966$	& $0.04695$ \\
		$6962.7440926$	& $-10.97398$	& $0.01573$	& $0.06803$	& $0.03146$	& $8.76906$	& $0.03697$ \\
		$7288.8624286$\tablenotemark{$\dagger$}	& $-10.87388$	& $0.04897$	& $0.03121$	& $0.09794$	& $8.65461$	& $0.07334$ \\
\enddata
\tablenotetext{\ast}{Datum excluded from our analysis, as the uncertainty on the measurement is $1.4$ times the standard deviation of the SOPHIE RV data.}
\tablenotetext{\dagger}{Datum obtained after the optical fibre of the CORALIE instrument was replaced, and excluded from our analysis.}
\end{deluxetable*}

\subsubsection{Photometric follow-up}
\label{sec:116phot}
Follow-up photometric observations of WASP-116 were carried out using TRAPPIST and EulerCam. Details of the observations can be found in Table\,\ref{tab:phot}. Two transits of the candidate planet were observed, on 2013 November 05 and 2013 November 25, with both events being covered by both telescopes. As the two instruments are fitted with different filters, this allows us to compare the simultaneous transit depth at different wavelengths, and to check for wavelength dependent activity signatures.

Unfortunately, none of our follow-up observations were able to capture a full transit of this planet owing to the long duration of the transits. This will be discussed further in Section\,\ref{sec:116model}.

\subsection{WASP-149}
\label{sec:149wasp}
$2{\rm MASS}$\,J$08161768-0841121$, from hereon referred to as WASP-149, was another equatorial candidate observed by both WASP telescopes; between 2009-11-20 and 2012-03-31 they obtained a combined $12466$ data. A potential planetary transit signal was identified in WASP data, but there was some ambiguity as to whether the putative signal had a period of $\sim4$\,days or $\sim1.33$\,days. The candidate was thus put forward for photometric and spectroscopic follow-up, with the aim of confirming the signal and resolving this degeneracy.

\subsubsection{Spectroscopic follow-up}
\label{sec:149spec}
Observations of this candidate were started in 2013 using the CORALIE instrument, and attempted to confirm the $4$\,day period. Two observations showed some CCF movement, but did not appear to phase with the ephemeris predicted by the WASP photometry. However, initial observations with SOPHIE in HE mode favoured the $1.33$\,day period; this was subsequently confirmed by further observations obtained by both instruments. Nine of our eleven CORALIE data were obtained after the instrument was upgraded to use octagonal fibres \citep{2013AA...549A..49B}; these are clearly marked in Table\,\ref{tab:149RVs}. 

\begin{figure}
\gridline{\fig{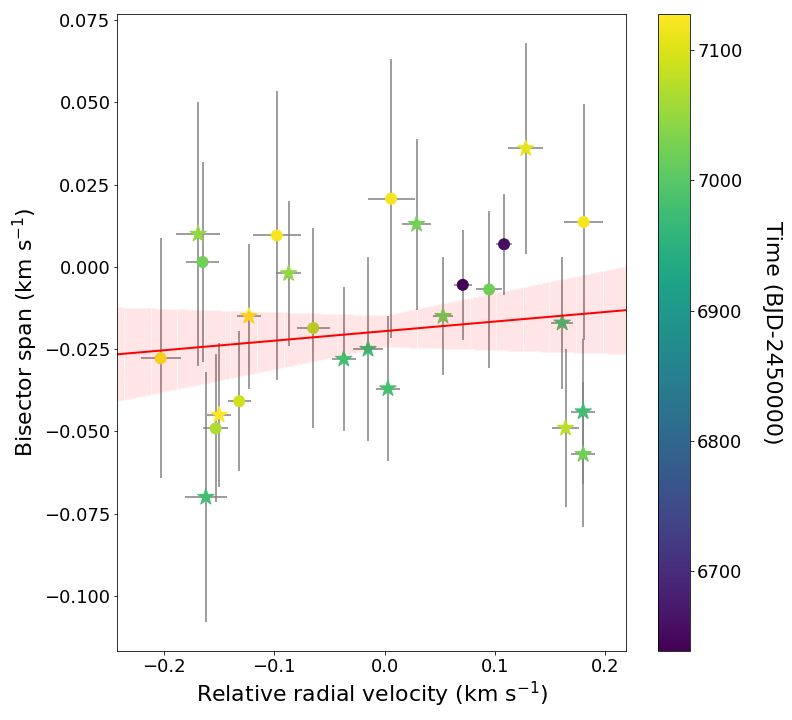}{0.48\textwidth}{}}
\gridline{\fig{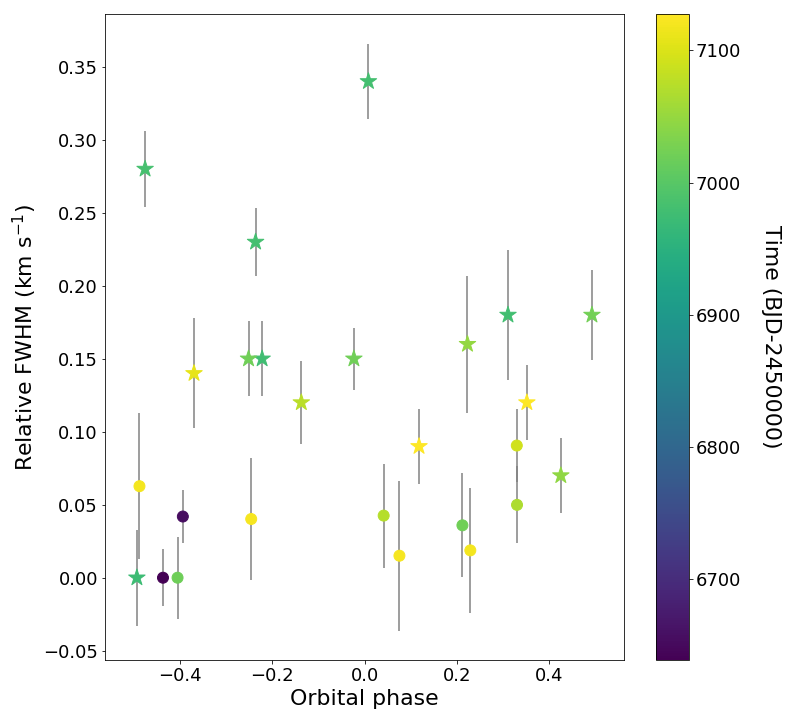}{0.48\textwidth}{}}
	\caption{\textit{Upper panel: }Radial velocity bisector span plotted as a function of relative radial velocity for the WASP-149 system. The uncertainties in the bisector measurements are taken to be $2.0\times\sigma_{\rm RV}$. CORALIE data are denoted by circles, SOPHIE data by stars. The shading indicates the observation date. No correlation with time is seen. The solid line represents an odr linear fit to the data, with the shaded region indicating the $1\sigma$ limits to the fit. No statistically significant correlation is found. \textit{Lower panel: } Radial velocity full width at half maximum (FWHM) as a function of orbital phase. The uncertainties in the FWHMs are taken to be $2.35\times\sigma_{\rm RV}$. The FWHM values have been offset from the minimum value for each data set to allow comparison. Legend as for the upper panel. There is no clear variation with orbital phase.}
	\label{fig:bisector149}
\end{figure}

The bisector spans of our data show no indication of a significant correlation with the RVs; we obtain a Spearman Rank Correlation coefficient of $0.11$ with a p-value of $0.61$, and a Pearson Rank Correlation coefficient of $0.14$ with a p-value of $0.49$, insufficient to reject the null hypothesis of no correlation. In addition, an odr linear fit returned a slope of $0.03$ with only $0.75\sigma$ significance, further supporting the null hypothesis. The FWHMs of our RV CCFs show no correlation with orbital phase (See Figure\,\ref{fig:bisector149}), allowing us to rule out background eclipsing binaries as a source of false positive. We list our RV data, bisector spans, and CCF FWHMs for WASP-149 in Table\,\ref{tab:149RVs}. Our data show no additional dependence on time, indicating that there is no additional component to the radial velocity curve within our detection capability.

\begin{deluxetable*}{lllllll}
\tabletypesize{\scriptsize}
\caption{Radial velocity data of WASP-149, obtained using SOPHIE and CORALIE. The uncertainty on the bisectors is taken to be twice the uncertainty on the RVs. The uncertainty on the FWHMs is taken to be $2.35$ times the RV uncertainty.\label{tab:149RVs}}
\centering
\tablehead{
		\colhead{${\rm BJD}_{\rm TDB}$} & \colhead{RV} & \colhead{$\sigma_{\rm RV}$} & \colhead{Bisector} & \colhead{$\sigma_{\rm bis}$} & \colhead{FWHM} & \colhead{$\sigma_{\rm FWHM}$} \\
		\colhead{$-2450000$} & \colhead{} & \colhead{} & \colhead{} & \colhead{} & \colhead{} & \colhead{} \\
		\colhead{(days)} & \colhead{(\kms)} & \colhead{(\kms)} & \colhead{(\kms)} & \colhead{(\kms)} & \colhead{(\kms)} & \colhead{(\kms)}
		}
\startdata
		\multicolumn{7}{l}{\textit{SOPHIE}} \\
		$6974.60728$	& $17.843$	& $0.014$	& $-0.025$	& $0.028$	& $ 9.98$	& $0.033$ \\
		$6977.63448$	& $18.038$	& $0.011$	& $-0.044$	& $0.022$	& $10.13$	& $0.026$ \\
		$6979.67708$	& $17.696$	& $0.019$	& $-0.070$	& $0.038$	& $10.16$	& $0.045$ \\
		$6980.60698$	& $17.821$	& $0.011$	& $-0.028$	& $0.022$	& $10.32$	& $0.026$ \\
		$6981.61398$	& $18.019$	& $0.010$	& $-0.017$	& $0.020$	& $10.21$	& $0.024$ \\
		$6982.62788$	& $17.861$	& $0.011$	& $-0.037$	& $0.022$	& $10.26$	& $0.026$ \\
		$7020.54928$	& $17.911$	& $0.009$	& $-0.015$	& $0.018$	& $10.13$	& $0.021$ \\
		$7021.57798$	& $18.038$	& $0.011$	& $-0.057$	& $0.022$	& $10.13$	& $0.026$ \\
		$7022.56938$	& $17.887$	& $0.013$	& $0.013$	& $0.026$	& $10.16$	& $0.031$ \\
		$7046.47018$	& $17.771$	& $0.011$	& $-0.002$	& $0.022$	& $10.05$	& $0.026$ \\
		$7047.53348$	& $17.689$	& $0.020$	& $0.010$	& $0.040$	& $10.14$	& $0.047$ \\
		$7076.37568$	& $18.022$	& $0.012$	& $-0.049$	& $0.024$	& $10.10$	& $0.028$ \\
		$7109.38608$	& $17.986$	& $0.016$	& $0.036$	& $0.032$	& $10.12$	& $0.038$ \\
		$7126.34078$	& $17.708$	& $0.011$	& $-0.045$	& $0.022$	& $10.10$	& $0.026$ \\
		$7127.36228$	& $17.735$	& $0.011$	& $-0.015$	& $0.022$	& $10.07$	& $0.026$ \\ \\
		\multicolumn{7}{l}{\textit{CORALIE}} \\
		$6638.8142676$	& $17.97631$	& $0.00839$	& $-0.00546$	& $0.01678$	& $8.90925$	& $0.01972$ \\
		$6654.8653156$	& $18.01372$	& $0.00764$	& $0.00688$	& $0.01528$	& $8.95116$	& $0.01795$ \\
		$7018.7074576$\tablenotemark{$\dagger$}	& $17.99631$	& $0.01200$	& $-0.00687$	& $0.02400$	& $8.83967$	& $0.02820$ \\
		$7020.8626416$\tablenotemark{$\dagger$}	& $17.73665$	& $0.01519$	& $0.00147$	& $0.03038$	& $8.87557$	& $0.03570$ \\
		$7067.6686716$\tablenotemark{$\dagger$}	& $17.74847$	& $0.01124$	& $-0.04912$	& $0.02248$	& $8.88957$	& $0.02641$ \\
		$7072.6149356$\tablenotemark{$\dagger$}	& $17.83694$	& $0.01517$	& $-0.01860$	& $0.03034$	& $8.88222$	& $0.03565$ \\
		$7091.6585966$\tablenotemark{$\dagger$}	& $17.76987$	& $0.01064$	& $-0.04084$	& $0.02128$	& $8.93022$	& $0.02500$ \\
		$7118.5582976$\tablenotemark{$\dagger$}	& $17.90752$	& $0.02122$	& $0.02073$	& $0.04244$	& $8.90237$	& $0.04987$ \\
		$7119.5134406$\tablenotemark{$\dagger$}	& $17.69858$	& $0.01821$	& $-0.02779$	& $0.03642$	& $8.85840$	& $0.04279$ \\
		$7120.6416236$\tablenotemark{$\dagger$}	& $17.80387$	& $0.02190$	& $0.00955$	& $0.04380$	& $8.85474$	& $0.05147$ \\
		$7121.5465266$\tablenotemark{$\dagger$}	& $18.08196$	& $0.01788$	& $0.01362$	& $0.03576$	& $8.87989$	& $0.04202$ \\
\enddata
\tablenotetext{\dagger}{Ddata obtained after the optical fibre of the CORALIE instrument was replaced. These are fit as a separate dataset to account for an offset between the pre- and post-upgrade baseline.}
\end{deluxetable*}

\subsubsection{Photometric follow-up}
\label{sec:149phot}
Additional transits of WASP-149 were observed on 2015-01-29 by the NITES telescope \citep{2014MNRAS.438.3383M}, on 2015-02-06, 2015-05-05, and 2015-11-26 by TRAPPIST, and on 2015-04-07 and 2015-12-20 by EulerCam. We summarise the observational details in Table\,\ref{tab:phot}. The NITES observations were affected by poor weather conditions, which led to the egress phase of the transit being missed. 

During the photometric reduction and quality control process for our NITES observations, we noticed that the depth of the transit appeared greater than in the WASP photometry. Variable transit depth across different wavelength bands can be an indicator of a stellar blend (the chromatic depth effect). We rule out this possibility however, as initial processing of the EulerCam and TRAPPIST lightcurves showed a depth consistent with the NITES data. We therefore carried out a visual inspection of the CCD images, photometric aperture sizes, and aperture positions for the different telescopes. We found that a nearby star approximately $14\arcsec$ from WASP-149 was within the photometric aperture of the WASP pipeline, but was excluded from the apertures of the follow-up telescopes. The additional light from this nearby star, which is three magnitude fainter in the Cousins V-band, dilutes the transit depth in the WASP lightcurve, making it appear more shallow, but does not affect the transit depth in the follow-up lightcurves.

We will return to the subject of WASP-149's transit depth in Section\,\ref{sec:149model}.

\begin{deluxetable}{lllll}
\tabletypesize{\scriptsize}
\caption{Summary of photometric follow-up observations for WASP-85, WASP-116, and WASP-149. \label{tab:phot}}
\centering
\tablehead{\colhead{Instrument} & \colhead{Date} & \colhead{Filter} & \colhead{$N_{\rm data}$} & \colhead{Cadence} \\ 
			\colhead{} & \colhead{} & \colhead{} & \colhead{} & \colhead{(s)}
}
\startdata
		\multicolumn{5}{l}{\textit{WASP-85}} \\
		JGT & 2009-03-20 & Cousins R & $237$ & $47$\\ 
		IO:O & 2013-01-12 & Sloan $z\prime$ & $146$ & $95$\\ 
		IO:O & 2013-01-28 & Sloan $z\prime$ & $146$ & $37$\\ 
		EulerCAM & 2013-03-23 & RG & $186$ & $67$\\  
		TRAPPIST & 2013-03-31 & Sloan $z\prime$ & $569$ & $23$\\ 
		TRAPPIST & 2013-04-16 & Sloan $z\prime$ & $615$ & $23$\\ 
		TRAPPIST & 2013-05-09 & Sloan $z\prime$ & $503$ & $21$\\ 
		TRAPPIST & 2013-05-25 & Sloan $z\prime$ & $626$ & $20$\\ 
		\multicolumn{5}{l}{\textit{WASP-116}} \\
		TRAPPIST & 2013-11-05	& Blue-blocking	& $1193$ & $18$\\ 
		EulerCam	& 2013-11-05		& NGTS	& $197$ & $78$\\ [2pt] 
		TRAPPIST & 2013-11-25	& Blue-blocking	& $1199$ & $18$\\ 
		EulerCam	& 2013-11-25		& NGTS	& $277$ & $78$\\ 
		\multicolumn{5}{l}{\textit{WASP-149}} \\
		NITES	& 2015-01-29	& $-$	& $444$ & $27$\\ 
		TRAPPIST	& 2015-02-06	& Sloan $z\prime$	& $532$ & $23$\\ 
		EulerCam		& 2015-04-07	& ZG		& $196$ & $57$\\ 
		TRAPPIST	& 2015-05-05	& Sloan $z\prime$	& $465$ & $24$\\ 
		TRAPPIST	& 2015-11-26	& Sloan $z\prime$	& $507$ & $25$\\ 
		EulerCam		& 2015-12-20	& ZG		& $265$ & $48$\\ 
		\hline \\
\enddata
\tablenotetext{}{The RG filter used in EulerCam is a modified broad Gunn-R filter, with a central wavelength of $660$\,nm. The NGTS filter is that used by the NGTS project \citep{2018MNRAS.475.4476W}; it has a custom wavelength range of $550-900$\,nm.}
\end{deluxetable}

\section{Stellar parameters}
\label{sec:star}
As in previous WASP discovery papers, we determine the parameters of the planets' host stars through detailed spectral analysis. Using standard data reduction pipeline products, we follow \citet{2013MNRAS.428.3164D} and co-add the spectra obtained from a single instrument 
to produce a single, master spectrum with improved SNR.  Stellar effective temperatures (\teff) are determined using the excitation balance of available Fe~{\sc i} lines. Na~{\sc i} D lines, the 6439 \AA\,Ca\,{\sc i} line, and the ionisation balance of Fe~{\sc i} and Fe~{\sc ii}, are used as diagnostics of surface gravity (\logg)\footnote{The exact list of lines used is dependent on the resolution of the instrument from which the spectra are taken}. The Fe~{\sc i} lines are also used to estimate a value for the stellar microturbulence velocity using the method of \cite{1984AA...134..189M}. Stellar iron abundance with respect to the Sun is determined from equivalent width measurements of several unblended lines. Where possible, we calculate the activity index, $\log(R'_{\rm HK})$, using the core of Ca\,{\sc ii}\,H+K lines from spectra with SNR greater than some threshold (dependent on the quality of available spectra; ${\rm SNR}>4$ for 85\,A, $>2$ for 85\,B).

We calculate the projected stellar rotation velocity (\vsini) by fitting the profiles of several unblended lines, dependent on the quality and source of the spectra, and estimate the stellar macroturbulence velocity using the calibration of \citet{Doyle14}. Finally, we cross-reference our value for \teff\ with Table\,B.1 in Gray (2008) to estimate the spectral type of the star, and use the \cite{2010AARv..18...67T} calibration to determine stellar mass and radius. 

The results of this analysis for the systems we present herein are summarised in Table\,\ref{tab:star_params}; the quoted uncertainties account for the errors in \teff\ and \logg\, as well as the additional scatter induced by measurement and atomic data uncertainties.

We also cross-match the coordinates of our target stars with the \textit{Gaia} Data Release 2 (DR2) catalogue \citep{2016AA...595A...1G, 2018AA...616A...1G}, and in Table\,\ref{tab:star_params} we include the Gaia magnitude, proper motions, and parallaxes. To derive the distances, we corrected the listed parallaxes for the global offset of $-82\pm33$\,$\mu{\rm as}$ found by \citet{2018ApJ...862...61S}. The parallax distance for WASP-85\,A agrees with the value of $125\pm80$\,pc that we derive using the infra-red flux method. We also determine radii for the stars following \citet{2018AA...616A...8A}, assuming zero extinction and using the \teff\ values derived from the HARPS spectra. The results agree well for WASP-85A ($R_{\star,{\rm G}}=0.94\pm0.02$\,$R_\sun$), but for both WASP-116 ($R_{\star,{\rm G}}=1.55\pm0.05$\,$R_\sun$) and WASP-149 ($R_{\star,{\rm G}}=1.02\pm0.04$\,$R_\sun$) the radius derived from \textit{Gaia} data is significantly smaller than the value in Table\,\ref{tab:star_params}. This implies that our assumption of zero extinction is incorrect. We estimate that extinction values of $0.16\leq{\rm A}_{\rm G}\leq0.65$ for WASP-116 and $0.40\leq{\rm A}_{\rm G}\leq0.86$ for WASP-149 are required to bring the Gaia radii in line with the values in Table\,\ref{tab:star_params}.

\begin{deluxetable*}{llllll}
\tabletypesize{\scriptsize}
\caption{Summary of the stellar parameters for our target stars. Unless otherwise noted these were determined through analysis of co-added spectra, or derived from the results of said analysis. \label{tab:star_params}}
\centering
\tablehead{\colhead{Parameter} & \colhead{WASP-85\,A} & \colhead{WASP-85\,B} & \colhead{WASP-116\,A} & \colhead{WASP-149\,A} & \colhead{Units}
}
\startdata
	RA				& \multicolumn{2}{c}{$11$h$43$m$38.01$s} 			& $02$h$20$m$51.75$s				& $08$h$16$m$17.67$s &  J2000 \\
	DEC				& \multicolumn{2}{c}{$+06^\circ33$\arcmin$49.4$\arcsec} & $-01^\circ49$\arcmin$33.7$\arcsec	& $-08^\circ41$\arcmin$12.0$\arcsec &  J2000 \\
	V				& $11.2$ 				& $11.9$ 					& $12.4$							& $11.7$		&  mag \\
	\bv				& $0.670\pm0.022$ 		& $0.828^{+0.034}_{-0.036}$ 	& $0.583$							& $0.690$		&  \\
	G\tablenotemark{$\dagger$}		& $10.62$				& $11.54$					& $12.26$							& $11.26$         & mag \\
	pmRA\tablenotemark{$\dagger$}	& $-77.24\pm0.08$		& $-$					& $8.63\pm0.10$					& $-1.57\pm0.06$ & ${\rm mas}\,{\rm yr}^{-1}$ \\
	pmDEC\tablenotemark{$\dagger$}	& $-11.02\pm0.07$		& $-$					& $1.86\pm0.11$					& $22.90\pm0.05$ & ${\rm mas}\,{\rm yr}^{-1}$ \\
	Parallax\tablenotemark{$\dagger$}	& $7.02\pm0.06$		& $-$					& $1.76\pm0.05$					& $4.69\pm0.04$ & ${\rm mas}$ \\
	Distance\tablenotemark{$\dagger$}	& \multicolumn{2}{c}{$141\pm1$} 					& $543\pm18$						& $209\pm2$	&  pc \\
	Angular Separation	& \multicolumn{2}{c}{$1.5\pm0.1$}					& $-$							& $-$		& \arcsec \\
	Separation		& \multicolumn{2}{c}{$210\pm22$}					& $-$							& $-$		&  AU \\
	Spectral type		& G5 				& K0 					& G0								& G6			&  \\
	\teff\				& $5685\pm65$ 		& $5250\pm90$ 			& $5950\pm100$					& $5750\pm100$	&  K \\
	Mass, $M_\star$	& $1.04\pm0.07$ 		& $0.88\pm0.07$ 			& $1.25\pm0.18$					& $1.12\pm0.17$	&  $M_{\sun}$ \\
	Radius, $R_\star$	& $0.96\pm0.13$ 		& $0.77\pm0.13$ 			& $2.08\pm0.15$					& $1.23\pm0.09$	&  $R_\sun$ \\
	\logg\			& $4.48\pm0.11$ 		& $4.61\pm0.14$ 			& $3.9\pm0.1$						& $4.3\pm0.2$	&  cgs \\
	\mictrb\			& $0.6\pm0.1$ 			& $0.9\pm0.1$ 				& $1.2\pm0.1$						& $1.0\pm0.1$	&  \kms\ \\
	\mactrb\			& $2.93\pm0.73$ 		& $2.20\pm0.73$ 			& $4.6\pm0.7$						& $3.43\pm0.73$	&  \kms\ \\
	\vsini\			& $3.41\pm0.89$ 		& $3.32\pm0.82$ 			& $1.7\pm1.1$						& $4.6\pm0.6$	&  \kms\ \\
	$P_{\rm rot}$		& $14.6\pm1.5$		& $7.50\pm0.03$			& $-$							& $-$	&  days \\
	$\log A({\rm Li})$	& $2.19\pm0.06$ 		& $<0.70\pm0.10$			& $2.35\pm0.08$					& $<0.8$	&  \\
	${\rm [Fe/H]}$		& $0.08\pm0.10$ 		& $0.00\pm0.15$ 			& $-0.28\pm0.1$					& $0.16\pm0.11$	&  \\
	$\log(R'_{\rm HK})$	& $-4.43^{+0.06}_{-0.02}$ & $-4.37^{+0.09}_{-0.04}$		& $-$							& $-$	&  \\
\enddata
\tablenotetext{}{Masses and radii estimated using the calibration of \cite{2010AARv..18...67T}. Spectral Types estimated from \teff\ using Table\,B.1 in \cite{2008oasp.book.....G}. Iron abundances are relative to the solar value obtained by \cite{2009ARAA..47..481A}. The rotation period for WASP-85\,A is the mean of the periods determined from a periodogram analysis of the WASP lightcurve. The rotation period for WASP-85\,B was determined through modelling of spot modulation in the K2 long-cadence lightcurve.}
\tablenotetext{\dagger}{Data from Gaia DR2. Note that the parallaxes listed here are from the Gaia catalogue, and do not account for the global offset found by \citet{2018ApJ...862...61S}.}
\end{deluxetable*}

\subsection{Stellar activity}
\label{sec:activity}
Using the HARPS spectra with ${\rm S/N}>4$ for the core of the Ca\,{\sc ii}\,H+K lines, we calculate the $\log(R'_{\rm HK})$ activity index using the emission in the cores of the Ca\,{\sc ii}\,H+K lines following \citet{1984ApJ...279..763N}. We find an activity index of $\log(R'_{\rm HK})=-4.43^{+0.06}_{-0.02}$ for WASP-85\,A. For WASP-85\,B we find $\log(R'_{\rm HK})=-4.37^{+0.09}_{-0.04}$ using spectra with ${\rm S/N}>2$.

Comparing the Ca\,{\sc ii}\,H+K indices for the two stellar components of WASP-85 to the sample of \citet{2006MNRAS.372..163J, 2008AA...485..571J, 2011AA...531A...8J}, we find that both stars fall within the `very active stars' region of Figure\,4 in \citet{2006MNRAS.372..163J}. This is confirmed by Figure\,10 of \citet{2008AA...485..571J} and Figure\,6 of \citet{2011AA...531A...8J}, where in both cases the two stars fall within the secondary, `active' peak in the  $\log(R'_{\rm HK})$ distribution. Comparison to the sample of \citet{1996AJ....111..439H} shows that both stars fall in the `active' class of stars.

Comparing to the sample of planet hosting stars examined by \citet{2010ApJ...720.1569K}, we see that WASP-85\,A is more active than all but one of the stars considered. Only CoRoT-2 is more active. It may be that we have simply observed the system at the peak of the activity cycle, leading to an apparently greater level of activity. However, if we consider the solar cycle then the typical variation is $\sim0.2$\,dex; converting the solar Ca\,{\sc ii}\,H+K values presented in \citet{2007ApJ...657.1137L} indicates values of $\log(R'_{\rm HK})=-4.978$ and $-4.803$ at solar minimum and maximum, respectively. Measurements of activity during the Maunder Minimum \citep{1998ASPC..154.....D} indicate $\log(R'_{\rm HK})=-5.102$ during that particularly inactive period in the Sun's life, giving a pessimistic variation from maximum of $\sim0.3$\,dex. If we apply this to WASP-85\,A, then even at stellar `minimum' it will be more active than the Sun at solar maximum, and will still be classified in the `active' class of \citet{1996AJ....111..439H}. It therefore seems that these stars are unusually active for solar-type stars.

\subsection{Starspot modulation}
\label{sec:modulation}
\subsubsection{WASP data}
\label{sec:WASPmodulation}
We carried out a search for stellar modulation in the WASP lightcurves of our targets using the method described in Section\,4 of \citet{2011PASP..123..547M}. The relatively short lifetime of surface magnetic features implies that star spot induced variability is incoherent on long timescales. We therefore separated the data according to the year of observation, and fit a simple transit model following \citet{2002ApJ...580L.171M} to remove the transit signatures. A modified, generalised Lomb-Scargle periodogram \citep[e.g.][]{2009AA...496..577Z}, computed using the method of \citet{1989ApJ...338..277P} over 4096 uniformly spaced frequencies from $0$ to $2.5$\,cycles/day, was used to search for significant periodicity in the lightcurves. False alarm probabilities (FAP) were calculated using a Monte Carlo bootstrap method \citep{2011PASP..123..547M}.

\begin{figure*}
	\includegraphics[width=0.98\textwidth]{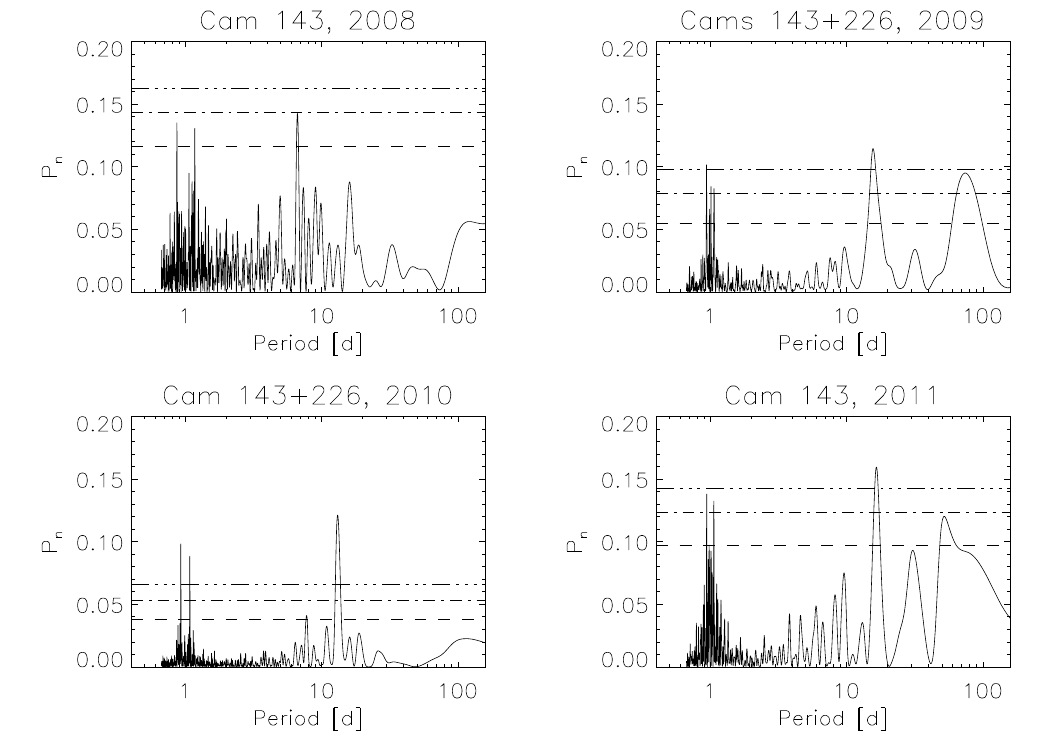}
	\caption{Periodograms for the four seasons of WASP data for the WASP-85 system. The year is given in the title of each panel. The horizontal lines indicate false-alarm probability levels of ${\rm FAP}=0.1$, $0.01$, and $0.001$. In 2009, 2010, and 2011, a significant peak is seen at approximately $15$\,days. The strongest peak in the 2008 data is at approximately half of this period.\\ \vspace{0.1cm}\\
	}
	\label{fig:pgram}
\end{figure*}

No signal was found in any of the yearly data sets for either WASP-116 or WASP-149. We place upper limits on the amplitude of any modulation of $2$\,mmag (95\% confidence) and $1.5$\,mmag (95\% confidence) for WASP-116 and WASP-149, respectively. However, significant periodicities were found in the WASP-85 data; the resulting periodograms are shown in Figure\,\ref{fig:pgram}, and the parameters for the most significant peaks are given in Table\,\ref{tab:pgram}. In the 2009, 2010, and 2011 data there are significant peaks at approximately $15$\,days. We associate this with the stellar rotation of one of the stellar components. The shorter period found in the 2008 data is approximately $P_{\rm rot}/2$, and easily explained by the presence of two active regions on opposite sides of the star. This would produce a photometric signature at twice the rotational frequency, and will contribute power in other seasons as well, but less dominantly. Note that there is significant power at periods close to $1$\,day in all four seasons. This likely arises as a result of the diurnal observing schedule forced upon WASP by its ground-based nature.

We estimate a mean rotation period of $14.6\pm1.5$\,days using the results from all four seasons (after doubling the 2008 period). This value is in good agreement with the rotation period of $14.2\pm4$\,d implied by the values of \vsini\ and the stellar radius of WASP-85\,A. However it is also in agreement with the rotation period of $11.7\pm4$\,days implied for WASP-85\,B, making it impossible to determine which star the rotational modulation originates from using this data alone. However, we are able to associate this rotation period with the planet hosting WASP-85\,A based on analysis of the \textit{K2} data (see below).  

We whiten the four sets of data by fitting a harmonic series with $P_0 = P_{\rm rot}/2$ as the fundamental period. The number of terms in the series was determined by minimising the Bayesian Information Criterion \citep{schwarz1978}. The resulting fit was divided-out, and the lightcurves concatenated to produce a whitened WASP lightcurve.

\begin{deluxetable}{lllll}
\tabletypesize{\scriptsize}
\caption{Results from the periodogram analysis of the four seasons of WASP data for the WASP-85 system. \label{tab:pgram}}
\centering
\tablehead{\colhead{Year} & \colhead{$N_{\rm data}$} & \colhead{Period} & \colhead{Amp} & \colhead{FAP} \\
		\colhead{} & \colhead{} & \colhead{(days)} & \colhead{} & \colhead{}
}
\startdata
        $2008$ & $2559$ & $6.644$ & $0.003$ & $0.0389$ \\
		$2009$ & $6760$ & $15.600$ & $0.003$ & $0.0011$ \\
		$2010$ & $8157$ & $13.130$ & $0.003$ & $0.0000$ \\
		$2011$ & $2442$ & $16.550$ & $0.003$ & $0.0002$ \\ \\
		Adopted & & $14.6\pm1.5$ & & \\
\enddata
\end{deluxetable}

\subsubsection{K2 data for WASP-85}
\label{sec:85activity}
The signature of stellar activity from one or other (or both) of the binary components of WASP-85\,A\,B is apparent in the upper panel of Figure\,\ref{fig:K2}. Such photometric modulation, which for WASP-85 has an amplitude approximately half of the transit depth, is a common feature of the lightcurves of young, low-mass stars, which often have large numbers of star spots \citep[e.g.][]{2000MNRAS.316..699D}. An initial Lomb-Scargle periodogram analysis of the lightcurve implies a period of $6.6$\,days for the activity component, though we note that the power at $13.3$\,days (double the period of the primary peak) is very similar. \citet{2016AJ....151..150M} analysed the photometric modulation in the \textit{K2} short cadence data, finding a stellar rotation period of $13.6\pm0.1$\,days, roughly commensurate with twice the period found by our periodogram, and in agreement with the period we detect in the WASP photometry. They also identified several in-transit anomalies that can be attributed to the presence of star spots, and used the recurrence of these spots in consecutive transits to propose that the planet's orbit is misaligned with the rotation axis of its host star. Future observations of the Rossiter-McLaughlin effect will be needed to confirm this however.

We carry out our own modelling of the modulation in the full \textit{K2} long-cadence lightcurve, following the methods described in \citet{2017AA...599A..27G}. Briefly, we use the general spot model of \citet{1987ApJ...320..756D}, adapted to work with both binary stars and an arbitrary number of spots on each star. This model indirectly accounts for differential rotation by making allowance for spots on the same star to have different properties. The parameters of the model were the size, temperature ($T_{\rm spot}$), latitude, longitude, and rotational period ($P_{\rm spot}$) of each spot. We assume large, circular spots, and $I_{\rm s}=45^o$ for both stars. We explore the potential parameter space of spot properties using the affine invariant MCMC code \textsc{emcee} \citep{emcee}, using 100 walkers and running for 5000 steps (with 3000 previous steps discarded as a 'burn-in' phase), thinning the chains based on the autocorrelation lengths of the model parameters, and construct the final model by marginalising over the parameters of the fit, i.e. we take the mean of 200 spot light curves drawn from the converged MCMC walkers.

\begin{figure*}
\gridline{\fig{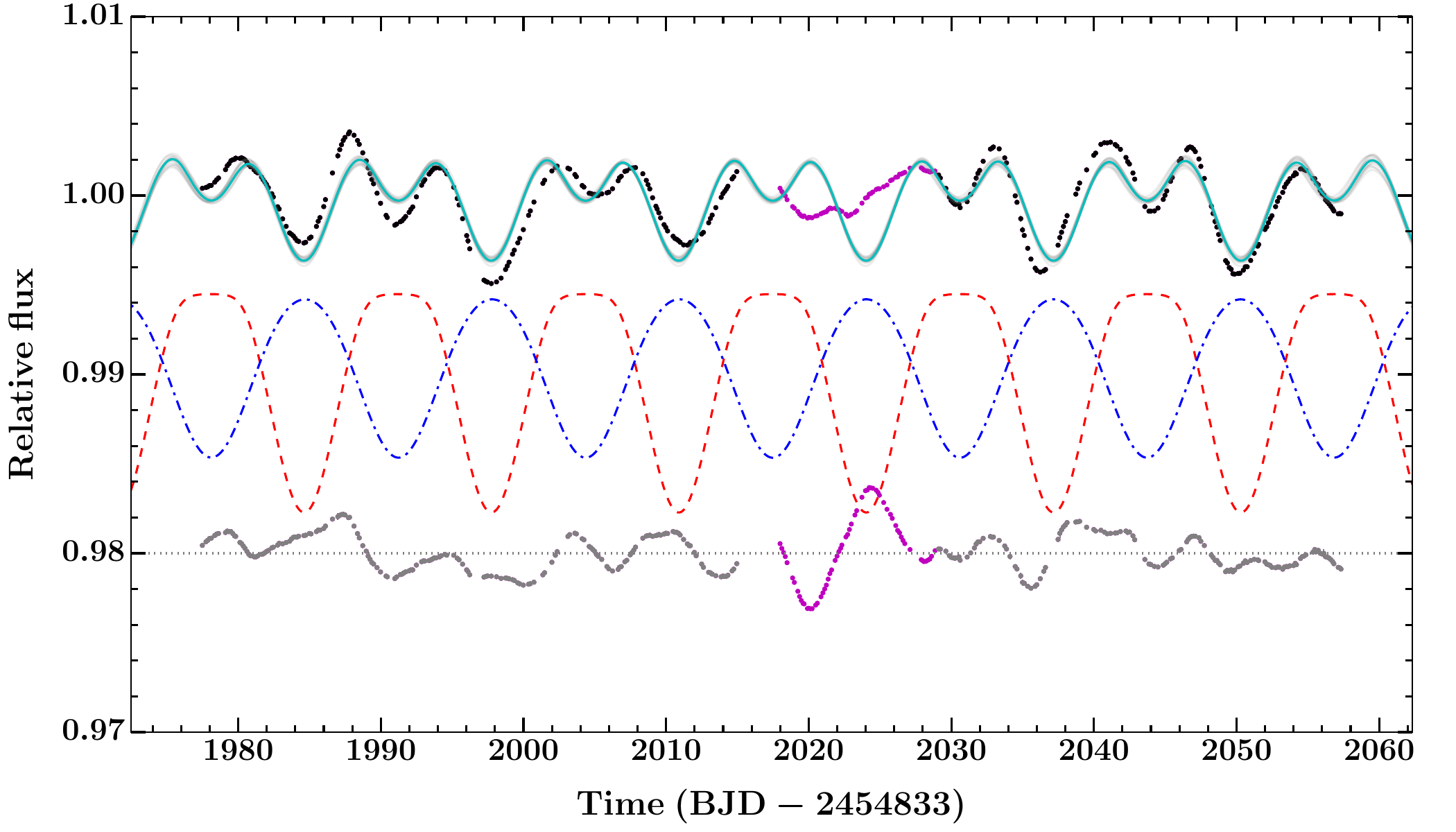}{0.5\textwidth}{}
            \fig{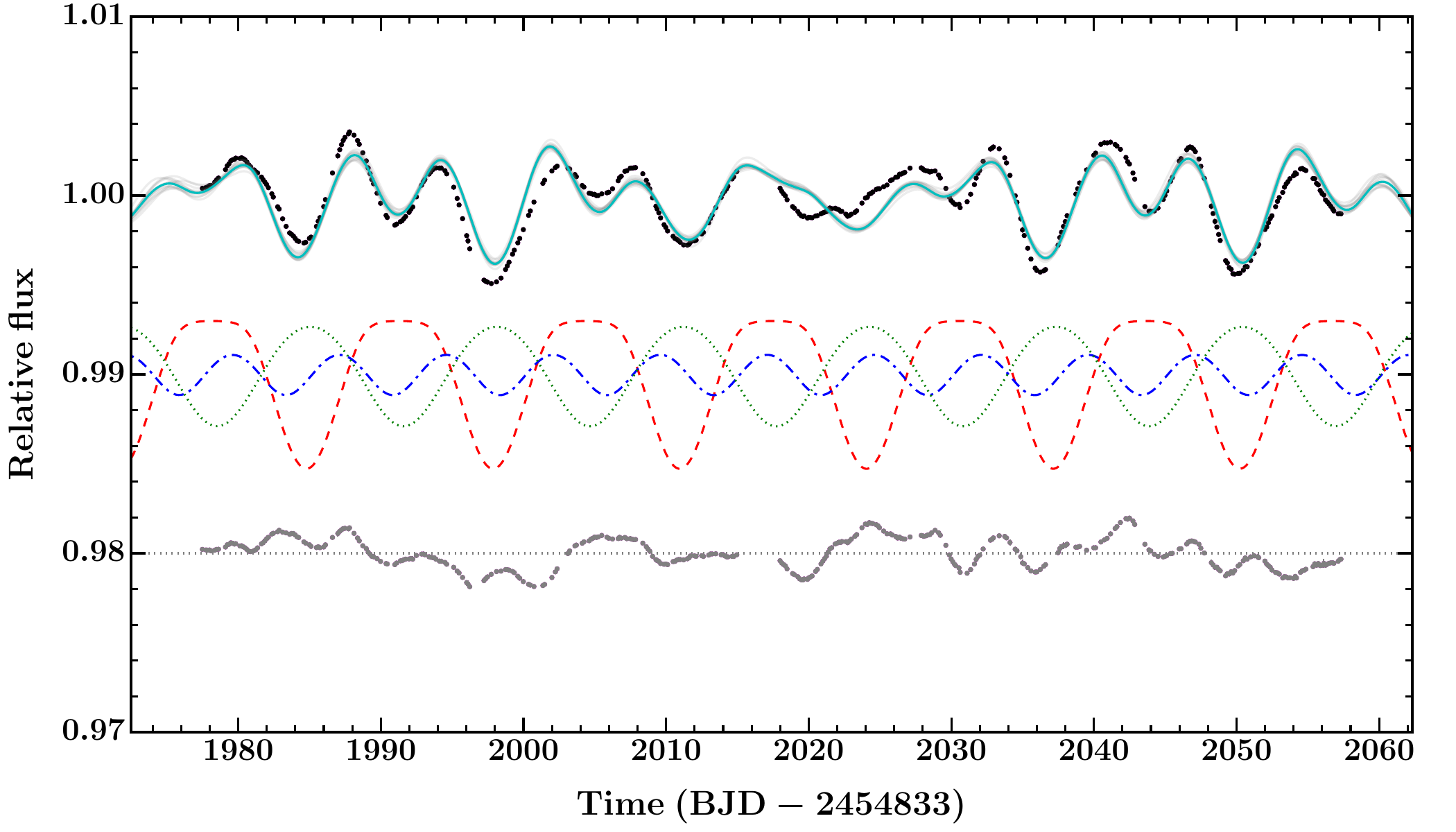}{0.5\textwidth}{}}
	\caption{Out of eclipse long cadence \textit{K2} lightcurve as output by the \textsc{everest} pipeline (black data). Masked data are shown in magenta. Overplotted are the combined spot model (cyan line) and individual spot models drawn from the posterior distribution (grey lines). Vertically offset below the lightcurve are the individual star spot signals (dashed lines), and the residuals from the star spot fit (grey data). \textit{Left: }Two spot model, with both spots on the primary star (dashed red and blue curves). \textit{Right: }Three spot model, with two spots on the primary star (dashed red and green curves) and one on the secondary star (dashed blue curve). Note that adding the third spot enables us to explain the broad characteristics of the previously masked section of data.}
	\label{fig:85spots}
\end{figure*}

Like the system described in \citet{2017AA...599A..27G}, WASP-85 consists of two stars that are both likely to have spots, and thus the modulation pattern could be explained by the two stars having different rotation periods, or having spots at different latitudes, leading to constructive / destructive interference of the signals from spots on the two stars. We initially mask out a roughly $5$\,day segment of the long cadence \textsc{everest} lightcurve \citep{2016AJ....152..100L}, just after the telescope rotation that takes place midway through each \textit{K2} field, as the characteristics of the data in this section of the lightcurve don't match the modulation through the rest of the lightcurve; a similar effect is seen in the \textsc{varcat} lightcurve of \citet{2015AA...579A..19A}. We first test a two-spot model (Figure\,\ref{fig:85spots}, upper panel), with two different size (and therefore temperature) spots on similar periods, located on opposite longitudinal hemispheres of the primary star. The upper panel of Figure\,\ref{fig:85spots} shows the results of fitting the out-of-transit \textsc{everest} lightcurve with the two-spot model. 

The two-spot model is able to recreate the general modulation pattern, but cannot explain the variation in the depth of the signal. We therefore test a three-spot model, retaining the parameters of the spots from the first model and adding a third spot on the second star (Figure\,\ref{fig:85spots}, lower panel). The effect of adding a third spot is strongly dependent on the rotation period of WASP-85\,B. Leaving the period of the secondary left unconstrained produced a rotation period of $28$\,days, but our Lomb-Scargle periodogram shows no power at this period. Moreover, this period is incompatible with our spectral analysis estimate of $v\sin I_{\rm s}=3.32\pm0.82$\,km\,s$^{-1}$ for the companion star. With a $28$\,day period and the limiting case of $I_{rm s}=90^o$, we expect $v\sin I_{\rm s}\sim1.37$\,km\,s$^{-1}$. 

We thus return to our Lomb-Scargle periodogram, which shows significant power at periods of $17.1$ and $7.5$\,days. Testing an initial period for the third spot of $17.1$\,days gave a good fit, but forced the secondary period out towards $19$\,days, which is not supported by the periodogram. Using an initial period of $7.5$\,days gives a model that returns a period supported by our other analyses, and that suitably fits the large-scale modulation features, including the previously masked section of data that we include in the three-spot fit. We therefore conclude that the rotation period of the binary companion is $\sim7.5$\,days (for full results see Table\,\ref{tab:85spots}).

Note that there are still significant residuals to the fit of our three-spot model, as our simple model is not able to fully reproduce the complexity of the modulation signal. We stress that the results of our spot model should not be interpreted literally, and are not meant to fully capture the intricacies of the spot distributions on the two stars; completely reproducing this would require information that we do not have. Instead, our model is designed to capture the overall representation of the effect of spots on the \textit{K2} light curve.

\begin{deluxetable*}{llllll}
\tabletypesize{\scriptsize}
\caption{Parameters of the three-spot model for the \textit{K2} long cadence, \textsc{everest} lightcurve of WASP-85\,A\,b\,B. Spots $1$ and $3$ are located on opposite hemispheres of the primary star (A), while spot 2 is located on the binary companion. \label{tab:85spots}}
\centering
\tablehead{\colhead{Parameter} & \colhead{symbol}& \multicolumn{3}{c}{3-spot model} & \colhead{Units} \\
            \colhead{} & \colhead{} & \colhead{Spot 1 (star A)} & \colhead{Spot 2 (star B)} & \colhead{Spot 3 (star A)} & \colhead{} 
}
\startdata
		Spot size					& $\alpha$							& $16.43^{+2.33}_{-2.43}$	& $12.86^{+3.05}_{-2.62}$	& $18.85^{+4.21}_{-4.12}$	& $^o$ \\
		Latitude					& $\beta_0$							& $25.67^{+8.71}_{-7.26}$	& $80.53^{+3.40}_{-6.28}$	& $69.11^{+8.64}_{-8.04}$		& $^o$ \\
		Longitude					& $\phi_0$							& $231.19^{+1.58}_{-1.74}$	& $288.02^{+3.41}_{-3.42}$	& $44.50^{+2.99}_{-3.12}$	& $^o$ \\
		Period					& $P_{\rm spot}$						& $13.09^{+0.03}_{-0.04}$	& $7.50\pm0.03$			& $13.08\pm0.05$			& days \\
		Temperature				& $T_{\rm spot}$						& $5519^{+48}_{-59}$		& $4779^{+163}_{-194}$		& $5540^{+41}_{-49}$		& K \\
		Total flux of unspotted stars	& $F_{\rm max}$						& \multicolumn{3}{c}{$1.008\pm0.002$}											& $-$ \\
		Jitter						& $s$								& \multicolumn{3}{c}{$0.0009\pm0.00004$}										& $-$ \\
		RMS of Everest data			& ${\rm rms}_{\rm raw}$					& \multicolumn{3}{c}{$0.00186$}												& $-$ \\
		RMS of residuals			& ${\rm rms}_{\rm residual}$				& \multicolumn{3}{c}{$0.00085$}												& $-$ \\
		RMS ratio					& ${\rm rms}_{\rm resid} / {\rm rms}_{\rm raw}$	& \multicolumn{3}{c}{$0.46$}													&  $-$ \\
\enddata
\end{deluxetable*}

Using the output of this model, we can constrain the stellar inclination of the two binary components. We use the stellar radii and $v\sin I_{\rm s}$ values obtained from spectral analysis, together with our estimated spot rotation periods, to estimate stellar inclinations of $I_{\rm A}=66.8^o\pm0.7$ and $I_{\rm B}=39.7^o\pm0.2$. Such a misalignment between the rotation axes of the two stellar components is unsurprising \citet{1994AJ....107..306H}

Our result for the primary star agrees with the assessment of \citet{2016AJ....151..150M}, who estimated a stellar inclination of $>50^o$ for the planet hosting WASP-85\,A. \citeauthor{2016AJ....151..150M} also estimated a projected obliquity (the angle between the planet's orbital axis and the host star's rotation axis) for WASP-85\,A\,b of $\lambda<14^o$, which we use with our estimates of the stellar and orbital inclinations to constrain the three-dimensional obliquity of WASP-85\,A\,b to be $\psi<27^o$. We therefore conclude that the system is very likely to be aligned.

\subsection{Stellar age}
\label{sec:age}
We use several independent methods to produce estimates of the ages of our planetary systems. The results are listed in Table\,\ref{tab:ages}.

\begin{deluxetable}{lllll}
\tabletypesize{\scriptsize}
\caption{Age estimates for the systems discussed herein, as determined using a variety of techniques.\\ We consider the GARSTEC results to be the most representative.\label{tab:ages}}
\centering
\tablehead{
		\colhead{Method} & \multicolumn{4}{c}{Ages} \\
		\colhead{} & \colhead{WASP-85\,A} & \colhead{WASP-85\,B} & \colhead{WASP-116\,A} & \colhead{WASP-149\,A} \\
		\colhead{} & \colhead{(Gyr)} & \colhead{(Gyr)} & \colhead{(Gyr)} & \colhead{(Gyr)}
}
\startdata
		\multicolumn{5}{l}{\textit{Isochrone placement}\tablenotemark{$\dagger$}} \\
		Padova			& $0.5^{+0.3}_{-0.1}$		& $1.1^{+10.4}_{-1.0}$		& $8.6^{+3.6}_{-1.6}$	& $1.6^{+0.4}_{-0.6}$	 \\
		YY				& $1.2^{+1.7}_{-0.1}$		& $1.2^{+9.0}_{-1.0}$		& $7.5^{+2.0}_{-1.4}$	& $2.2^{+2.9}_{-2.0}$	\\
		DSED 			& $2.4^{+0.7}_{-2.2}$		& $>0.1$					& $>8.2$				& $3.2^{+0.7}_{-1.5}$	 \\
		GARSTEC 		& $0.3\pm0.3$ 				& $7.3\pm4.4$				& $7.0\pm0.9$			& $2.7\pm1.5$	 \\ \\
		\multicolumn{5}{l}{\textit{Gyrochronology}} \\
		$P_{\rm rot, WASP}$ & $1.53^{+0.32}_{-0.28}$		& $-$					& $-$				& $-$	\\
		$P_{\rm rot, K2}$	& $1.24^{+0.07}_{-0.06}$ 		& $0.40\pm0.01$			& $-$				& $-$	\\
		\vsini\ 			& $<1.45^{+1.25}_{-0.58}$	& $<0.81^{+0.61}_{-0.30}$	& $<8.3$				& $<1.3$ \\ \\
		\multicolumn{5}{l}{\textit{Lithium abundance}} \\
		$\log {\rm A(Li)}$	& $>0.6$					& $-$ 					& $>0.6$				& $-$	\\
\enddata
\tablenotetext{\dagger}{For WASP-85\,B, $\rho_\star$ was taken from spectral analysis as there is no independent measurement.}
\end{deluxetable}

\subsubsection{Model fitting}
\label{sec:ageModel}
We place constraints on stellar age using two different model fitting methods. First, we use the method of \citet{2014MNRAS.442.1844B}, applying it in $[\rho_\star^{-1/3}, \teff]$ parameter space as suggested by \citet{2007ApJ...664.1190S}. We make use of three different set of stellar models when using this method: the Yonsei-Yale (YY) isochrones of \citet{2004ApJS..155..667D}; the Padova models of \citet{2008AA...482..883M,2010ApJ...724.1030G}, models from the Dartmouth Stellar Evolution Database (DSED; \citealt{2008ApJS..178...89D}). To improve the accuracy of our age estimates, we make use of independent measurements of $\rho_\star$ and \teff; the former derived directly from the photometric transits (see Section\,\ref{sec:orbit}), the latter estimated from stellar spectra (see Section\,\ref{sec:star} above). In the case of WASP-85\,B, as there is no independent measure of $\rho_\star$ we use the value derived from the HARPS spectra. We account for the uncertainty in [Fe/H] by carrying out model fits at both the central values and the $1\sigma$ limits given in Table\,\ref{tab:star_params}. 

Our second method use the Bayesian fitting process described in \citet{2015AA...575A..36M}, and available as the open source \textsc{BAGEMASS}\footnote{\url{https://sourceforge.net/projects/bagemass/}} code. This uses the GARSTEC models of \citet{2008ApSS.316...99W}, as computed by \citet{2013MNRAS.429.3645S}, and works in $[\log(L_\star), \teff\ ]$. Again, we use independent measurements of the two parameters (except for WASP-85\,B), with $L_{\star}$ computed from the results of our global modelling (see Section\,\ref{sec:orbit}).

\begin{figure*}
\gridline{\fig{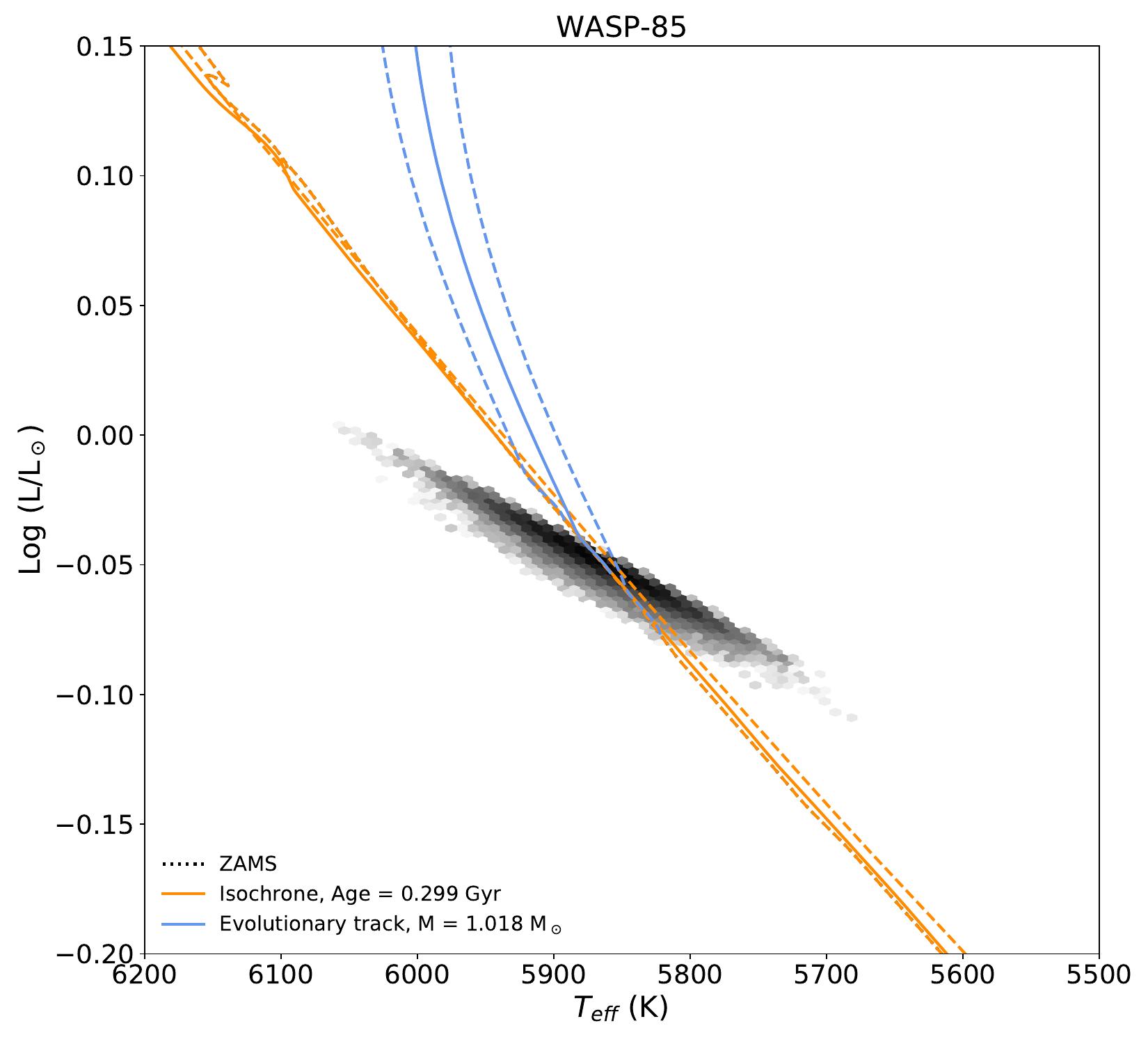}{0.48\textwidth}{}
          \fig{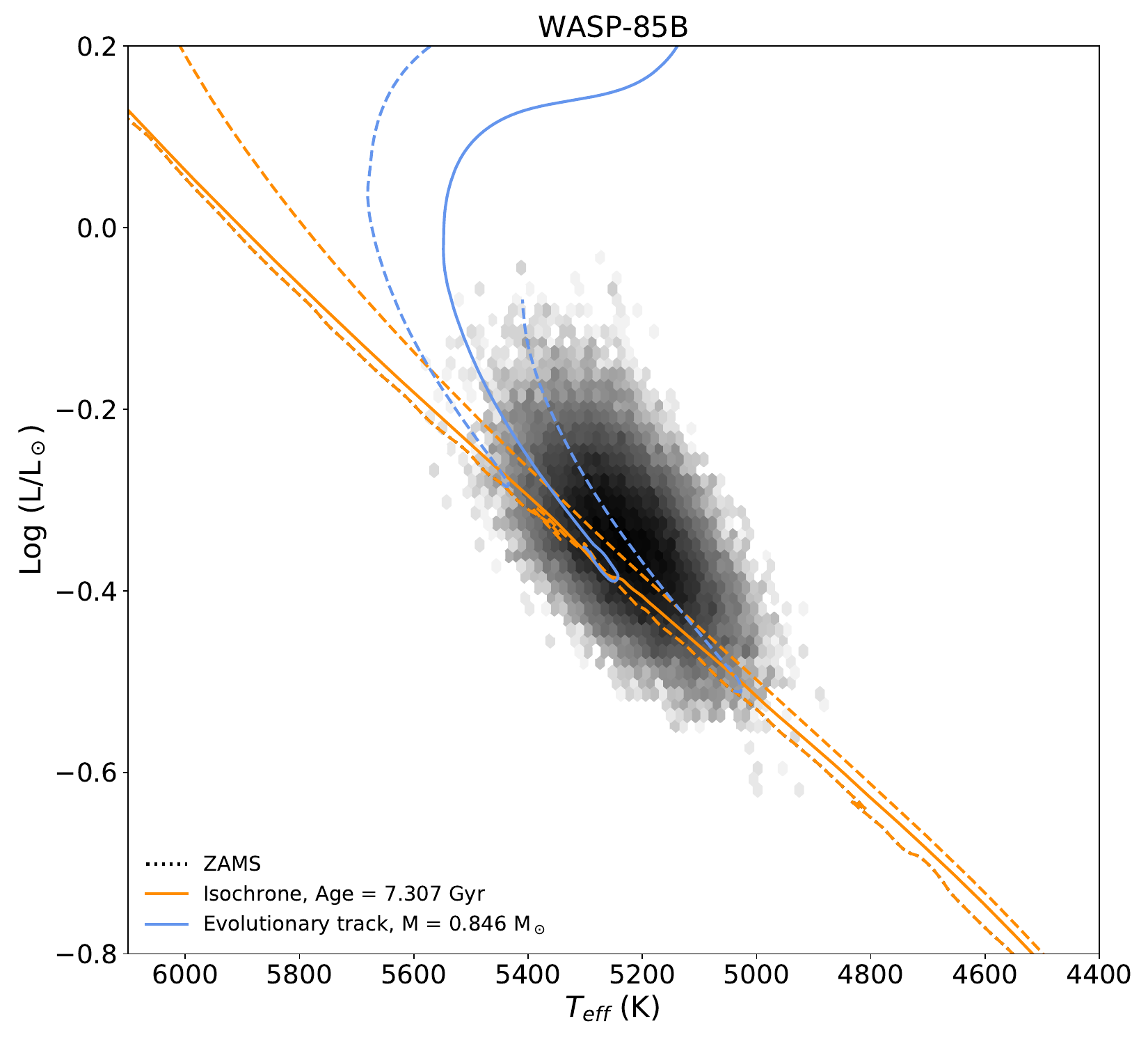}{0.48\textwidth}{}
          }
\gridline{\fig{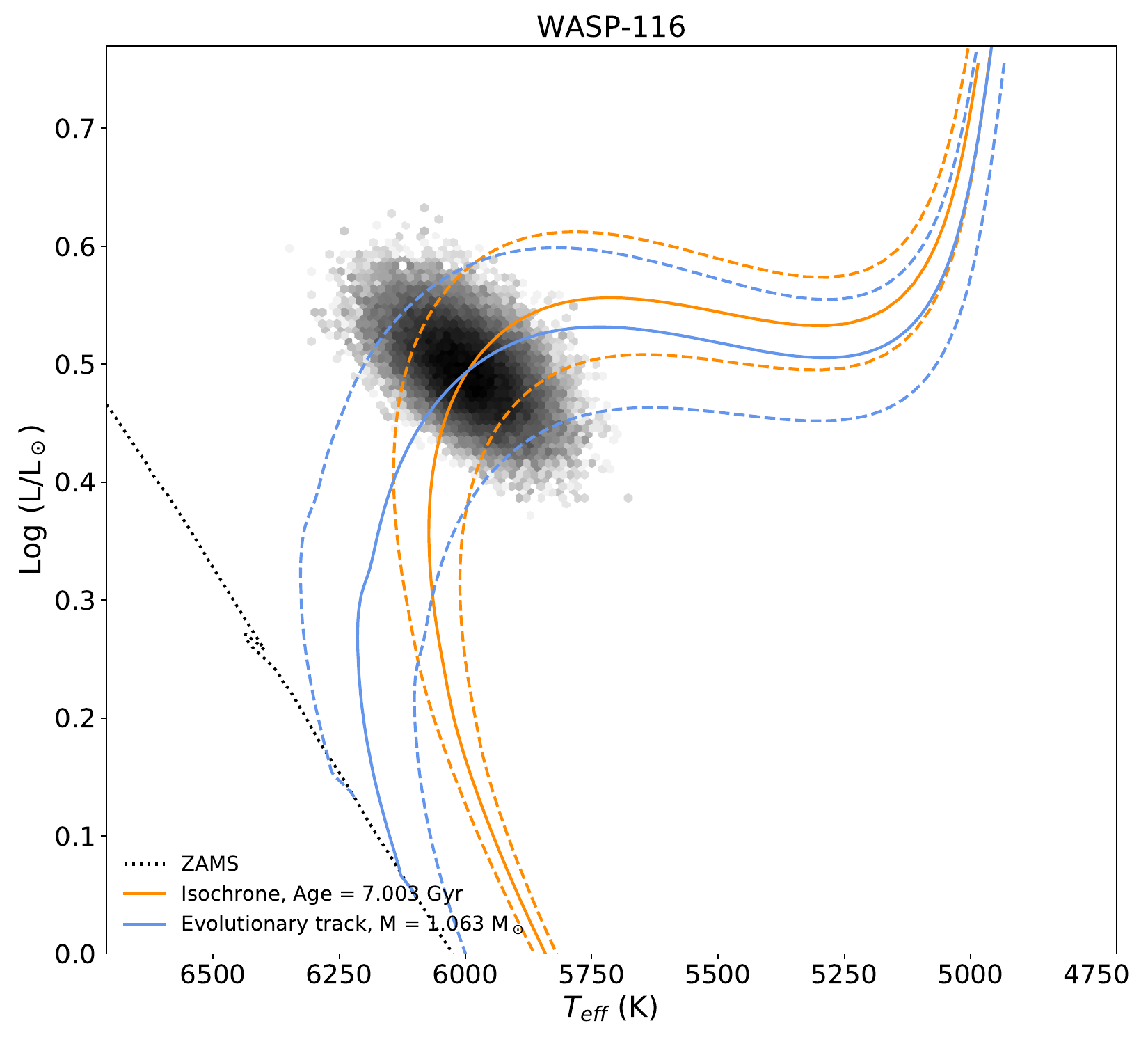}{0.48\textwidth}{}
          \fig{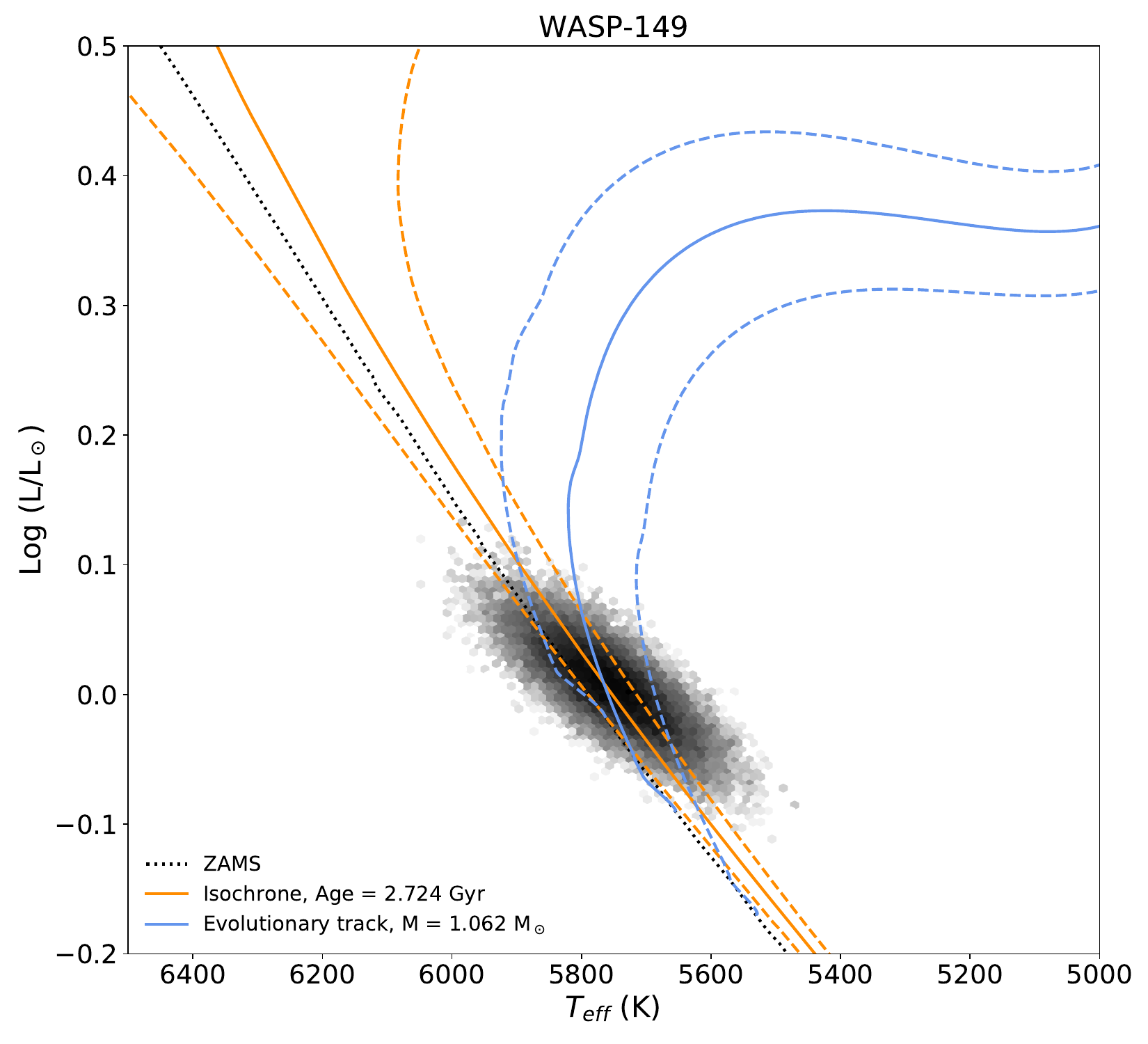}{0.48\textwidth}{}
          }
	\caption{The results from stellar model fitting using \textsc{BAGEMASS}, showing the best-fitting stellar isochrones and evolutionary tracks, along with the posterior probability distribution of the MCMC fitting process. The ZAMS is shown as a dotted black line. The solid blue line is the best-fitting stellar evolutionary tracks, with the blue dashed lines representing evolutionary tracks for the $1\sigma$ limits on stellar mass. The solid orange line is the stellar isochrone, with the orange dashed lines representing isochrones for the $1\sigma$ limits on stellar age. The posterior probability distribution of the MCMC fitting process is shown with the colour scale representing the density of points.}
	\label{fig:bagemass}
\end{figure*}

\subsubsection{Gyrochronology}
\label{sec:ageGyro}
For each of our target stars, we also estimate the ages of the host stars using their estimated rotation periods. We draw $10^4$ samples from skewed Gaussian distributions in the projected stellar rotation, $v\sin I_s$, the stellar radius, $R_s$, and the (B-V) colour.  We assume that the stellar rotation axis is perpendicular to the line of sight. $v\sin I_s$ and $R_s$ values are taken from our spectral analysis results (see Section\,\ref{sec:star}), while the broad band (B-V) colour indices are derived from AAVSO Photometric All-Sky Survey (APASS; \citealt{2012JAVSO..40..430H}) data. For each set of samples we derive an upper limit on the rotation period using the sampled values of $v\sin I_s$ and $R_s$, and calculate the age of the system using the gyrochronology formulation of \citet{2010ApJ...722..222B}, calculating the convective turnover timescale using Table\,1 of \citet{2010ApJ...721..675B} and setting $P_0=1.1$\,d. We take the median value of the resulting age distribution as our result. Since the derived rotation period is an upper limit, this age is also an upper limit, assuming that the star has spun-down in isolation.

For the WASP-85 system, we are also able to derive gyrochronological ages for both stars using our measured rotation periods from the WASP and K2 photometry.

\subsubsection{Lithium abundance}
\label{sec:lithium}
Lithium is detected in WASP-85\,A, with an equivalent width of $53$\,m\AA, corresponding to an abundance of $\log A{\rm (Li)}=2.19\pm0.06$. This implies an age of $>0.6$\,Gyr \citep{2005AA...442..615S}; note that we are unable to place an upper limit on age using the lithium abundance, as the measured value falls on the lithium `plateau' exhibited by some solar-type stars older than $1$\,Gyr \citep{2009IAUS..258..133R}. Similarly, we detect lithium in the spectrum of WASP-116\,A with an equivalent width corresponding to $\log A{\rm (Li)}=2.35\pm0.08$. This also implies an age of $>0.6$\,Gyr, and again falls on the `plateau' for older solar-type stars.

There is no significant detection of lithium in WASP-85\,B, with an equivalent width upper limit of $2$\,m\AA, corresponding to an abundance upper limit of $\log A{\rm (Li)}<0.70\pm0.10$. No lithium was identified in the spectrum of WASP-149\,A either; we place an upper limit on the abundance for that star of $\log A{\rm (Li)}<0.8$. Thus, we cannot place a limit on the ages of WASP-85\,B or WASP-149\,A using this method.

\subsubsection{Comparing methods}
\label{sec:compare}
The ages that we obtain using our different methods are generally consistent with each other, though as expected there is some variation between the different sets of stellar models. This spread of age estimates arises from the differing input physics and assumptions that are used to construct the models. In general, we gravitate towards the results obtained using the combination of the GARSTEC models and \textsc{BAGEMASS} as they use the most comprehensive set of models, and the MCMC approach of \textsc{BAGEMASS} leads to a more representative age estimate that more rigorously accounts for the shape of the stellar models and isochrones.

The ages for WASP-85\,A that we derive using the four different isochrone analyses are broadly consistent, though the Yonsei-Yale models suggest a slightly older age that is in agreement with the gyrochronology ages that we derive. This suggests that the star has not been spun up through tidal interactions. However, WASP-85\,B's age is less clear. If we assume that the two stars are gravitationally bound then we would anticipate them to be co-evolving and thus to show similar ages. However, the age constraints from stellar model fitting for WASP-85\,B are very loose, easily accomodating a much older star than WASP-85\,A; note the relevant GARSTEC results in particular. It is possible that these seemingly incompatible ages arise because we have no independent measurement of $\rho_\star$ for WASP-85\,B, and thus used the stellar parameters from our spectral analysis. We note that the gyrochronology results for the binary companion, in contrast with the model fitting results, imply that WASP-85\,B is of similar age to the planet hosting WASP-85\,A.

It is also possible that the temperate estimate for WASP-85\,A is biasing the model fitting towards a younger age. We note that there is a discrepancy between the temperature derived from the HARPS spectrum and that derived from our global modelling. However, using the latter temperature has a negligible effect on the estimated age for WASP-85\,A.

WASP-116 is significantly older than WASP-85\,A, and appears to be evolving off the main sequence (see Figure\,\ref{fig:bagemass}). The age estimates from all of the model fitting agree on this point, but they do not all agree with each other; the lower limit provided by the DSED models is at odds with the GARSTEC result. The upper limit on age that we obtain from gyrochronology is very similar to the DSED limit as well, and together these age estimates leave only a small range of possible ages for the system, a range that disagrees with the GARSTEC result. However, the gyrochronology result was obtained using the stellar $v\sin I_{\rm s}$ rather than a rotation period measurement, and thus is not entirely reliable. 

For WASP-149, the upper age limit provided by gyrochronology (again derived from $v\sin I_{\rm s}$) disagrees with the result obtained using the DSED models. We note, however, that the four stellar model fitting results are in agreement. 

\section{Modelling approach}
\label{sec:orbit}
We simultaneously model all of the available photometric (both WASP discovery data and follow-up transit lightcurves) and spectroscopic data for each system. In this way we can fully characterise inter-parameter correlations and intra-parameter uncertainties. We use a Markov Chain Monte Carlo (MCMC) code based on the approach described in \citet{cameron2007} and \citet{2008MNRAS.385.1576P}, and which has since been used in a range of publications concerning WASP planetary systems \citep[e.g.][]{2017MNRAS.464..810B,2016arXiv160804225F,2018arXiv181209264A,2018MNRAS.480.5307T,2019MNRAS.482.1379H}. Our jump parameters (see Table\,\ref{tab:params}) are chosen to minimise inter-parameter correlations, and to maximise the mutual orthogonality of the parameter set. We use $\sqrt{e}\sin(\omega)$ and $\sqrt{e}\cos(\omega)$ to impose a uniform prior on $e$ and avoid bias towards higher values \citep{2006ApJ...642..505F, 2011ApJ...726L..19A}. As standard practice we apply Gaussian priors on stellar effective temperature, \teff\ and [Fe/H] using the results from our spectral analysis; prior values are listed in Table\,\ref{tab:results}. We also have a set of optional constraints that we can choose to adopt, namely: 
\begin{enumerate}
	\item Applying a Gaussian prior on $v\sin I_{\rm s}$.
	\item Forcing the planet's orbit to be circular, $e\overset{!}{=}0$. This indirectly controls the jump parameters $e\sin w$ and $e\cos w$, setting them equal to zero.
	\item Forcing the barycentric system RV to be constant with time, $\dot{\gamma}{\overset{!}{=}}0$, neglecting long-term trends that are indicative of third bodies.
	\item Forcing the stellar radius, $R_{\rm s}$, to follow a main sequence relationship with $M_{\rm s}$, \textit{or} use the result from spectral analysis as a prior on $R_{\rm s}$.
\end{enumerate}
These constraints are all \textit{optional}, and are not necessarily applied to each system being analysed. Note that we prefer to use the stellar radius from spectral analysis for the prior rather than the value derived from \textit{Gaia} data owing to the extinction issue discussed previously.

We compare solutions from the equivalent eccentric and non-eccentric combinations using the F-test of \citet{1971AJ.....76..544L}, with the circular solution as the null hypothesis \citep{2012MNRAS.422.1988A}. We adopt this approach on the basis that tidal circularisation timescales for hot Jupiters are often significantly shorter than their expected lifetimes \citep{2011MNRAS.414.1278P}. We similarly adopt $\dot{\gamma}=0$ as our null hypothesis, testing solutions with non-zero $\dot{\gamma}$ for significance using the reduced $\chi^2$ as calculated for the spectroscopic data only. 

We discuss these analyses in the following sections. Once all combinations have been examined, we identify the most suitable combination by selecting that which provides the minimal value of the reduced chi-squared statistic, $\chi^2_{\rm red}$. This combination is then reported as the \textit{final solution} for each system.

At each MCMC step we calculate both a photometric transit model (using the approach of \citealt{2002ApJ...580L.171M}) and a Keplerian radial velocity curve. We remove systematic trends from our photometric data using a linear decorrelation with time, and account for limb darkening using a non-linear, four-component model; we derive wavelength appropriate coefficients by interpolating the tables of \citet{2000AA...363.1081C,2004AA...428.1001C}. We calculate stellar radius using the value of $R_{\rm s}/a$ derived from the transit model, calculating the semi-major axis from the orbital period using Kepler's third law. Stellar mass is determined using the $T_{\rm eff}$--$M_{\rm s}$ calibration of \citet{2010AARv..18...67T}, with updated parameters from \citet{southworth2011}. Quality of fit  is determined using the $\chi^2$ statistic, and the Metropolis--Hastings decision maker \citep{1953JChPh..21.1087M,Hastings1970} is used to accept or reject each MCMC step.

We use a burn-in phase with a minimum length of $5\times10^3$ accepted steps. Once this minimum is reached, we check for convergence using the simple method of \citep{2008ApJ...673..526K}, which compares the value of $\chi^2$ for the current step to the median of all previous values in the chain. We also carry out secondary checks for convergence using trace plots, autocorrelation statistics, inspection of one- and two-dimensional parameter distributions, and the \citet{gelman1992} statistic. If convergence is not achieved within $10^4$ steps, we restart the chain.

Following convergence, we use a set of $10^3$ accepted steps to rescale the uncertainties on the primary jump parameters. We then run the chain for a further $2\times10^4$ accepted steps, taking the last $10^4$ steps as the production runs. We run five separate chains, and concatenate the production runs to produce a final chain of length $5\times10^4$ steps. We then test this final chain for full convergence using the approach of \citet{geweke1992}, re-running chains as necessary until our test indicates a fully-converged final chain. Our reported parameters are the median values from this chain, with the listed uncertainties taken to be the values enclosing the $68.3$\,percent confidence interval.

We use the TDB time standard in conjunction with Barycentric Julian Dates (BJD), as recommended by \citet{2010PASP..122..935E}. We take the equatorial Solar and Jovian radii and masses taken from Allen's Astrophysical Quantities as our standard values.

\begin{deluxetable*}{llll}
\tabletypesize{\scriptsize}
\caption{Details of the jump parameters that we use for our MCMC analysis. These parameters have been selected to maximise mutual orthogonality, and minimise correlations. We note those parameters controlled by Gaussian priors in column 4. $e\sin w$ and $e\cos w$ are controlled by the prior on orbital eccentricity, $e$. Priors listed as `Yes/No' are controlled through the optional constraints listed in the text. \label{tab:params}}
\tablehead{\colhead{Parameter} & \colhead{Units} & \colhead{Symbol} & \colhead{Prior?} 
}
\startdata
		Epoch					& ${\rm BJD}_{\rm TDB}-2450000$	& $t_0$				& No							\\
		Orbital period				& days						& $P_{\rm orb}$		& No							\\
		Transit width				& days						& $W$				& No							\\
		Transit depth				& --							& $d$				& No							\\
		Impact parameter			& Stellar radii					& $b$				& Yes/No						\\
		Effective temperature		& K							& $T_{\rm eff}$			& Yes						\\
		`Metallicity'				& dex						& $[{\rm Fe}/{\rm H}]$	& Yes						\\
		RV semi-amplitude			& km s$^{-1}$					& $K$				& No							\\
		$\sqrt{e}\sin(\omega)$		& --							& $e\sin w$			& indirectly; Yes/No				\\
		$\sqrt{e}\cos(\omega)$		& --							& $e\cos w$			& indirectly; Yes/No 				\\
		Long-term RV trend			& --							& $\dot{\gamma}$		& Yes/No						\\
		Stellar rotation velocity		& (km s$^{-1}$)$^{-1/2}$			& $v\sin (I_{\rm s})$		& Yes/No						\\		
\enddata
\end{deluxetable*}

\section{Results for WASP-85}
\label{sec:85model}
As noted in Section\,\ref{sec:85phot}, owing to the proximity of the binary companion all of the photometric observations of the WASP-85 system contain light from both stars. Following the examples of \citet{2014MNRAS.445.1114A}, \citet{2014AA...572A..49N}, and \citet{2013PASP..125...48M}, we corrected for this dilution by manually increasing the depth of the WASP-85 transits, accounting for the third light contribution following the model of \citet{2010MNRAS.408.1689S}. For observations in the in the $z$ band, we use the flux ratio of $0.54\pm0.02$ measured during the course of the IO:O observations. For observations in the Cousins\,R, RG, and \textit{Kepler} filters we scale this ratio assuming perfect blackbody emission for the two stars, giving a flux ration of 0.50 for all cases. In addition, owing to the presence of an additional trend in the LT / IO:O lightcurves (probably secondary extinction) we found it necessary to use a quadratic decorrelation in phase for the two lightcurves obtained with that instrument. Finally, during each of the TRAPPIST transit observations the telescope underwent a meridian flip to allow it to continue observing the system. To reduce potential systematics associated with these events, the data pre- and post-flip were fit as separate lightcurves.

The CORALIE and SOPHIE RVs are also affected by contamination caused by the presence of the companion star. Our HARPS observations show that WASP-85\,B is relatively constant in RV, and thus one of the CCF peaks will be constant in velocity space, while the second CCF peak representing WASP-85\,A will shift with orbital phase. Owing to the very similar systemic velocities for the two stars (as expected if they are bound), however, these peaks merge to form a single, mildly asymmetric CCF. The measured velocity will be biased towards the stationary position of WASP-85\,B, effectively reducing the shift in velocity induced by the planet, and leading to a smaller semi-amplitude. We attempted to correct for this by modelling the CCFs as a pair of Gaussian functions. We used the flux ratio, \vsini\ values, and mean RV for star B to place priors on the relative amplitudes, widths, and positions, respectively, of the two peaks, but were unable to disentangle the two stars' CCFs. We therefore elect not to include the RV data obtained using CORALIE and SOPHIE in our final fit, using only the uncontaminated HARPS data for WASP-85\,A. We add an additional stellar ``jitter'' term of $2$\,\ms to the formal RV uncertainties in order to obtain a reduced spectroscopic $\chi^2\approx1$.

\begin{figure}
	\includegraphics[width=0.48\textwidth]{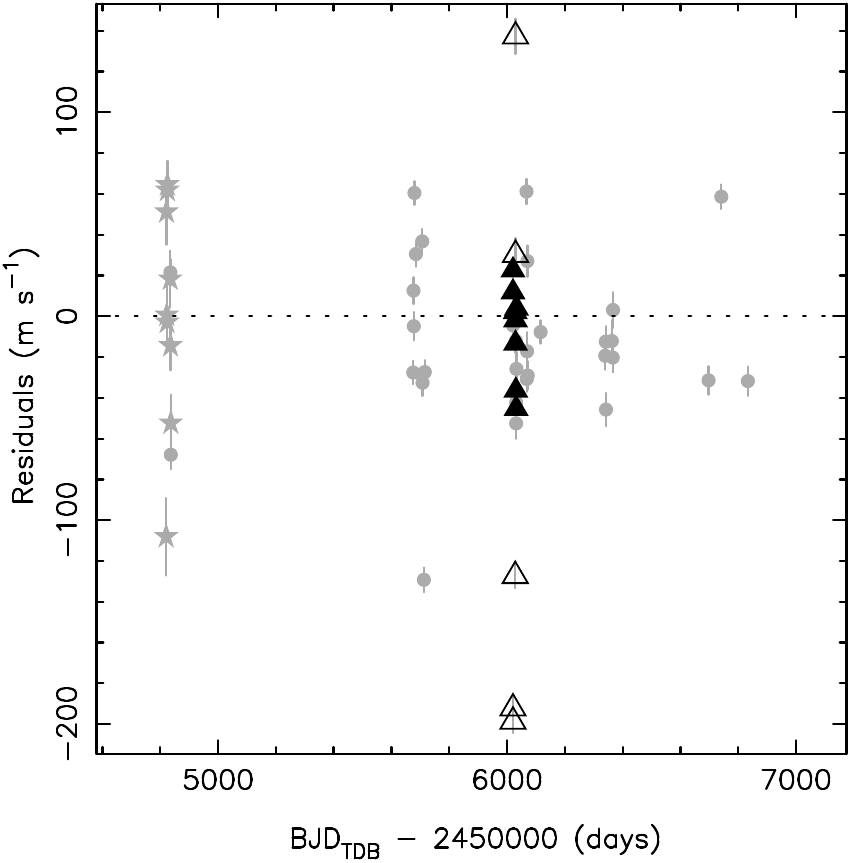}
	\caption{Radial velocity residuals for WASP-85 as a function of time. Residuals are calculated compared to the RV model from our MCMC global analysis. HARPS data for WASP-85\,A are denoted by solid, black triangles, data for WASP-85\,B by open, black triangles.  The CORALIE and SOPHIE data, which were not used in our global modelling, are denoted by solid grey circles and stars, respectively. It is clear that the HARPS data used for the global analysis cover too short a baseline to constrain any linear trend in RV. Furthermore, there is no evidence for a long-term trend over the full timespan of our observations.\\ \vspace{0.1cm}\\
	}
	\label{fig:85rvtime}
\end{figure}

We find a significant improvement in the spectroscopic fit (parameterised using $\chi^2_{\rm reduced}$) when including an RV trend, which we measure at $-110\pm3$\,m\,s$^{-1}$\,yr$^{-1}$. However, we are sceptical that such a trend truly exists. The timespan of the HARPS observations is only twelve days, covering just over $4.5$ orbits. This is insufficient to truly constrain any trend that might be present in the systemic velocity. Moreover, the CORALIE and SOPHIE data show no signs of any shift in $\gamma$ over time (see Figure\,\ref{fig:85rvtime}). Even with the dilution caused by WASP-85\,B, a trend of $-110\pm3$\,m\,s$^{-1}$\,yr$^{-1}$ would be easily identifiable, as those data cover nearly $2000$\,days, over five years, and the postulated trend is on the order of the detected semi-amplitude. We therefore do not fit a trend in our final solution.

We find eccentricities of between $0.0763$ and $0.1774$, depending on the optional constraints that we apply. All are consistent with zero at $<2\sigma$, suggesting that our null hypothesis is correct. We test this further by using the F-test of \citet{1971AJ.....76..544L} to compare the eccentric and circular fits, finding that the probability of the eccentricity being detected by chance varies between $0.79$ to $0.06$. Furthermore, the short timespan of the HARPS observations provides insufficient constraint even if the eccentricity was found to be significant.We therefore adopt the null hypothesis of a circular orbit.

One of the HARPS RV data has an uncertainty three times that of the other measurements, and a bisector span that is twice as large as the next greatest value. We carried out runs omitting this datum to check that it was not biasing our results. We found that removing this point produced an insignificant increase in the value of the fitted trend in $\gamma$, and that the eccentricity of the orbit tended to increase with the removal of the highly uncertain HARPS measurement, though we note that eccentricity was still not significantly detected. We attribute this to the spacing of the RV data in orbital phase, which leads to a fit dominated by the three data between phase $0$ and phase $0.2$.

\begin{figure}
	\includegraphics[width=0.48\textwidth]{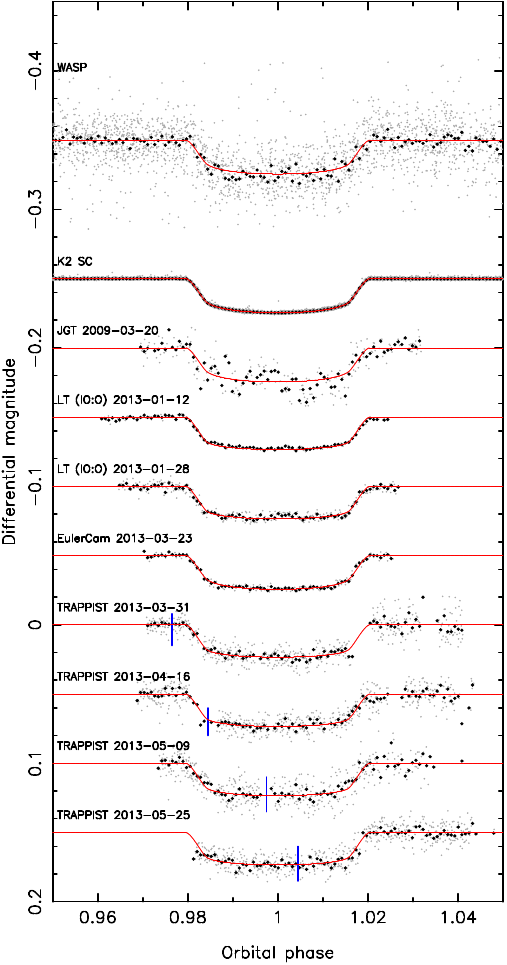}
	\caption{Photometric transit lightcurves of the WASP-85 system. The data have been phase folded using the best-fitting orbital period and epoch for WASP-85\,A\,b, and binned using a bin width equivalent to $180$\,s. We plot the original data in grey, and the best-fitting transit model in red. The lightcurves have been offset for clarity; for the same reason, we omit the error bars. The telescope, instrument, and date of observation are listed alongside each lightcurve. A slewing problem is responsible for the gaps in the lightcurve obtained by TRAPPIST on 2013-03-31. The vertical blue lines denote the timing of the meridian flips undergone by TRAPPIST. \\ \vspace{0.1cm}\\
	}
	\label{fig:85photometry}
\end{figure}

The system parameters that we adopt are listed in Table\,\ref{tab:results}, along with the prior values. We list the \teff\ and ${\rm [Fe/H]}$ results from our MCMC for completeness, and to verify consistency with the spectroscopically-derived values for these particular parameters. Though ${\rm [Fe/H]}$ is consistent between the MCMC and the spectral analysis, we note that the global analysis finds a significantly hotter star than is implied by the HARPS data. A similar discrepancy was noted by \citet{2016AJ....151..150M} in reference to an earlier version of this paper with the same spectroscopic \teff\ , though our MCMC temperature increases the apparent discrepancy. As suggested by \citeauthor{2016AJ....151..150M}, this may imply that the HARPS data are not completely uncontaminated, though any contamination would be at a sufficiently low level to have negligible effect on the orbital solution. The rest of our solution agrees with the results in table\,1 of \citeauthor{2016AJ....151..150M}, as is to be expected given that the same dataset was used for both analyses.

\begin{figure}
	\includegraphics[width=0.48\textwidth]{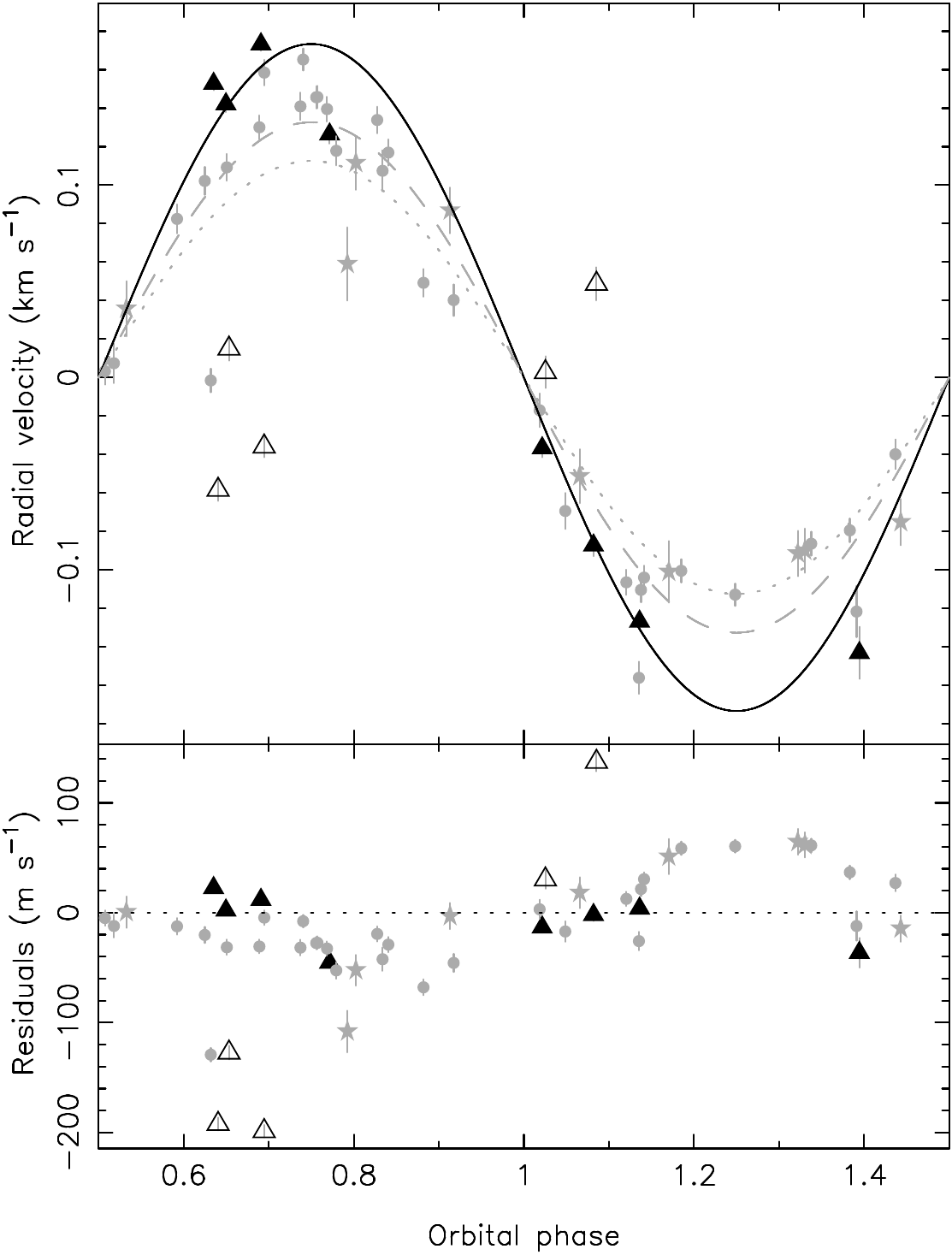}
	\caption{\textit{Upper panel: }Radial velocity data for WASP-85, phase folded using the best-fitting orbital period and epoch for WASP-85\,A\,b. HARPS data for WASP-85\,A are denoted by solid, black triangles, data for WASP-85\,B by open, black triangles.  The CORALIE and SOPHIE data, which were not used in our global modelling, are denoted by solid grey circles and stars, respectively. The best-fitting $\gamma_i$ value for each set of data has been subtracted to allow comparison. Overplotted is our best-fitting orbital solution as derived from our MCMC global analysis using only the HARPS data to constrain the RV curve. Also shown are the Keplerian solutions derived using $K$ for the CORALIE data (dashed line) and SOPHIE data (dotted line), which clearly show the effect of contamination in reducing the semi-amplitude. The data for WASP-85\,B show variation that is uncorrelated with the phase of the planet's orbital solution. \textit{Lower panel: }Radial velocity residuals as compared to the best-fitting global model (solid line in the upper panel). The effect of contamination gives a sinusoidal form to the residuals for the SOPHIE and CORALIE data.\\ \vspace{0.1cm}\\
	}
	\label{fig:85rvcurve}
\end{figure}

In Figure\,\ref{fig:85photometry} we display the phase-folded photometric lightcurves, overlaid with the best-fitting transit model. There is a notable difference in scatter between the two LT / IO:O lightcurves. This is likely to be a result of the comparison stars that were available for the data reduction process. For the lightcurve obtained on the night of 2013-01-12 there were two suitable comparison stars, which were fainter than than WASP-85 by factors of $1.8$ and $2.25$. For the lightcurve obtained on the night of 2013-01-28, only one comparison star was available, and it was four times fainter than our target. In Figure\,\ref{fig:85rvcurve} we plot our HARPS RV data, overlaid with the best-fitting Keplerian orbit model for HARPS only analysis. We show all of our RV data to illustrate the effect of the dilution on the CORALIE and SOPHIE observations, with the data from these instruments greyed-out to indicate that it was not fit. The primary effect is a reduction in the semi-amplitude of the data (see also Table\,\ref{tab:results}), though there is also increased scatter compared to the HARPS observations. We also show the residuals of the data compared to the plotted HARPS model. The sinusoidal structure present in the residuals for the CORALIE and SOPHIE data clearly indicates the reduced semi-amplitude.

\subsection{Diluted observations}
\label{sec:dilution}
We tested the effect of the dilution on our derived parameters by carrying out several different analyses. We first fit the uncorrected photometry and HARPS RVs using the same initial conditions and prior set, finding a transit depth of $0.0149\pm0.0001$, some $20$\,percent smaller than the depth when using the contamination-corrected photometry.

To characterise the effect of the third light dilution on the CORALIE and SOPHIE RV data, we carry out an additional fit using the same initial conditions, but including all of our RV data. We find that this makes little difference to the reported solution, but does provide additional useful information on the level of dilution. For this analysis, we use a number of $K_i$ jump parameters equal to the number of datasets. $K_{\rm SOPHIE}$ is significantly smaller than $K_{\rm CORALIE}$, which is in turn significantly smaller than $K_{\rm HARPS}$. This is as expected given the different instrument specifications, and the different levels of dilution that are expected in the three sets of data.

As expected, the inclusion of the CORALIE and SOPHIE data reduces the magnitude of the trend in systemic velocity to zero.

We also test the effect of the diluted RVs on the planet mass that we derive by carrying out runs using the CORALIE data or SOPHIE data in place of the HARPS RVs. When using the CORALIE RVs we find a planet mass of $0.93\pm0.01$\,$M_{\rm Jup}$, $26$\,percent smaller than the value we find using the HARPS data. With the SOPHIE data, we find $M_{\rm p}=0.79\pm0.05$\,$M_{\rm Jup}$, $37$\,percent smaller than the HARPS result. These discrepancies match the differences in $K_i$ that we find when fitting all of the RV data simultaneously, as expected.

We estimate the RV signature that would be induced in WASP-85\,A by the binary companion. We calculate an expected RV semi-amplitude of $K_{\rm binary}=1180$\,m\,s$^{-1}$ using the orbital period implied by the change in position angle. The greatest negative rate of change of RV occurs at phase 0; adopting this as our initial condition, we calculate $\Delta{\rm RV}=-0.04$\,m\,s$^{-1}$ over the course of one year.

\section{Results for WASP-116}
\label{sec:116model}
We approximate the NGTS and blue-blocking filters as the Cousins $I$ and Cousins $Z$ filters, respectively, for the purposes of limb darkening coefficients. We adopt the null hypothesis of a circular orbit with no long-term velocity trend, and apply an additional Gaussian prior on $v\sin I_{\rm s}$ to combat the tendency for the MCMC algorithm to boost the rotation velocity to speeds that are unphysical for the physical stellar parameters measured from our co-added CORALIE spectra. We find that the overall fit to our data is significantly worse when forcing the stellar parameters to follow a main-sequence mass-radius relationship, and that doing so also forces the MCMC towards hotter stars with either super-Solar metallicity, or metallicity that is closer to the Solar value than our spectral analysis implies. However, allowing the stellar mass and radius to vary freely leads to a significantly more massive, significantly larger star than is suggested by our analysis of the CORALIE spectra, albeit one with an effective temperature and metallicity consistent with our spectrally derived values.

Since the temperature and metallicity are derived directly from the CORALIE spectra, while the mass and radius are derived from the calibration of \cite{2010AARv..18...67T}, we prefer to use a combination of priors that return consistent values of the former two parameters. We therefore allow the stellar mass and radius to vary freely during the MCMC chain. Our age estimates for WASP-116 support this choice of prior; the left panel of Figure\,\ref{fig:bagemass} indicates that the host star is slightly evolved, and using the stellar mass in Table\,\ref{tab:results} we estimate the main sequence lifetime to be $4.2$\,Gyr ($5.0$\,Gyr with the mass from spectral analysis), significantly shorter than the $7.0\pm0.9$\,Gyr estimated by \textsc{BAGEMASS}.

As before, we show our final, adopted system parameters, and our prior values, in Table\,\ref{tab:results}, listing the \teff\ and ${\rm [Fe/H]}$ results from our MCMC for completeness, and to verify consistency with our adopted spectroscopically-derived values.

WASP-116\,b has a relatively long orbital period that results in a transit duration of nearly six hours. This makes it very difficult to observe a complete transit from the ground whilst also acquiring sufficient out-of-transit data to give a secure baseline. As noted in Section\,\ref{sec:116phot}, this meant that our photometric follow-up efforts were unable to secure a complete transit observation.  This adversely affects the quality of our solution for the system, particularly in terms of the uncertainty on the orbital inclination and impact parameter, and likely also contributes to the problems encountered with the masses and radii discussed above. We note however, that WASP-116 lies in \textit{TESS} sector 4, so further photometry will be available in the near future that will enable the system's parameters to be refined.

\begin{figure}
	\includegraphics[width=0.48\textwidth]{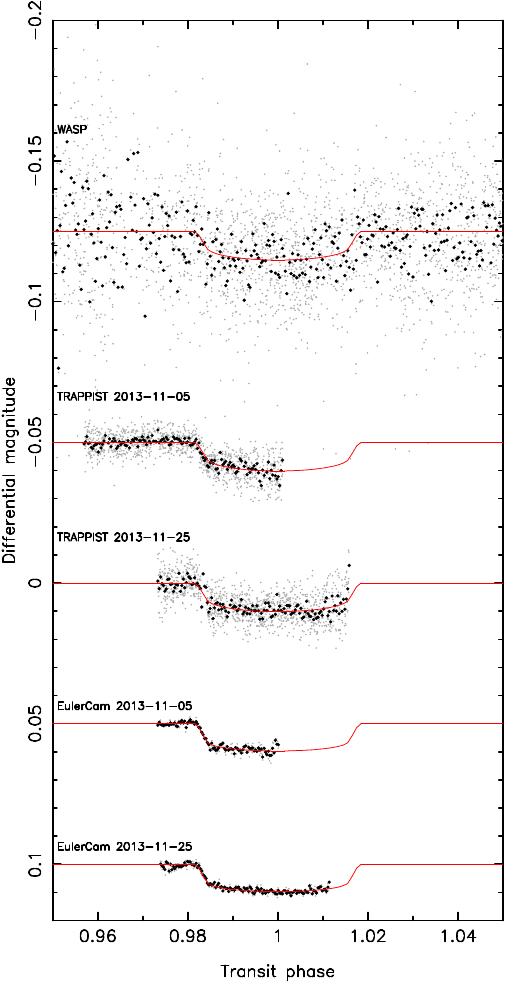}
	\caption{Photometric transit lightcurves of the WASP-116 system. The data have been phase folded using the best-fitting orbital period and epoch for WASP-116\,b, and binned using a bin width equivalent to $180$\,s. We plot the original data in grey, and the best-fitting transit model in red. The lightcurves have been offset for clarity; for the same reason, we omit the error bars. The telescope, instrument, and date of observation are listed alongside each lightcurve. \vspace{0.1cm}\\
	}
	\label{fig:116photometry}
\end{figure}

\begin{figure}
	\includegraphics[width=0.48\textwidth]{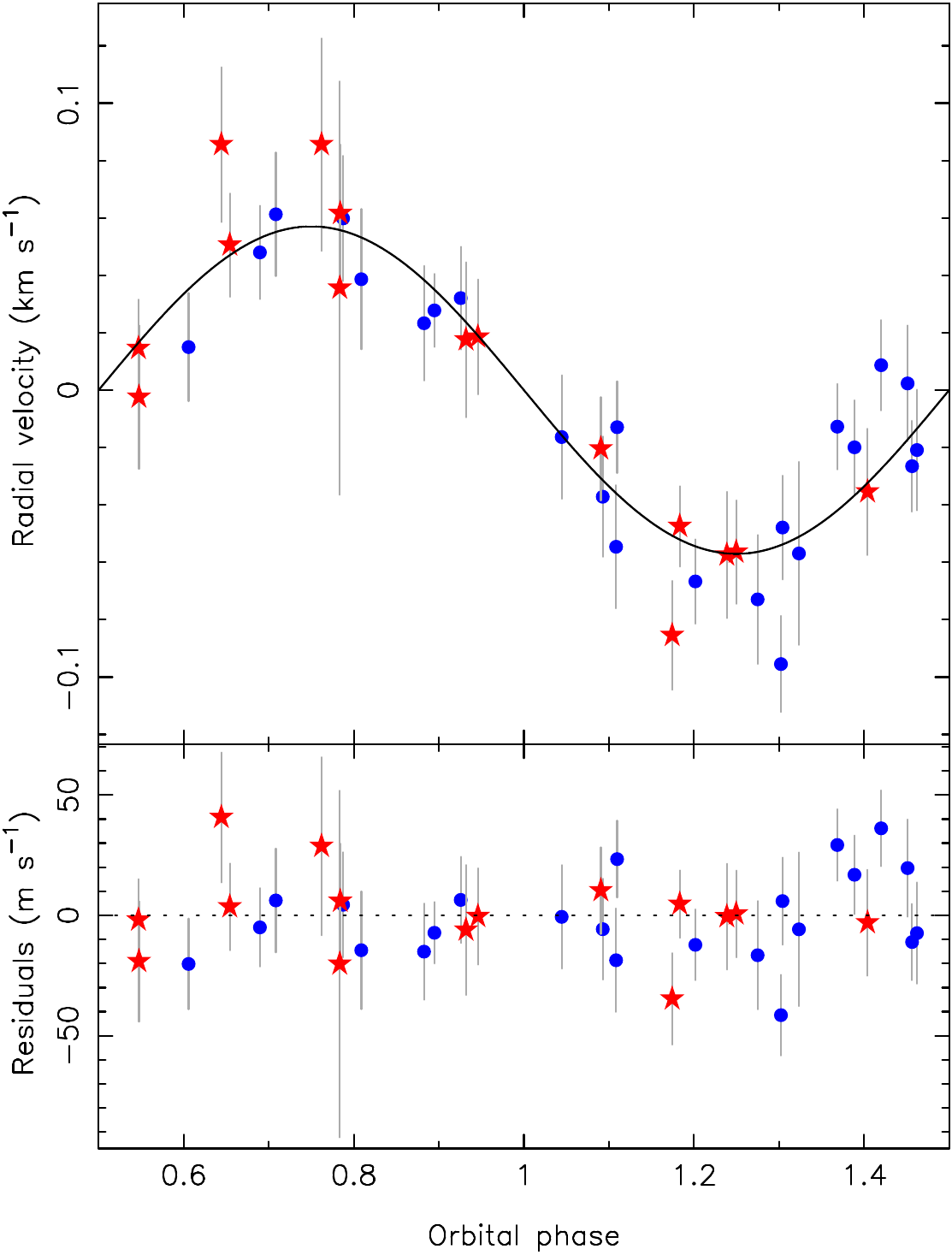}
	\caption{\textit{Upper panel: }Radial velocity data for WASP-116, phase folded using the best-fitting orbital period and epoch for WASP-116\,b. CORALIE data are denoted by blue circles, while SOPHIE data are denoted by red stars. The best-fitting $\gamma_i$ value for each set of data has been subtracted. Overplotted is our best-fitting orbital solution as derived from our MCMC analysis. \textit{Lower panel: }Radial velocity residuals as compared to the best-fitting model shown in the upper panel.\\ \vspace{0.1cm}\\
	}
	\label{fig:116rvcurve}
\end{figure}

\section{Results for WASP-149}
\label{sec:149model}
We approximate the clear filter used by NITES as the Cousins $R$ filter for the purposes of limb darkening, and similarly approximate the ZG filter as Sloan $z\prime$. We initially carried out two separate fits to the WASP-149 data in an effort to characterise the dilution in the WASP photometry. The first fit used all of the photometric and spectroscopic data, while the second used only the WASP photometry alongside the full set of spectroscopic data. We found that the transit depth that we obtained using only the WASP photometry was $0.0115^{+0.0007}_{-0.0005}$, compared to $0.0170\pm0.0001$ when we used all of the photometry, a decrease of $32\%$.

We corrected for this dilution by manually increasing the depth of the WASP-149 transits. We took the best-fitting ephemeris and transit width from the full data fit, using these to identify data that fell inside a transit window. These data were multiplied by the ratio of our two fitted depths; this increases the depth of the WASP transits to match those of the follow-up photometry, but does increase the in-transit scatter in the WASP data. However, the fit is dominated by the better quality follow-up photometry (as evidenced by the fact that our initial all-data fit provided a depth that matched their transits), so the effect on the final results of this increased scatter is minimal.

We again apply an additional Gaussian prior on $v\sin I_{\rm s}$, as without this the MCMC was boosting the rotational velocity of the star to unphysical values. We also again adopt the null hypothesis of a circular orbit with no long-term velocity trend. The eccentric fits that we explored were not significantly detected, and the trends that we derived when allowing them in our fits were all consistent with $0$. We found no significant differences between the results obtained when allowing the stellar physical parameters to vary freely, and those obtained when the parameters were constrained by the main sequence mass-radius relation. Moreover, all of our exploratory analyses returned effective temperatures and metallicities in agreement with the spectral analysis results. We therefore allow the stellar mass and radius to vary freely in our final fit, which shows a larger, more massive star than is suggested by the spectral analysis.

As for our other two systems, Table\,\ref{tab:results} displays the prior values and our adopted system parameters, as well as listing list the \teff\ and ${\rm [Fe/H]}$ results from our MCMC for completeness and consistency verification purposes.

We note that WASP-149 lies in \textit{TESS} sector 7, so space-based photometry will be available in the near future that will allow for refinement of our solution.

\begin{figure}
	\includegraphics[width=0.48\textwidth]{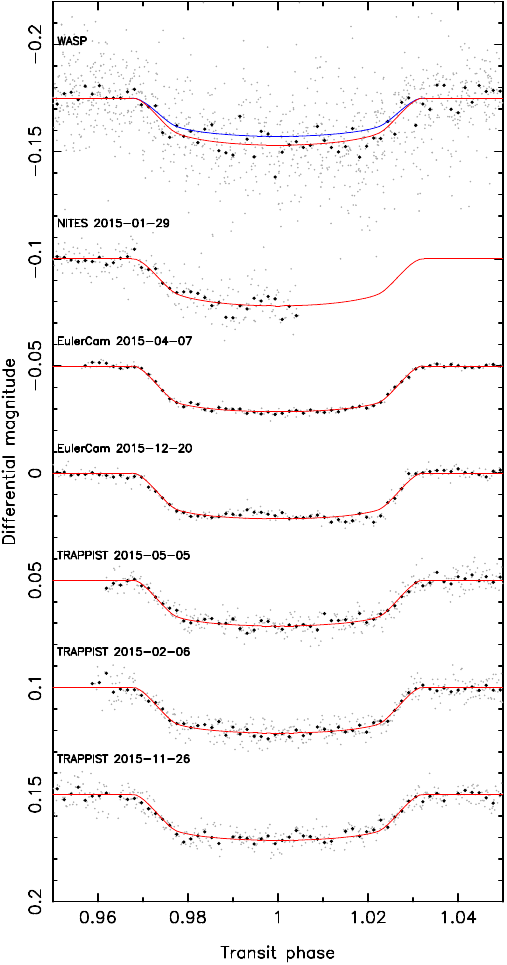}
	\caption{Photometric transit lightcurves of the WASP-149 system. The data have been phase folded using the best-fitting orbital period and epoch for WASP-149\,b, and binned using a bin width equivalent to $180$\,s. We plot the original data in grey, and the best-fitting transit model in red. Note that the WASP data have had the transit depth artificially increased to match the follow-up photometry; we plot the best-fitting transit model for the original WASP data in blue. The lightcurves have been offset for clarity; for the same reason, we omit the error bars. The telescope, instrument, and date of observation are listed alongside each lightcurve.\vspace{0.1cm}\\
	}
	\label{fig:149photometry}
\end{figure}

\begin{figure}
	\includegraphics[width=0.48\textwidth]{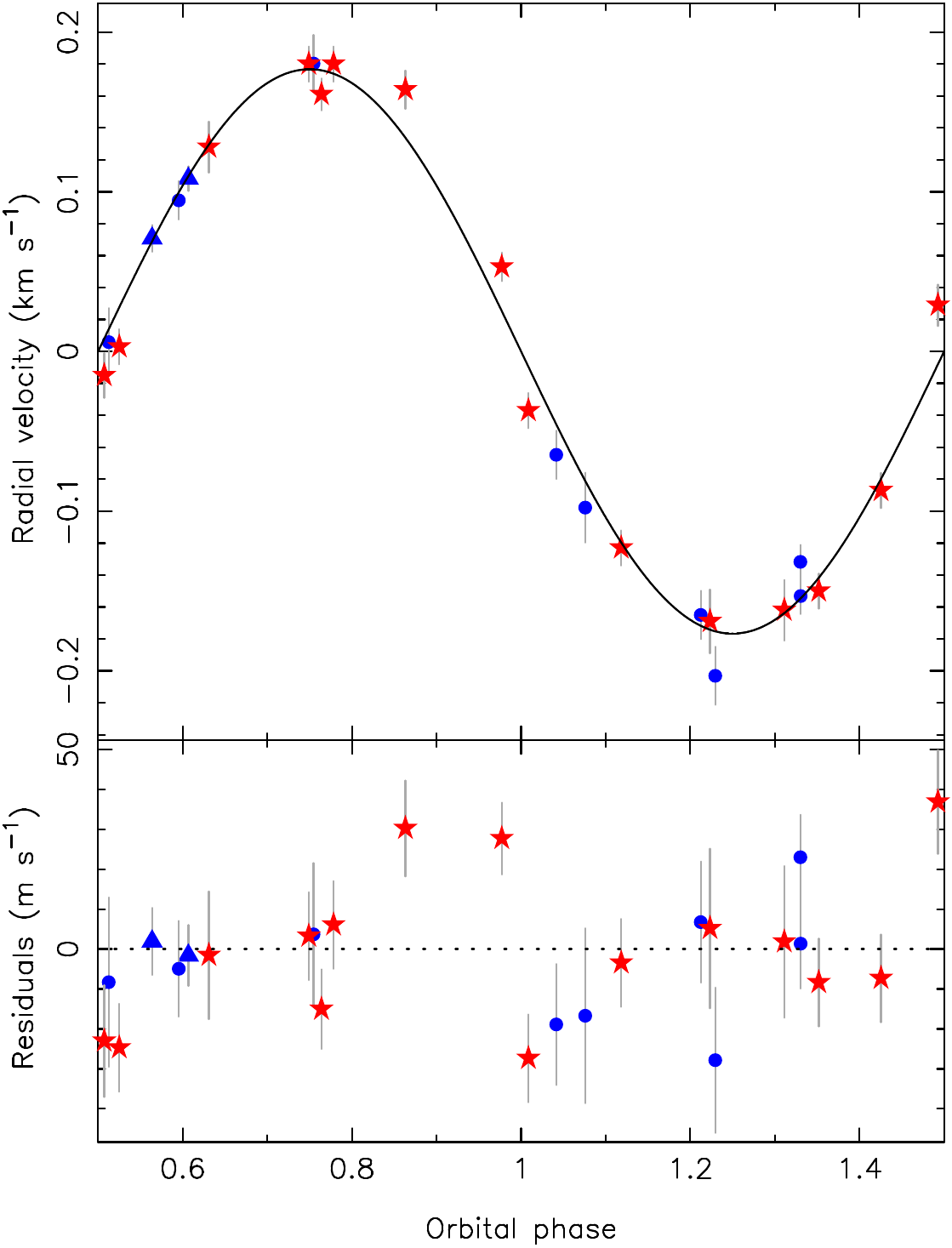}
	\caption{\textit{Upper panel: }Radial velocity data for WASP-149, phase folded using the best-fitting orbital period and epoch for WASP-149\,b. Pre-upgrade CORALIE data are denoted by blue triangles, while post-upgrade data are denoted by blue circles. SOPHIE data are denoted by red stars. The best-fitting $\gamma_i$ value for each set of data has been subtracted. Overplotted is our best-fitting orbital solution as derived from our MCMC analysis. \textit{Lower panel: }Radial velocity residuals as compared to the best-fitting model shown in the upper panel.\\ \vspace{0.1cm}\\
	}
	\label{fig:149rvcurve}
\end{figure}

\section{Discussion}
\label{sec:discuss}
\begin{deluxetable*}{llllll}
\tabletypesize{\scriptsize}
\caption{System parameters for WASP-85\,A and WASP-85\,A\,b, WASP-116\,A and WASP-116\,b, and WASP-149\,A and WASP-149\,b. We also list the Bayesian priors applied in the course of the MCMC analysis. The flux ratios listed for WASP-85 were used to manually correct the photometry prior to starting the analysis.	\label{tab:results}}
\centering
\tablehead{\colhead{Parameter} & \colhead{Symbol} & \multicolumn{3}{c}{System values for:} & \colhead{Units} \\
			\colhead{} & \colhead{} & \colhead{WASP-85} & \colhead{WASP-116} & \colhead{WASP-149} & \colhead{}
}
\startdata
		\multicolumn{4}{l}{\textit{Priors}} \\
		Projected rotation velocity		& \vsini\					& $3.41\pm0.89$				& $1.7\pm1.1$ 						& $4.6\pm0.6$				& \kms\ \\
		Effective temperature 		& \teff\					& $5685\pm65$				& $5950\pm100$ 					& $5750\pm100$			& K \\
		Metallicity 					& [Fe/H] 					& $0.08\pm0.10$ 				& $-0.28\pm0.10$ 					& $0.16\pm0.11$			& dex \\
		Flux ratio 					& $L_3$ 					& $0.50$ (adopted) 				& $-$ 							& $-$					& Cousins R filter \\
		Flux ratio 					& $L_3$ 					& $0.50$ (adopted)				& $-$ 							& $-$					& RG filter \\
		Flux ratio 					& $L_3$ 					& $0.54$ (adopted) 				& $-$ 							& $-$					& Sloan $z'$ filter\\
		Flux ratio 					& $L_3$					& $0.50$ (adopted) 				& $-$ 							& $-$					& \textit{Kepler} filter\\
		\hline \\
		\multicolumn{4}{l}{\textit{Model parameters}} \\
		Orbital period 				& $P$ 					& $2.6556763\pm0.0000003$ 		& $6.61321\pm0.00002$				& $1.332811\pm0.000001$	& days \\
		Epoch of mid-transit 			& $t_0$ 					& $6844.81719\pm0.00001$ 		& $6602.85352^{+0.0039}_{-0.0029}$ 	& $7160.52490\pm0.00007$	& ${\rm BJD}_{\rm TDB}-2450000$ \\
		Transit duration 			& $T_{\rm dur}$			& $0.10817\pm0.00002$ 			& $0.243^{+0.009}_{-0.006}$	 		& $0.0842\pm0.0006$		& days \\
		Planet:star area ratio 		& $R_{\rm p}^2/R_\star^2$	& $0.018718^{+0.000004}_{-0.000010}$		& $0.0077\pm0.0002$				& $0.0176\pm0.0002$		& \\
		Impact parameter 			& $b$ 					& $0.042\pm0.002$ 				& $0.103^{+0.089}_{-0.066}$			& $0.49\pm0.03$			& \\
		RV semi-amplitude 			& $K$					& $-$						& $57.7\pm4.4$					& $177.6^{+4.2}_{-4.8}$		& m\,s$^{-1}$ \\
								& $K_{\rm HARPS}$ 		& $173.3\pm1.8$		 		& $-$							& $-$					& m\,s$^{-1}$ \\
								& ${K_{\rm CORALIE}}$ 		& $132.8\pm1.7$\tablenotemark{$\ast$} 			& $-$							& $-$					& m\,s$^{-1}$ \\
								& ${K_{\rm SOPHIE}}$ 		& $112.5^{+6.2}_{-6.5}$\tablenotemark{$\ast$}	 	& $-$							& $-$					& m\,s$^{-1}$ \\
		Systemic velocity 			& $\gamma_{\rm HARPS}$ 				& $13.5946\pm0.0018$ 	& $-$ 				& $-$					& \kms \\
								& ${\gamma_{\rm CORALIE}}$ pre-upgrade	& $-$ 						& $-10.9827\pm0.0038$	& $17.9046\pm0.0006$		& \kms\ \\
								& ${\gamma_{\rm CORALIE}}$ post upgrade 	& $13.5335\pm0.0013$\tablenotemark{$\dagger$} 	& $-$				& $17.9021\pm0.0005$		& \kms\ \\
								& ${\gamma_{\rm SOPHIE}}$ 				& $13.4610\pm0.0045$\tablenotemark{$\dagger$} 	& $-11.0255\pm0.0054$	& $17.8579\pm0.0003$		& \kms\ \\
		Effective temperature 		& \teff\ 					& $6150\pm15$ 			& $5967^{+97}_{-75}$				& $5781^{+106}_{-90}$		& K \\
		Metallicity 					& [Fe/H]		 			& $0.10^{+0.13}_{-0.10}$			& $-0.28\pm0.10$					& $0.15\pm0.12$			& dex \\
		Limb darkening	(R-band)		& $R_1$					& $0.57$						& $0.41$							& $0.42$					& \\
								& $R_2$					& $0.05$						& $0.20$							& $0.28$					& \\
								& $R_3$					& $0.31$						& $0.20$							& $0.002$					& \\
								& $R_4$					& $-0.21$						& $-0.11$							& $0.007$					& \\
		Limb darkening	(I-band)		& $I_1$					& $-$						& $0.65$							& $-$					& \\
								& $I_2$					& $-$						& $-0.24$							& $-$					& \\
								& $I_3$					& $-$						& $0.42$							& $-$					& \\
								& $I_4$					& $-$						& $-0.17$							& $-$					& \\
		Limb darkening	(z'-band)		& $z_1$					& $0.66$						& $-$							& $0.74$					& \\
								& $z_2$					& $-0.32$						& $-$							& $-0.50$					& \\
								& $z_3$					& $0.55$						& $-$							& $0.66$					& \\
								& $z_4$					& $-0.28$						& $-$							& $-0.24$					& \\
		Limb darkening	(\textit{Kepler}-band)		& $k_1$			& $0.57$						& $-$							& $-$					& \\
								& $k_2$					& $0.01$						& $-$							& $-$					& \\
								& $k_3$					& $0.41$						& $-$							& $-$					& \\
								& $k_4$					& $-0.26$						& $-$							& $-$					& \\
		\multicolumn{4}{l}{\textit{Derived parameters}} \\
		Ingress / egress duration 		& $T_{12}=T_{34}$ 			& $0.013037\pm0.000002$ 			& $0.0202\pm0.0001$				& $0.0123\pm0.0001$		& days\\
		Orbital inclination 			& $i_{\rm p}$ 				& $89.73\pm0.01$		 		& $89.38^{+0.40}_{-0.54}$			& $84.68\pm0.31$			& $^\circ$ \\
		Orbital eccentricity 			& $e$ 					& $0$ (adopted) 				& $0$ (adopted)					& $0$ (adopted)			& \\
		Stellar mass 				& $M_\star$ 				& $1.09\pm0.09$ 				& $1.18\pm0.05$					& $1.09\pm0.06$			& $M_\sun$ \\
		Stellar radius 				& $R_\star$ 				& $0.94\pm0.03$ 				& $1.68\pm0.07$					& $1.02\pm0.03$			& $R_\sun$ \\
		Stellar surface gravity 		& \logg\ 					& $4.533\pm0.011$ 				& $4.074\pm0.003$					& $4.478\pm0.001$			& (cgs) \\
		Stellar density 				& $\rho_\star$ 				& $1.330\pm0.001$ 				& $0.25\pm0.02$					& $1.03\pm0.06$			& $\rho_\sun$ \\
		Planet mass 				& $M_{\rm p}$ 				& $1.25\pm0.03$ 				& $0.59\pm0.05$					& $1.02\pm0.04$			& $M_{\rm Jup}$ \\
		Planet radius 				& $R_{\rm p}$ 				& $1.25\pm0.07$ 				& $1.43\pm0.07$					& $1.32\pm0.04$			& $R_{\rm Jup}$ \\
		Planet surface gravity 		& $\log g_{\rm p}$ 			& $3.267\pm0.005$ 				& $2.839\pm0.004$					& $3.164\pm0.001$			& (cgs) \\
		Planet density 				& $\rho_{\rm p}$ 			& $0.646\pm0.018$ 				& $0.213\pm0.003$					& $0.497\pm0.001$			& $\rho_{\rm Jup}$ \\
		Scaled stellar radius 			& $R_\star/a$ 				& $0.11264\pm0.00002$ 			& $0.1071^{+0.0036}_{-0.0030}$		& $0.1945\pm0.0037$		& \\
		Semi-major axis 			& $a$ 					& $0.0386\pm0.0010$ 			& $0.0730\pm0.0011$				& $0.0247\pm0.0004$		& AU \\
		Planet equilibrium temperature & $T_{{\rm eq}=0}$ 			& $1459\pm4$ 				& $1400\pm4$						& $1870\pm1$				& K \\
\enddata
\tablenotetext{\ast}{Values of $K$ for the CORALIE and SOPHIE observations of WASP-85\,AB were determined through MCMC runs using only the sets of RV data obtained by those instruments.}
\tablenotetext{\dagger}{Values of $\gamma$ for the CORALIE and SOPHIE observations of WASP-85\,AB were determined through an MCMC run using all three sets of RV data. This run did not significantly change the other model or derived parameters.}
\end{deluxetable*}

\subsection{WASP-85\,B - the binary companion}
\label{sec:binary}
WASP-85 is listed as a known binary in the Washington Double Star catalogue (WDS). We obtained the full set of WDS measurements for the system, which stretch back to $1881$. We also measured the position angle and angular separation of the companion star through careful analysis of EulerCAM observations taken on 2012 February 7th.

\begin{figure}
	\includegraphics[width=0.48\textwidth]{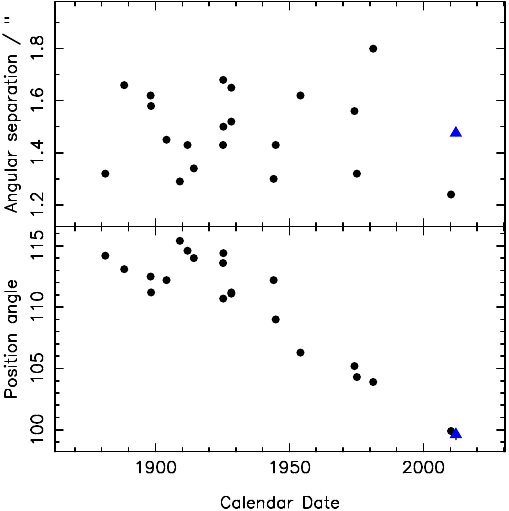}
	\caption{\textit{Upper panel: }Angular separation of the two binary components as a function of calendar date. There appears to be short term variation with a period of approximately $30$\,years. The minimum separation implies an orbital distance of $\sim150$\,AU. \textit{Lower panel: }Binary position angle as a function of calendar date. There is a clear, long-term trend for decreasing position angle. This suggests an orbital period for the binary of $\gtrsim3000$\,years. \\
	In both panels, black circles indicate data obtained from the Washington Double Star (WDS) database, while blue triangles indicate data derived from EulerCam observations. The WDS data have no associated uncertainties. The uncertainty on the EulerCam separation measurement is smaller than the symbol size.\\ \vspace{0.1cm}\\}
	\label{fig:PAbinary}
\end{figure}

In Figure\,\ref{fig:PAbinary} we show both the binary angular separation and the binary position angle as a function of calendar date. A Lomb-Scargle periodogram reveals no significant periodicity in the angular separation of the two stellar components, though there is substantial scatter about the mean. Future data releases from \textit{Gaia} will be of great help in reducing the uncertainty in the binary separation.

\begin{deluxetable}{lll}
\tabletypesize{\scriptsize}
\caption{Position angles and angular separations for the binary star BD+07$^{\circ}$2474 (WASP-85\,A\,B). The majority of the measurements were obtained from the Washington Double Star catalogue. The most recent datum (marked with $^\ast$) was obtained through analysis of images from EulerCAM, taken on 2012 February 7th. \label{tab:PA}}
\centering
\tablehead{\colhead{Date} & \colhead{Position Angle} & \colhead{Angular Separation} \\
		    \colhead{(year)} & \colhead{(degrees)} & \colhead{(arc sec)}
}
\startdata
		$1881.32$ & $114.2$ & $1.32$ \\
		$1888.351$ & $113.1$ & $1.66$ \\
		$1898.215$ & $112.5$ & $1.62$ \\
		$1898.40$ & $111.2$ & $1.58$ \\
		$1904.138$ & $112.2$ & $1.45$ \\
		$1909.14$ & $115.4$ & $1.29$ \\
		$1911.98$ & $114.6$ & $1.43$ \\
		$1914.32$ & $114$ & $1.34$ \\
		$1925.22$ & $113.6$ & $1.43$ \\
		$1925.31$ & $110.7$ & $1.68$ \\
		$1925.36$ & $114.4	$ & $1.50$ \\
		$1928.31$ & $111.1$ & $1.52$ \\
		$1928.35$ & $111.2$ & $1.65$ \\
		$1944.11$ & $112.2$ & $1.30$ \\
		$1944.87$ & $109$ & $1.43$ \\
		$1954.11$ & $106.3$ & $1.62$ \\
		$1974.286$ & $105.2$ & $1.56$ \\
		$1975.16$ & $104.3$ & $1.32$ \\
		$1981.24$ & $103.9$ & $1.80$ \\
		$2010.287$ & $99.9$ & $1.24$ \\
		$2012.107^\ast$ & $99.62\pm0.41$ & $1.48\pm0.01$ \\
\enddata
\end{deluxetable}

The position angle of the binary shows a clear, long-term, decreasing trend. This is a clear indication of orbital motion for the binary, and suggests an orbital period of $\gtrsim3000$\,years. This change of position angle suggests that the planetary orbit is misaligned from the plane of the binary orbit. 

Using the distance listed in Table\,\ref{tab:star_params}, the mean separation of $1.5\pm0.1$\,\arcsec corresponds to a distance between the stars of $210\pm22$\,AU. Using the masses that we estimate for the two components from our spectral analysis, and assuming a circular orbit, this indicates a period of $2190\pm340$\,years, less than the lower limit of $3000$\,years suggested by the position angle change. The reverse transformation, from $P_{\rm binary}=3000$\,years, gives a binary distance of $259$\,AU, discrepant with the separation derived value. This mismatch suggest that the binary orbit is inclined relative to our line of sight by $\sim45^\circ$, again indicating a misalignment with the planet's orbital plane. Such misalignment between the binary orbital plane and the planetary orbital plane has been observed in young protoplanetary discs \citet{2014Natur.511..567J, 2019NatAs.tmp..189K}.

The presence of the binary companion at $<259$\,AU from the planet host star suggests that the protoplanetary disc was likely truncated, limiting the quantity of material available for planet formation. Exploring this possibility is beyond the scope of this paper, but could be an interesting subject for future work.

\subsection{Examining the population}
\label{sec:population}
We compare our new discoveries to the existing population of transiting exoplanets. From the \url{exoplanets.org} database\footnote{Accessed on 2019-01-31} we select all systems with measured planetary radius, planetary mass,  orbital inclination, and stellar effective temperature. These cuts provides a sample of $366$.

We calculate the stellar luminosity for the host stars of our sample, and use this to calculate the incident flux on each of the exoplanets. We then calculate a simple estimate of the equilibrium temperature, $T_{\rm eq}$, of each planet in our sample, assuming an albedo of $0.503$ (equal to the Bond albedo of Jupiter, \citealt{bondalbedoJupiter}) and a circular orbits for all planets. We also assume that each planet has a uniform day-side temperature and a nightside temperature of $0$\,K, with no recirculation, such that the area ratio for absorption / radiation of energy is $0.5$ \citep{2011ApJ...729...54C}. 

\begin{figure}
	\includegraphics[width=0.48\textwidth]{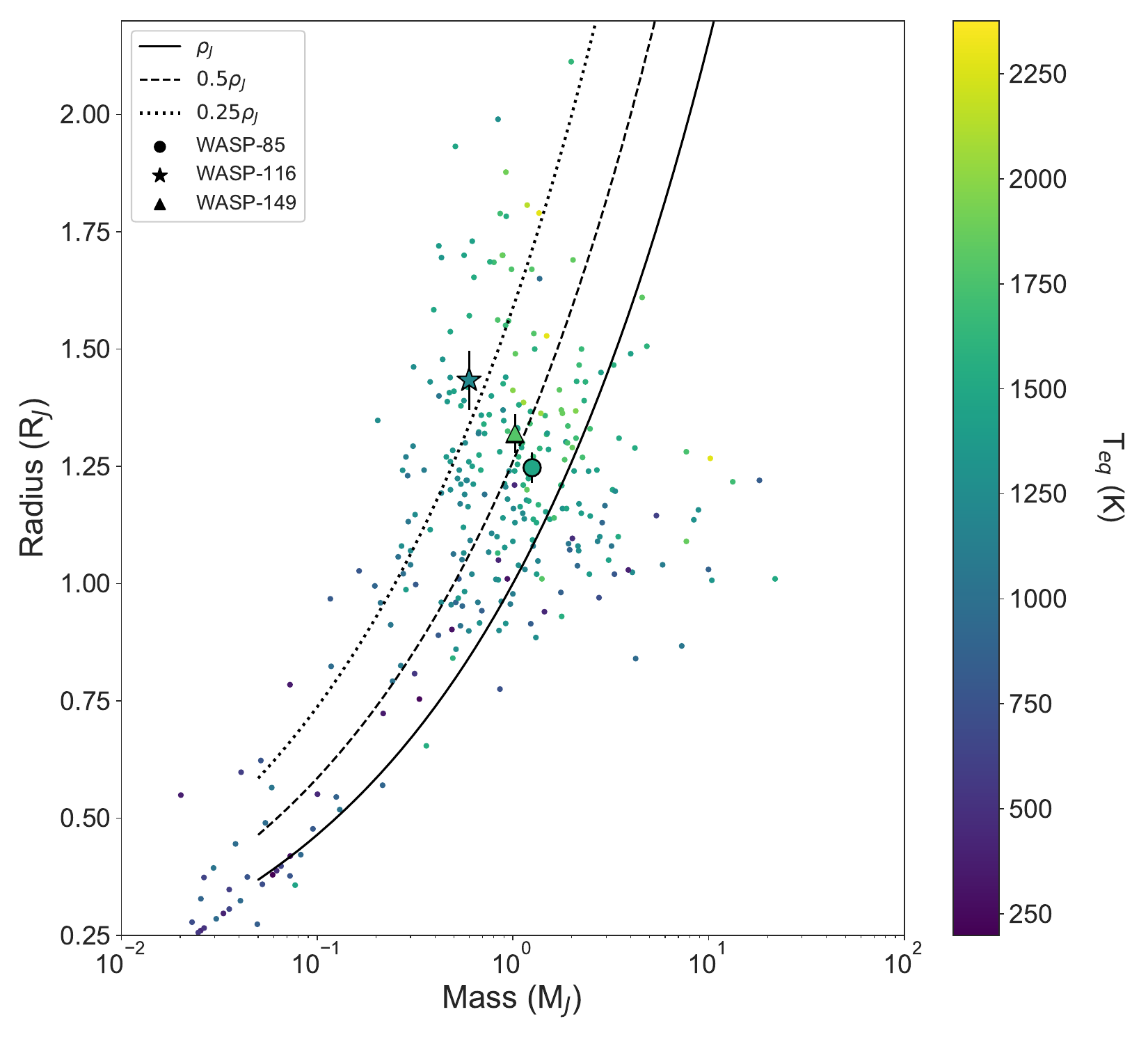}
	\caption{Planetary radius as a function of planetary mass for the existing population of transiting systems. Data are colour-scaled by planetary equilibrium temperature (calculated assuming an albedo of $0.343$ and tidal locking). We also plot isodensity lines for $1.0$, $0.5$, and $0.25$\,$\rho_{\rm Jupiter}$.
	}
	\label{fig:massradius}
\end{figure}

We plot planetary radius as a function of planetary mass in Figure\,\ref{fig:massradius}, focusing on systems with masses greater than $0.01$\,M$_{\rm Jup}$ and radii greater than $0.25$\,R$_{\rm Jup}$. We colour-scale the data for the existing population according to the calculated $T_{\rm eq}$, then overplot our three new systems. We also plot isodensity lines for $1.0$, $0.5$, and $0.25$\,$\rho_{\rm Jupiter}$. Our solution for WASP-116\,b sits just above the $0.25$\,$\rho_{\rm Jupiter}$ isodensity line, suggesting that it is mildly inflated, though we note that the addition of \textit{TESS} photometry covering the full planetary transit may alter this conclusion

Comparing to systems with mass within $1\sigma$ of WASP-116\,b, we find that the planet falls between the 10th and 15th percentiles of the radius distribution. This is skewed by a small number of strongly inflated planets; if we ignore the two systems with radius greater than $1.7$\,$R_{\rm Jup}$ then WASP-116\,b is in the largest $5\%$ of the radius distribution. Similarly comparing to systems with radius within $1\sigma$ of WASP-116\,b reveals that the planet lies between the 25th and 39th percentiles of the mass distribution. Finally, comparing to planets with mass \textit{and} radius within $3\sigma$ of WASP-116\,b places the planet between the 17th and 28th percentiles of the density distribution. The position of WASP-116\,b in mass-radius space compared to objects with similar physical properties thus supports the idea that the planet is mildly inflated.

\begin{figure}
	\includegraphics[width=0.48\textwidth]{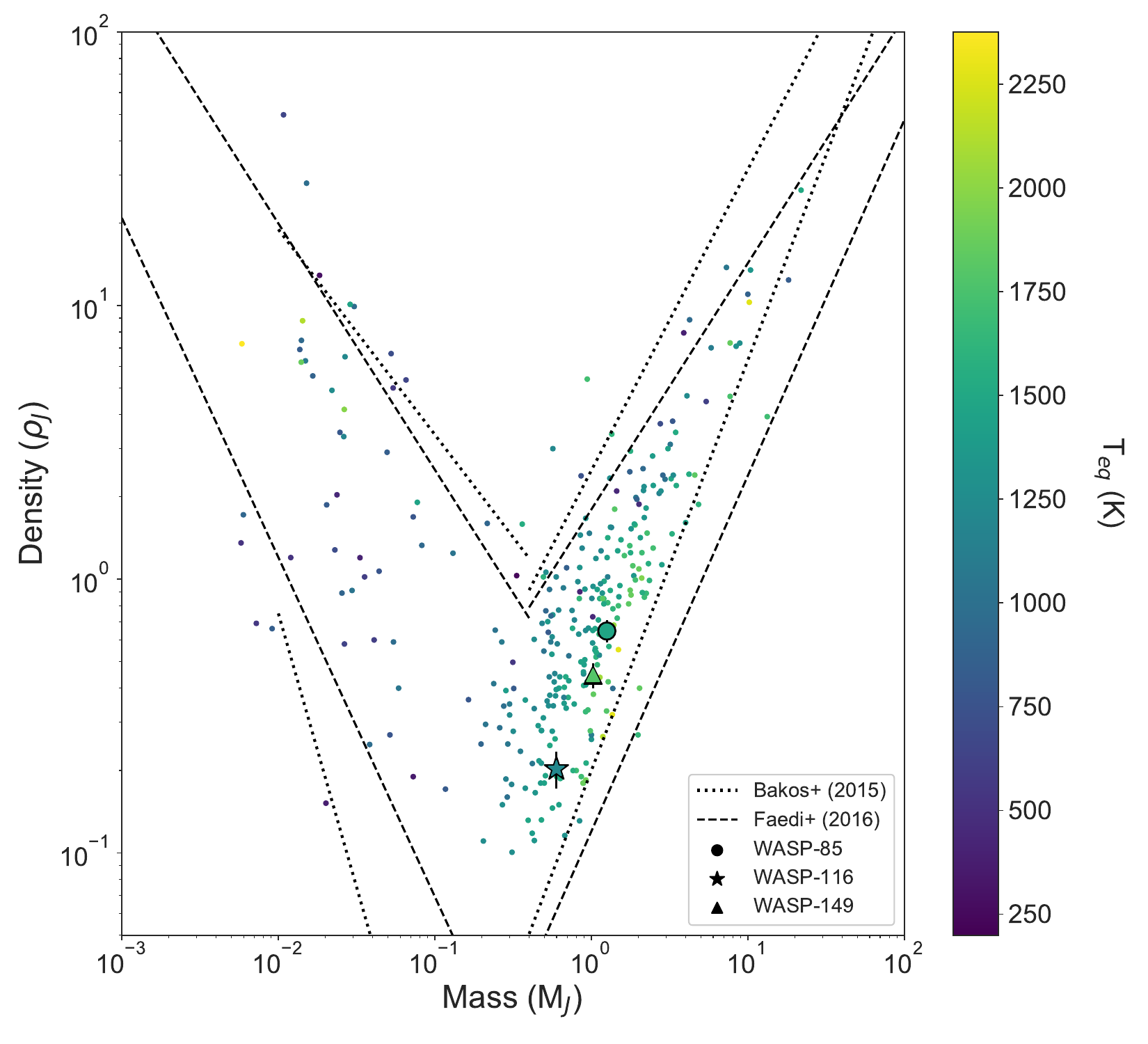}
	\caption{Planetary density as a function of planetary mass for the existing population of transiting systems. Data are colour-scaled by planetary equilibrium temperature (calculated assuming an albedo of $0.343$ and tidal locking). We also plot the envelopes of the distributions for ice / gas giants, and for Neptunes / Saturns; we plot these as defined by both \citet{2015ApJ...813..111B} and \citet{2016arXiv160804225F}.}
	\label{fig:massdensity}
\end{figure}

In Figure\,\ref{fig:massdensity}, we plot planetary density as a function of planetary mass, colour-scaling the data by $T_{\rm eq}$ as before. We also plot envelopes for density-mass distributions in Figure\,\ref{fig:massdensity}. We plot the envelopes for ice / gas giants, and for Neptunes / Saturns, as defined by both \citet{2015ApJ...813..111B} and \citet{2016arXiv160804225F}. These two formulations are subtly different, but both encompass our three new systems within the envelope of the ice / gas giant population.

\begin{figure}
	\includegraphics[width=0.48\textwidth]{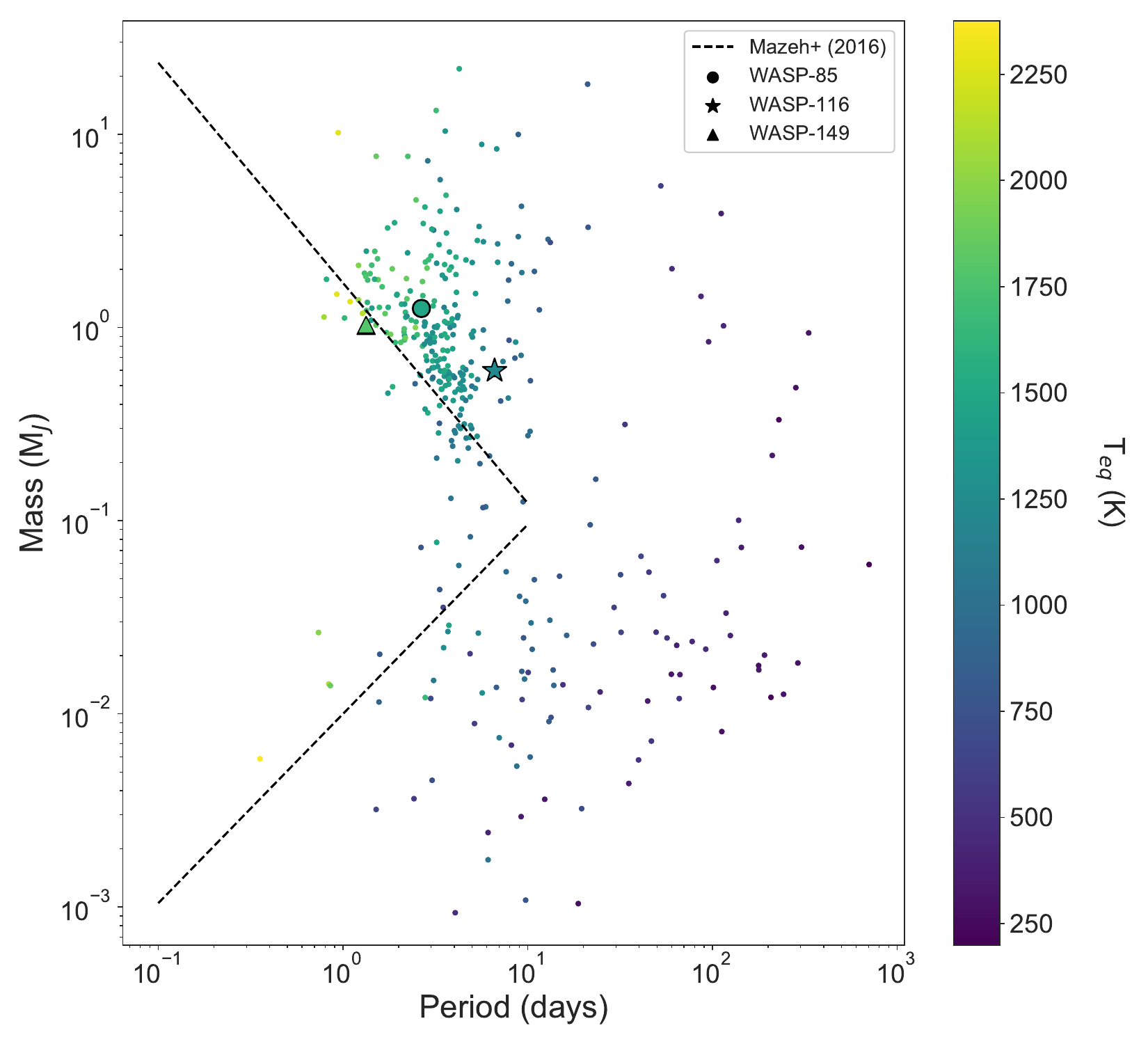}
	\caption{Planetary mass as a function of orbital period for the existing population of transiting systems. Data are colour-scaled by planetary equilibrium temperature (calculated assuming an albedo of $0.343$ and tidal locking). We also plot the envelope of the `Neptune desert' defined by \citet{2016AA...589A..75M}. WASP-149 seems to lie just inside the Neptune desert. Uncertainties on the data for WASP-85, WASP-116, and WASP-149 are smaller than the symbols.}
	\label{fig:massperiod}
\end{figure}

Finally, we plot planetary mass as a function of orbital period in Figure\,\ref{fig:massperiod}, again using $T_{\rm eq}$ to colour-scale the data. We also show the edges of the `Neptune desert' defined by \citet{2016AA...589A..75M}. WASP-149\,b lies within this underpopulated region of parameter space, but this is by no means unique as recent discoveries have slowly eroded the edges of the `desert'.  Using the formulation for the desert edge as defined by \citeauthor{2016AA...589A..75M}, and accounting for both the uncertainty on its location and the uncertainty in our system parameters, we find that WASP-149 lies less than $1\sigma$ within the desert in period space, but just under $4\sigma$ inside the desert in planetary mass space.

Comparing the equilibrium temperatures of our three new planets (see Table\,\ref{tab:results}) to the existing population, we find that WASP-116\,b, and WASP-149\,b are in the bottom $5-15\%$ and the top $90-95\%$, respectively of stars with masses and radii within $3\sigma$ of their measured values. We attribute both of these outlying positions to the orbital periods of the two planets, which are the third longest and third shortest, respectively, of the groups of stars with immediately comparable mass and radius.

\section{Summary}
\label{sec:summary}
In this paper we have presented the discovery of three new transiting exoplanets by the WASP project: WASP-85\,A\,b, WASP-116\,b, and WASP-149\,b. WASP-85\,A\,b is a hot Jupiter orbiting the brighter, solar-type component of a close visual binary, BD+07$^{\circ}$2474, that has an orbital period of $\sim2000-3000$\,years. The binary companion is cooler than the host star, but of similar magnitude, and contaminates both the photometric and spectroscopic data for the system.WASP-116\,b is a warm, mildly inflated super-Saturn orbiting an extremely metal-poor star that seems to be mildly evolved, while WASP-149\,b is a hot Jupiter orbiting a late G-dwarf. 

We broadly recreate the modulation visible in the \textit{K2} lightcurve of WASP-85 using a simple three-spot model comprising two spots on WASP-85\,A, and one spot on WASP-85\,B. The complexity of the detailed spot structure is not captured by our model given the degeneracy of the problem, but we capture the essential features of the spot-induced brightness variations.
From this model we determine rotation periods for the two stars of $13.1\pm0.1$ and $7.5\pm0.03$\,days for WASP-85\,A and B, respectively; the result for the primary star is in strong agreement with previously published results, and our own estimates from modulation found in the WASP photometry. 

Using the output of our spot model, we estimate stellar inclinations of $I_{\rm A}=66.8^o\pm0.7$ and $I_{\rm B}=39.7^o\pm0.2$, and constrain the obliquity of WASP-85\,A\,b to be $\psi<27^o$. We therefore conclude that the system is very likely to be aligned.

We carried out an MCMC analysis of the full set of photometric and spectroscopic data for our three systems to determine their orbital and physical characteristics, determined stellar parameters through spectral analysis, and estimated the ages for the systems by fitting their parameters to sets of stellar models. We place our systems in the context of the existing population of planets with both mass and radius measurements, finding that WASP-149\,b lies within the `Neptune desert'.


\acknowledgments

The WASP Consortium consists of representatives from the Universities of Keele, Leicester, The Open University, Queens University Belfast and St Andrews, along with the Isaac Newton Group (La Palma) and the Instituto de Astrof\'isica de Canarias (Tenerife). WASP-South is hosted by the SAAO and SuperWASP by the Isaac Newton Group and the Instituto de Astrof\'isica de Canarias; we gratefully acknowledge their ongoing support and assistance. The SuperWASP and WASP-South cameras are operated with funds made available from Consortium Universities and the STFC. TRAPPIST is funded by the Belgian Fund for Scientific Research (Fond National de la Recherche Scientifique, FNRS) under the grant FRFC 2.5.594.09.F, with the participation of the Swiss National Science Foundation (SNF). The Liverpool Telescope is operated on the island of La Palma by Liverpool John Moores University in the Spanish Observatorio del Roque de los Muchachos of the Instituto de Astrof\'isica de Canarias with financial support from the UK Science and Technology Facilities Council. D.J.A.B and D.J.A. acknowledge funding from the European Union Seventh Framework programme (FP7/2007- 2013) under grant agreement No. 313014 (ETAEARTH). D.J.A.B acknowledges funding from the UK Space Agency and the University of Warwick. M.G. and E.J. are FRS-FNRS Senior Research Associates. L.D. acknowledges support from the Gruber Foundation Fellowship. A.H.M.J.T is a Swiss National Science Foundation fellow under grant number P300P2-147773. K.W.F.L. acknowledges the support of the DFG priority program SPP 1992 ``Exploring the Diversity of Exoplanets in the Mass-Density Diagram'' (RA 714/14-1). The research leading to these results has received funding from the ARC grant for Concerted Research Actions, financed by the Wallonia-Brussels Federation. This research has made use of NASA's Astrophysics Data System Bibliographic Services, the ArXiv preprint service hosted by Cornell University, the Washington Double Star Catalog maintained at the U.S. Naval Observatory, the VizieR catalogue access tool, CDS, Strasbourg, France, and Astropy,\footnote{http://www.astropy.org} a community-developed core Python package for Astronomy \citep{astropy:2013, astropy:2018}. The original description of the VizieR service was published in A\&AS 143, 23. This work uses observations collected with the SOPHIE spectrograph on the 1.93-m telescope at Observatoire de Haute-Provence (CNRS), France. This work has made use of data from the European Space Agency (ESA) mission {\it Gaia} (\url{https://www.cosmos.esa.int/gaia}), processed by the {\it Gaia} Data Processing and Analysis Consortium (DPAC, \url{https://www.cosmos.esa.int/web/gaia/dpac/consortium}). Funding for the DPAC has been provided by national institutions, in particular the institutions participating in the {\it Gaia} Multilateral Agreement.

%

\vspace{5mm}
\facilities{SuperWASP, WASP-South, Euler1.2, Liverpool:2m, ESO:3.6m, OHP:1.93m, TRAPPIST:0.6m
            }


\software{astropy  
          }

\end{document}